%% file: nuc.tex
\documentclass[twoside,fleqn,epj,numbook]{svjour}
\usepackage{ifthen}                   
\usepackage{exscale}                  
\usepackage[intlimits]{amsmath}       
\usepackage{amsfonts}
\usepackage{amssymb,amscd}
\usepackage[dvips]{epsfig}                   
\usepackage{array}
\usepackage{wrapfig}
\input{newcommands.tex}

\input{FTeX/FEYNMAN.tex}

\begin{document}

\psfull

\include{TitlePage}

\pagenumbering{arabic}

\input Sec1

\input Sec2

\input Sec3

\input Sec4

\input Sec5

\input Sec6

\input Sec7

\input Sec8

\input Acknowledgement.tex


\begin{appendix}
\input Appendix0

\input Appendix

\newpage 

\input References
\end{document}

%% file: newcommands.tex
\def\Quadrat#1#2{{\vcenter{\hrule height #2
  \hbox{\vrule width #2 height #1 \kern#1
    \vrule width #2}
  \hrule height #2}}}
\def\dAlemb{\mathop{\kern 1pt\hbox{$\Quadrat{8pt}{0.4pt}$} \kern1pt}}
\def\dAlember{\mathop{\kern 1pt\hbox{$\Quadrat{4pt}{0.4pt}$} \kern1pt}}
\def\Chi{{\mathop{\kern 2pt\vcenter{\hbox{$\chi $}}\kern2pt}}}
\def\dslash{\partial\kern-.5em\slash}
\def\kslash{k\kern-.5em\slash}
\def\slh#1{#1\kern-.5em\slash}
\def\fslash#1{#1 \kern-.5em\slash}

\def\fbar#1{#1\kern-.5em\raise6pt\hbox{\footnotesize /}}

\newcommand{\lsim}{\mathrel{\rlap{\lower4pt\hbox{\hskip1pt$\sim$}}
\raise1pt\hbox{$<$}}}

\newcommand{\nn}{\nonumber}        
\newcommand{\tr}{\hbox{tr}}

\newcommand{\be}{\begin{eqnarray}}
\newcommand{\ee}{\end{eqnarray}}

\newcommand{\fourint}[1]{\int\!\frac{d^4 #1}{(2\pi)^4}}

%% file: FTeX/FEYNMAN.tex
\gdef\Feynmanlength{\setlength{\unitlength}{0.01pt}}  

\global\newcount\LINETYPE                     
\global\newcount\LINEDIRECTION
\global\newcount\LINECONFIGURATION
\newcommand{\LTYPE}{\LINETYPE}
\newcommand{\LDIR}{\LINEDIRECTION}
\newcommand{\LCONFIG}{\LINECONFIGURATION}
\global\LINETYPE=1  \global\LINEDIRECTION=0  \global\LINECONFIGURATION=0
\global\newcount\fermion    \fermion=1
\global\newcount\scalar     \scalar=2
\global\newcount\photon     \photon=3
\global\newcount\gluon      \gluon=4
\global\newcount\especial   \especial=5
\gdef\N{0}  \gdef\NE{1}  \gdef\E{2}   \gdef\SE{3}
\gdef\S{4}  \gdef\SW{5}  \gdef\W{6}   \gdef\NW{7}
\global\newcount\REG            \global\REG=0
\global\newcount\FLIPPED        \global\FLIPPED=1
\global\newcount\CURLY          \global\CURLY=2
\global\newcount\FLIPPEDCURLY   \global\FLIPPEDCURLY=3
\global\newcount\FLAT           \global\FLAT=4
\global\newcount\FLIPPEDFLAT    \global\FLIPPEDFLAT=5
\global\newcount\CENTRAL        \global\CENTRAL=6
\global\newcount\FLIPPEDCENTRAL \global\FLIPPEDCENTRAL=7
             
\global\newcount\SQUASHEDGLUON  \global\SQUASHEDGLUON=8

\newcount\adjx \adjx=0
\newcount\adjy \adjy=0
\global\newdimen\BIGPHOTONS     \BIGPHOTONS=0pt  
\global\newdimen\THICKPHOTONS     \THICKPHOTONS=0pt  
\global\newdimen\THICKPHOTONSWITCH    \THICKPHOTONSWITCH=0pt
\gdef\THICKPHOTONTEST{
\THICKPHOTONSWITCH=0pt
\ifdim\THICKPHOTONS=0pt \relax
  \else \ifnum\LTYPE=3
           \ifnum\LDIR=2 \THICKPHOTONSWITCH=1pt \fi 
           \ifnum\LDIR=6 \THICKPHOTONSWITCH=1pt \fi 
        \fi
\fi
}  

\global\newcount\phantomswitch   \global\phantomswitch=0
\global\newcount\stemlength   \global\stemlength=275   
\global\newcount\absstemlength        
\global\newcount\stemlengthx          
\global\newcount\stemlengthy          
\newdimen\FRONTSTEM  \FRONTSTEM=0pt   
\newdimen\BACKSTEM   \BACKSTEM=0pt    
\newdimen\EITHERSTEM \EITHERSTEM=0pt  
\gdef\backstemmed{\BACKSTEM=1pt}              
\global\newcount\arrowlength                
\global\newdimen\ATTIP   \global\ATTIP=0pt  
\global\newdimen\ATBASE  \global\ATBASE=1pt 
\global\newcount\unitboxnumber  
\global\newcount\unitboxnumberpo  
\global\newcount\particlelengthx  
\gdef\plengthx{\particlelengthx}
\global\newcount\particlelengthy  
\gdef\plengthy{\particlelengthy}  
\global\newcount\boxlengthx  
\global\newcount\boxlengthy  
\global\newcount\particleadjustx  
\global\newcount\particleadjusty  
\global\newcount\particlelength   
\global\newcount\particlefrontx
\gdef\pfrontx{\particlefrontx}
\global\newcount\PFRONTx
\global\newcount\particlefronty
\gdef\pfronty{\particlefronty}
\global\newcount\PFRONTy
\global\newcount\particlebackx
\gdef\pbackx{\particlebackx}
\global\newcount\particlebacky
\gdef\pbacky{\particlebacky}
\global\newcount\particlemidx
\gdef\pmidx{\particlemidx}
\global\newcount\particlemidy
\gdef\pmidy{\particlemidy}
\global\newcount\seglength  \global\newcount\gaplength
\global\gaplength=850  
\global\seglength=1416  
\global\newcount\Xone    \global\newcount\Yone    
\global\newcount\Xtwo    \global\newcount\Ytwo    
\global\newcount\Xthree  \global\newcount\Ythree  
\global\newcount\Xfour   \global\newcount\Yfour   
\global\newcount\Xfive   \global\newcount\Yfive   
\global\newcount\Xsix    \global\newcount\Ysix    
\global\newcount\Xseven  \global\newcount\Yseven  
\global\newcount\Xeight  \global\newcount\Yeight  
\newsavebox{\lastline}  
\global\newcount\numlineparts   
\global\newcount\upperlineadjx  \upperlineadjx=0  
\global\newcount\upperlineadjy  \upperlineadjy=0  
\global\newcount\lowerlineadjx  \lowerlineadjx=0  
\global\newcount\lowerlineadjy  \lowerlineadjy=0  
\global\newcount\thirdlineadjx  \thirdlineadjx=0  
\global\newcount\thirdlineadjy  \thirdlineadjy=0  
\global\newcount\fourthlineadjx \fourthlineadjx=0  
\global\newcount\fourthlineadjy \fourthlineadjy=0  
\global\newcount\unitboxwidth   \unitboxwidth=1000
\global\newcount\unitboxheight  \unitboxheight=0  
\global\newcount\numupperunits  \numupperunits=8  
\global\newcount\numlowerunits  \numlowerunits=8  
\global\newcount\numthirdunits  \numthirdunits=8  
\global\newcount\numfourthunits \numfourthunits=8  
\global\newcount\fermioncount   \global\fermioncount=0    
\global\newcount\scalarcount    \global\scalarcount=0    
\global\newcount\photoncount    \global\photoncount=0    
\global\newcount\gluoncount     \global\gluoncount=0    
\global\newcount\especialcount  \global\especialcount=0    
\global\newcount\vertexcount    \global\vertexcount=-1
\global\newcount\XDIR
\global\newcount\YDIR
\gdef\SETDIR{  
\ifcase\LDIR 
     \global\XDIR=0  \global\YDIR=1   
\or  \global\XDIR=1  \global\YDIR=1   
\or  \global\XDIR=1  \global\YDIR=0   
\or  \global\XDIR=1  \global\YDIR=-1  
\or  \global\XDIR=0  \global\YDIR=-1  
\or  \global\XDIR=-1 \global\YDIR=-1  
\or  \global\XDIR=-1 \global\YDIR=0   
\or  \global\XDIR=-1 \global\YDIR=1   
\else\DIRECTERROR 
\fi}  
\gdef\moduloeight#1{
\ifnum#1>7 \global\advance #1 by -8 
\relax
\moduloeight#1 
\relax
\else \relax  
\fi}
\gdef\multroothalf#1{\global\multiply #1 by 7071 \global\divide #1 by 10000}
\gdef\negate#1{\global\multiply #1 by -1}
\gdef\double#1{\global\multiply #1 by 2}
\gdef\slanttest(#1,#2){ 
\ifodd\LDIR
\multiply #1 by 7071  \divide #1 by 10000
\multiply #2 by 7071  \divide #2 by 10000
\fi
}
\gdef\gslanttest(#1,#2){
\ifodd\LDIR
\multroothalf#1
\multroothalf#2
\fi
}
\gdef\setplength{ 
\global\particlelengthx=\unitboxwidth
\global\particlelengthy=\unitboxheight
\global\multiply \particlelengthx by \unitboxnumber
\global\multiply \particlelengthy by \unitboxnumber
\global\advance \particlelengthx by \particleadjustx
\global\advance \particlelengthy by \particleadjusty
}
\gdef\boxlengthdefault{  
\global\boxlengthx=\plengthx
\global\boxlengthy=\plengthy
\ifnum\plengthx<0 \global\multiply\boxlengthx by -1 \fi
\ifnum\plengthy<0 \global\multiply\boxlengthy by -1 \fi
}
\gdef\rearcoords{  
\global\particlebacky=\particlefronty 
\global\particlebackx=\particlefrontx 
\global\advance \particlebackx by \particlelengthx
\global\advance \particlebacky by \particlelengthy
}
\gdef\midcoords{  
\global\particlemidy=\particlefronty
\global\particlemidx=\particlefrontx
\global\stemlengthx=\particlelengthx  
\global\stemlengthy=\particlelengthy  
\global\divide\stemlengthx by 2
\global\divide\stemlengthy by 2
\global\advance \particlemidx by \stemlengthx
\global\advance \particlemidy by \stemlengthy
}
\gdef\setparticle{\setplength\rearcoords\midcoords\boxlengthdefault}  
\gdef\setcoords(#1,#2,#3)(#4,#5,#6)[#7,#8]{  
\global\upperlineadjx=#1
\global\lowerlineadjx=#2
\global\thirdlineadjx=#3
\global\upperlineadjy=#4
\global\lowerlineadjy=#5
\global\thirdlineadjy=#6
\global\unitboxwidth=#7
\global\unitboxheight=#8
}
\gdef\drawoldpic#1(#2,#3){  
\global\particlefrontx=#2
\global\particlefronty=#3
\rearcoords  
\midcoords
\put(#2,#3){\usebox{#1}}
}
\gdef\drawsavedline`#1' as #2[#3#4](#5,#6)[#7]{
\global\LINETYPE=#2
\global\LINEDIRECTION=#3
\global\LINECONFIGURATION=#4
\global\particlefrontx=#5
\global\particlefronty=#6
\global\unitboxnumber=#7  
\selectcase
\rearcoords
\midcoords
\ifnum\phantomswitch=0 \drawas{#1}\fi
}

\gdef\drawas#1{
\global\savebox{#1}(\boxlengthx,\boxlengthy){
\setlength{\unitlength}{0.01pt}
\begin{picture}(\boxlengthx,\boxlengthy)
\multiput(\upperlineadjx,\upperlineadjy)(\unitboxwidth,\unitboxheight)
{\numupperunits}{\upperunitbox}
\ifnum\numlineparts > 1  
\multiput(\lowerlineadjx,\lowerlineadjy)(\unitboxwidth,\unitboxheight)
{\numlowerunits}{\lowerunitbox}  
\fi
\ifnum\numlineparts > 2  
\multiput(\thirdlineadjx,\thirdlineadjy)(\unitboxwidth,\unitboxheight)
{\numthirdunits}{\thirdunitbox}  
\fi
\ifnum\numlineparts > 3  
\multiput(\fourthlineadjx,\fourthlineadjy)(\unitboxwidth,\unitboxheight)
{\numfourthunits}{\lowerunitbox}  
\fi
\end{picture} }
\global\PFRONTx=\pfrontx  \global\PFRONTy=\pfronty   
\SETFRONTSTEM
\THICKPHOTONTEST
\ifdim\THICKPHOTONSWITCH=1pt\global\advance\PFRONTy by 20  \fi
\put(\PFRONTx,\PFRONTy) {\usebox{#1}}   
\ifdim\THICKPHOTONSWITCH=1pt
\global\advance\PFRONTy by -40
\put(\PFRONTx,\PFRONTy) {\usebox{#1}}   
\global\advance \PFRONTy by 20  
\fi  
\SETBACKSTEM
\seglength=1416   \gaplength=850   
}
\gdef\drawandsaveline`#1' as #2[#3#4](#5,#6)[#7]{
\global\newsavebox{#1}
\drawsavedline`#1' as #2[#3#4](#5,#6)[#7]
}

\gdef\drawline#1[#2#3](#4,#5)[#6]{   
\drawsavedline`\lastline' as #1[#2#3](#4,#5)[#6]}

\gdef\TYPEERROR{\message{*** ERROR IN PARTICLE TYPE SELECTION ***}
\message{+++ Try with line type \fermion,\scalar,\photon,\gluon 
(see manual) +++}\SETERR}
\gdef\DIRECTERROR{\SETERR\message{*** ERROR IN PARTICLE DIRECTION SELECTION ***}
\message{+++ Try again with direction N, NE, E, SE  etc. or see manual +++}}
\gdef\UNIMPERROR{\message{*** ERROR IN PARTICLE OPTIONS SELECTION ***}
\message{
+++ The requested options combination has not yet been implemented +++}\SETERR}
\gdef\SETERR{\gdef\upperunitbox{{\tiny Error}}  
\gdef\lowerunitbox{\relax}
\gdef\thirdunitbox{\relax}
}
\gdef\neglengthcheck{\ifnum\unitboxnumber < 1 
\message{   *** ERROR:  PARTICLE OF NEGATIVE OR ZERO LENGTH REQUESTED. ***   }
\message{   ***         TAKING ABSOLUTE VALUE. ***   }\negate\unitboxnumber \fi}
\gdef\selectcase{  
\neglengthcheck   
\SETDIR  
\ifcase\LINETYPE
\TYPEERROR  
\or \selectfermion  
\or \selectscalar   
\or \selectphoton   
\or \selectgluon    
\or \selectespecial 
\else \TYPEERROR \fi  }
\gdef\selectfermion{
\ifnum\fermioncount=0 \input FTeX/FERMIONSETUP \fi   
\global\advance\fermioncount by 1  
\ALLfermion   
}
\gdef\selectscalar{
\ifnum\scalarcount=0 \input FTeX/SCALARSETUP \fi   
\global\advance\scalarcount by 1  
\ALLscalar
}
\gdef\selectphoton{   
\ifnum\photoncount=0 \input FTeX/PHOTONSETUP  \fi   
\selectphoton
}
\gdef\selectgluon{   
\ifnum\gluoncount=0 \input FTeX/GLUONSETUP  \fi
\selectgluon
}
\gdef\selectespecial{\UNIMPERROR}
\gdef\checkvertex{ 
\ifnum\vertexcount=-1   \input FTeX/VERTEX  \fi}
\gdef\drawvertex#1[#2#3](#4,#5)[#6]{\checkvertex\drawvertex#1[#2#3](#4,#5)[#6]}
\gdef\vertexcap#1{\checkvertex\vertexcap#1}
\gdef\vertexcaps{\checkvertex\vertexcaps}
\gdef\vertexlink#1{\checkvertex\vertexlink#1}
\gdef\vertexlinks{\checkvertex\vertexlinks}
\gdef\stemvertex#1{\checkvertex\stemvertex#1}
\gdef\stemvertices{\checkvertex\stemvertices}
\gdef\flipvertex{\checkvertex\flipvertex}
\global\arrowlength=349  
\gdef\drawarrow[#1#2](#3,#4){
\global\LDIR=#1
\SETDIR
\global\boxlengthx=#3  
\global\boxlengthy=#4  
\ifdim#2=1pt  
\adjx=\arrowlength      \adjy=\arrowlength
\multiply\adjx by \XDIR \multiply\adjy by \YDIR  
\slanttest(\adjx,\adjy)
\global\advance\boxlengthx by \adjx    \global\advance\boxlengthy by \adjy
\fi
\ifnum\phantomswitch=0\put(\boxlengthx,\boxlengthy){\vector(\XDIR,\YDIR){0}}\fi
}  
\gdef\SETFRONTSTEM{
\EITHERSTEM=\FRONTSTEM   \advance\EITHERSTEM by \BACKSTEM
\ifdim\EITHERSTEM>0pt
\global\stemlengthx=\stemlength   \global\stemlengthy=\stemlength   
\global\absstemlength=\stemlength   
\SETDIR
\gslanttest(\stemlengthx,\stemlengthy)
\gslanttest(\absstemlength,\REG)  
\ifnum\XDIR=0 \stemlengthx=0 \fi
\ifnum\YDIR=0 \stemlengthy=0 \fi
\global\multiply\stemlengthx by \XDIR
\global\multiply\stemlengthy by \YDIR
\ifdim\FRONTSTEM=1pt 
\ifnum\phantomswitch=0
          \put(\pfrontx,\pfronty){\line(\XDIR,\YDIR){\absstemlength}}\fi
\global\advance\plengthx by \stemlengthx
\global\advance\plengthy by \stemlengthy
\global\advance\PFRONTx by \stemlengthx   
\global\advance\PFRONTy by \stemlengthy
\global\advance\pmidx by \stemlengthx
\global\advance\pmidy by \stemlengthy
\global\advance\pbackx by \stemlengthx
\global\advance\pbacky by \stemlengthy
\ifnum\LTYPE=3
\global\photonfrontx=\PFRONTx  \global\photonfronty=\PFRONTy
\global\photonbackx=\pbackx    \global\photonbacky=\pbacky
\fi  
\ifnum\LTYPE=4
\global\gluonfrontx=\PFRONTx  \global\gluonfronty=\PFRONTy
\global\gluonbackx=\pbackx    \global\gluonbacky=\pbacky
\fi  
\fi  
\fi  
}    
\gdef\SETBACKSTEM{
\ifdim\BACKSTEM=1pt 
\ifnum\phantomswitch=0
       \put(\pbackx,\pbacky){\line(\XDIR,\YDIR){\absstemlength}}\fi
\global\advance\plengthx by \stemlengthx
\global\advance\plengthy by \stemlengthy
\global\advance\pbackx by \stemlengthx
\global\advance\pbacky by \stemlengthy
\fi  
\global\stemlength=275  \FRONTSTEM=0pt  \BACKSTEM=0pt 
}    
\gdef\drawloop#1[#2#3](#4,#5){  
\input FTeX/LOOPS  
\drawloop#1[#2#3](#4,#5)}
\Feynmanlength  

%% file: FTeX/FERMIONSETUP.tex
\global\newcount\fermionlength  
\global\newcount\fermionlengthx
\global\newcount\fermionlengthy
\global\newcount\fermionfrontx  
\global\newcount\fermionfronty  
\global\newcount\fermionbackx
\global\newcount\fermionbacky
\gdef\ALLfermion{  
\global\fermionfrontx=\particlefrontx \global\fermionfronty=\particlefronty
\ifnum\unitboxnumber > 50000
\message{   *** WARNING *** Fermion of length
\the\unitboxnumber\space requested ***   }
\ifnum\unitboxnumber > 80000
\message{   *** Reducing fermion length to 30000 (max 80000) ***   }
\global\unitboxnumber=30000 \fi \fi  
\global\fermionlength=\unitboxnumber 
\global\particleadjustx=0   \global\particleadjusty=0 
\global\numlineparts = 1    \global\numupperunits=1
\global\upperlineadjx=-200  \global\upperlineadjy=0
\global\fermionlengthx=\fermionlength    \global\fermionlengthy=\fermionlength
\gslanttest(\fermionlengthx,\fermionlengthy)  
\global\multiply\fermionlengthx by \XDIR  
\global\multiply\fermionlengthy by \YDIR  
\global\unitboxheight=\fermionlengthy   \global\unitboxwidth=\fermionlengthx   
\global\advance \fermionlengthx by \particleadjustx
\global\advance \fermionlengthy by \particleadjusty
\global\particlelengthx=\fermionlengthx
\global\particlelengthy=\fermionlengthy  
\boxlengthdefault    \rearcoords    \midcoords
\global\fermionbackx=\particlebackx     \global\fermionbacky=\particlebacky
\ifcase\LINECONFIGURATION  
\ifnum\XDIR=0 
\gdef\upperunitbox{\line(\XDIR,\YDIR){\boxlengthy}} 
\else
\gdef\upperunitbox{\line(\XDIR,\YDIR){\boxlengthx}}
\fi
\else \UNIMPERROR
\fi
}

%% file: FTeX/SCALARSETUP.tex
\newcount\scalarlength
\newcount\scalarlengthx
\newcount\scalarlengthy
\newcount\scalarfrontx  
\newcount\scalarfronty  
\newcount\scalarbackx
\newcount\scalarbacky
\gdef\ALLscalar{
\global\scalarfrontx=\particlefrontx   
\global\scalarfronty=\particlefronty   
\numlineparts = 1      \numupperunits=\unitboxnumber
\ifcase\LINECONFIGURATION
\global\upperlineadjx=-200     \global\upperlineadjy=0 
\slanttest(\seglength,\gaplength)   
\gdef\upperunitbox{\line(\XDIR,\YDIR){\seglength}}
\else \UNIMPERROR 
\fi
\global\unitboxwidth=\seglength  \global\advance\unitboxwidth by \gaplength
\global\multiply \unitboxwidth by \XDIR
\global\unitboxheight=\seglength  \global\advance\unitboxheight by \gaplength
\global\multiply \unitboxheight by \YDIR
\global\particleadjustx=\gaplength \global\multiply\particleadjustx by \XDIR 
\global\particleadjusty=\gaplength \global\multiply\particleadjusty by \YDIR
\negate\particleadjustx   \negate\particleadjusty   
\setparticle  
\global\scalarlengthx=\particlelengthx  
\global\scalarlengthy=\particlelengthy  
\ifnum\boxlengthx > 50000
\message{   *** WARNING *** Scalar of length in excess of 50000cp requested!}\fi
\ifnum\boxlengthy > 50000
\message{   *** WARNING *** Scalar of length in excess of 50000cp requested!}\fi
\global\scalarbackx=\pbackx      \global\scalarbacky=\pbacky   
}

%% file: FTeX/PHOTONSETUP.tex
\newcount\numwiggles    \newcount\numwigglespo
\global\newcount\photonlengthx
\global\newcount\photonlengthy
\global\newcount\photonfrontx  
\global\newcount\photonfronty  
\global\newcount\photonbackx
\global\newcount\photonbacky
\newcount\halfwigglelength
\global\font\Twelverom=cmr12
\global\font\Tenrom=cmr10
\gdef\Lbr{{\Twelverom(}}   \gdef\Rbr{{\Twelverom)}}
\gdef\SLbr{{\Tenrom(}}     \gdef\SRbr{{\Tenrom)}}
\gdef\Smile{{\large$\smile$}}  
\gdef\Frown{{\large$\frown$}}  
\ifdim\BIGPHOTONS>0pt  \gdef\Smile{$\smile$} \gdef\Frown{$\frown$} \fi
%
\gdef\selectphoton{   
\global\advance\photoncount by 1  
\global\photonfrontx=\particlefrontx   
\global\photonfronty=\particlefronty   
\ifnum\unitboxnumber > 50
\message{   *** WARNING *** Photon with 
\the\unitboxnumber\space half-wiggles requested ***   }
\ifnum\unitboxnumber > 150
\message{   *** Reducing photon length to 10 half-wiggles (max 150) ***   }
\ifnum\unitboxnumber > 1000
\message{   *** Probable Cause:  Photon selected instead of Fermion ***   }
\fi \global\unitboxnumber=10 \fi \fi  
\numwiggles=\unitboxnumber
\divide\numwiggles by 2
\global\unitboxnumberpo=\numwiggles 
\global\multiply \unitboxnumberpo by -1
\numwigglespo=\unitboxnumber
\advance\numwigglespo by \unitboxnumberpo 
\global\numlineparts = 2  
\global\numupperunits=\numwigglespo  
\global\numlowerunits=\numwiggles  
\particleadjustx=0  
\particleadjusty=0  
\ifcase\LINEDIRECTION
     \Nphoton    
\or  \NEphoton   
\or  \Ephoton    
\or  \SEphoton   
\or  \Sphoton    
\or  \SWphoton   
\or  \Wphoton    
\or  \NWphoton   
\else\DIRECTERROR \fi
\setplength
\global\divide\plengthx by 2  \global\divide\plengthy by 2
\rearcoords  \boxlengthdefault   \midcoords
\global\photonbackx=\pbackx  
\global\photonbacky=\pbacky  
\global\photonlengthx=\plengthx  
\global\photonlengthy=\plengthy  
}
\gdef\SETUNITBOX(#1)[#2][#3]{ 
\gdef\upperunitbox{\oval(#1,#1)[#2]}
\gdef\lowerunitbox{\oval(#1,#1)[#3]}
}
\gdef\Nphoton{  
\ifcase\LINECONFIGURATION  
\setcoords(-490,-250,0)(260,1250,0)[0,2000]
\gdef\upperunitbox{\SLbr}   \gdef\lowerunitbox{\SRbr}
\particleadjusty=10
\or 
\setcoords(-271,-501,0)(250,1250,0)[0,2000]   
\gdef\upperunitbox{\SRbr}   \gdef\lowerunitbox{\SLbr}
\or 
\particleadjusty=0
\setcoords(-501,-351,0)(300,1400,0)[0,2200]
\gdef\upperunitbox{\Lbr}   \gdef\lowerunitbox{\Rbr}
\or 
\setcoords(-353,-499,0)(300,1400,0)[0,2200]
\gdef\upperunitbox{\Rbr}   \gdef\lowerunitbox{\Lbr}
\or 
\setcoords(-481,-371,0)(280,1300,0)[0,2000]
\gdef\upperunitbox{\Lbr}   \gdef\lowerunitbox{\Rbr}
\particleadjusty=150
\ifnum\numwiggles=\number\numwigglespo \particleadjustx=-50 \fi
\or 
\setcoords(-321,-391,0)(280,1300,0)[0,2000]
\gdef\upperunitbox{\Rbr}   \gdef\lowerunitbox{\Lbr}
\particleadjusty=150
\ifnum\numwiggles=\number\numwigglespo \particleadjustx=80 \fi
\or 
\setcoords(-490,-260,0)(300,1500,0)[0,2400]
\gdef\upperunitbox{\Lbr}   \gdef\lowerunitbox{\Rbr}
\or 
\setcoords(-301,-531,0)(300,1500,0)[0,2400]
\gdef\upperunitbox{\Rbr}   \gdef\lowerunitbox{\Lbr}
\else \UNIMPERROR
\fi
}
\gdef\NEphoton{    
\ifcase\LINECONFIGURATION  
\setcoords(425,425,0)(1250,0,0)[1250,1250]       \SETUNITBOX(1250)[br][tl]  
\ifnum\numwigglespo > \number \numwiggles \particleadjustx=15 \fi
\or 
\setcoords(1050,-200,0)(625,625,0)[1250,1250]    \SETUNITBOX(1250)[tl][br]
\ifnum\numwigglespo > \number \numwiggles \particleadjustx=25 \fi
\or 
\setcoords(500,500,0)(1400,0,0)[1400,1400]       \SETUNITBOX(1400)[br][tl]
\or 
\setcoords(1200,-200,0)(700,700,0)[1400,1400]    \SETUNITBOX(1400)[tl][br]  
\or 
\setcoords(400,400,0)(1200,0,0)[1200,1200]       \SETUNITBOX(1200)[br][tl]  
\or 
\setcoords(1000,-200,0)(600,600,0)[1200,1200]    \SETUNITBOX(1200)[tl][br]
\else \UNIMPERROR
\fi
\numupperunits=\numwiggles   \numlowerunits=\numwigglespo
}
\gdef\Ephoton{    
\ifcase\LINECONFIGURATION  
\setcoords(-285,715,0)(-150,-400,0)[2005,0]
\gdef\upperunitbox{\Frown}   \gdef\lowerunitbox{\Smile}
\or  
\setcoords(-285,715,0)(-420,-170,0)[2005,0]
\gdef\upperunitbox{\Smile}   \gdef\lowerunitbox{\Frown}
\else \UNIMPERROR
\fi
\particleadjustx=-15 
}
\gdef\SEphoton{   
\ifcase\LINECONFIGURATION  
\setcoords(-200,1050,0)(-625,-625,0)[1250,-1250] \SETUNITBOX(1250)[tr][bl]
\ifnum\numwigglespo > \number \numwiggles \particleadjustx=25 \fi
\or 
\setcoords(425,425,0)(0,-1250,0)[1250,-1250]     \SETUNITBOX(1250)[bl][tr]
\ifnum\numwigglespo > \number \numwiggles \particleadjustx=15 \fi
\or 
\setcoords(-200,1200,0)(-700,-700,0)[1400,-1400] \SETUNITBOX(1400)[tr][bl]  
\or 
\setcoords(500,500,0)(0,-1400,0)[1400,-1400]     \SETUNITBOX(1400)[bl][tr]  
\or 
\setcoords(-200,1000,0)(-600,-600,0)[1200,-1200] \SETUNITBOX(1200)[tr][bl]
\particleadjustx=-20
\or 
\setcoords(420,420,0)(0,-1200,0)[1200,-1200]     \SETUNITBOX(1200)[bl][tr]
\particleadjustx=40
\else \UNIMPERROR
\fi
}
\gdef\Sphoton{  
\ifcase\LINECONFIGURATION  
\setcoords(-252,-490,0)(-740,-1740,0)[0,-2000]
\gdef\upperunitbox{\SRbr}   \gdef\lowerunitbox{\SLbr}
\or 
\setcoords(-490,-260,0)(-740,-1740,0)[0,-2002]
\gdef\upperunitbox{\SLbr}   \gdef\lowerunitbox{\SRbr}
\or 
\setcoords(-299,-449,0)(-870,-1970,0)[0,-2200]
\gdef\upperunitbox{\Rbr}    \gdef\lowerunitbox{\Lbr}
\particleadjusty=-95
\or 
\setcoords(-517,-371,0)(-900,-2000,0)[0,-2200]
\gdef\upperunitbox{\Lbr}    \gdef\lowerunitbox{\Rbr}
\particleadjusty=-165
\or 
\setcoords(-299,-409,0)(-885,-1905,0)[0,-2000]
\gdef\upperunitbox{\Rbr}   \gdef\lowerunitbox{\Lbr}
\particleadjustx=50     \particleadjusty=-380
\ifodd\unitboxnumber\relax\else\particleadjustx=250 \particleadjusty=-400 \fi
\or 
\setcoords(-519,-449,0)(-900,-1920,0)[0,-2000]
\gdef\upperunitbox{\Lbr}   \gdef\lowerunitbox{\Rbr}
\particleadjusty=-370
\ifodd\unitboxnumber\relax\else\particleadjustx=-240 \particleadjusty=-400 \fi
\or 
\gdef\upperunitbox{\Rbr}   \gdef\lowerunitbox{\Lbr}
\setcoords(-325,-555,0)(-900,-2100,0)[0,-2400]
\particleadjusty=-40
\or 
\setcoords(-505,-275,0)(-900,-2100,0)[0,-2400]
\gdef\upperunitbox{\Lbr}   \gdef\lowerunitbox{\Rbr}
\particleadjusty=-30  
\else \UNIMPERROR
\fi
}
\gdef\SWphoton{  
\ifcase\LINECONFIGURATION  
\setcoords(-825,-825,0)(0,-1250,0)[-1250,-1250]     \SETUNITBOX(1250)[br][tl]  
\or 
\setcoords(-175,-1425,0)(-625,-625,0)[-1250,-1250]  \SETUNITBOX(1250)[tl][br]  
\or 
\setcoords(-900,-900,0)(0,-1410,0)[-1400,-1400]     \SETUNITBOX(1400)[br][tl]  
\or 
\setcoords(-200,-1600,0)(-700,-700,0)[-1400,-1400]  \SETUNITBOX(1400)[tl][br]  
\or 
\setcoords(-800,-800,0)(0,-1200,0)[-1200,-1200]     \SETUNITBOX(1200)[br][tl]  
\or 
\setcoords(-200,-1400,0)(-600,-600,0)[-1200,-1200]  \SETUNITBOX(1200)[tl][br]  
\else \UNIMPERROR
\fi
}
\gdef\Wphoton{
\ifcase\LINECONFIGURATION 
\setcoords(-2245,-1245,0)(-150,-400,0)[-2005,0]
\gdef\upperunitbox{\Frown}   \gdef\lowerunitbox{\Smile}
\or 
\setcoords(-2245,-1245,0)(-400,-150,0)[-2005,0]
\gdef\upperunitbox{\Smile}   \gdef\lowerunitbox{\Frown}
\else \UNIMPERROR
\fi
\particleadjustx=57 
\ifnum\numwigglespo=\number\numwiggles \particleadjustx=0  \fi
\numlowerunits=\numwigglespo   \numupperunits=\numwiggles
}
\gdef\NWphoton{  
\ifcase\LINECONFIGURATION  
\setcoords(-200,-1425,0)(625,625,0)[-1250,1250]   \SETUNITBOX(1250)[bl][tr]
\or 
\setcoords(-825,-825,0)(0,1250,0)[-1250,1250]     \SETUNITBOX(1250)[tr][bl]
\ifnum\numwigglespo > \number \numwiggles \particleadjusty=-15 \fi
\or 
\setcoords(-200,-1600,0)(700,700,0)[-1400,1400]   \SETUNITBOX(1400)[bl][tr]
\or 
\setcoords(-900,-900,0)(0,1400,0)[-1400,1400]     \SETUNITBOX(1400)[tr][bl]
\or 
\setcoords(-200,-1400,0)(600,600,0)[-1200,1200]   \SETUNITBOX(1200)[bl][tr]  
\or 
\setcoords(-800,-800,0)(0,1200,0)[-1200,1200]     \SETUNITBOX(1200)[tr][bl]  
\else \UNIMPERROR
\fi
}

%% file: FTeX/GLUONSETUP.tex
\global\newcount\gluonlength
\global\newcount\gluonlengthx
\global\newcount\gluonlengthy
\global\newcount\gluonfrontx  
\global\newcount\gluonfronty  
\global\newcount\gluonbackx
\global\newcount\gluonbacky
%
\gdef\setunitbox(#1)[#2][#3](#4)[#5]{
\gdef\upperunitbox{\oval(#1,#1)[#2]}
\gdef\lowerunitbox{\oval(401,401)[#3]}
\gdef\thirdunitbox{\oval(#4,#4)[#5]}
}
\gdef\selectgluon{  
\global\advance\gluoncount by 1  
\global\gluonfrontx=\particlefrontx   
\global\gluonfronty=\particlefronty   
\global\particleadjustx=0     \global\particleadjusty=0
\ifnum\unitboxnumber > 40
\message{   *** WARNING *** Gluon with 
\the\unitboxnumber\space loops requested ***   }
\ifnum\unitboxnumber > 85
\message{   *** Reducing gluon length to 6 loops (max 85) ***   }
\ifnum\unitboxnumber > 1000
\message{   *** Probable Cause:  Gluon selected instead of Fermion ***   }
\fi \global\unitboxnumber=6 \fi \fi  
\global\unitboxnumberpo=\unitboxnumber  
\global\advance\unitboxnumberpo by 1 
\global\numlineparts = 3
\global\numupperunits=\unitboxnumber
\global\numlowerunits=\unitboxnumber
\global\numthirdunits=\unitboxnumber
\ifcase\LINEDIRECTION
\Ngluon    
\or  \NEgluon  
\or  \Egluon   
\or  \SEgluon
\or  \Sgluon
\or  \SWgluon 
\or  \Wgluon
\or  \NWgluon
\else\DIRECTERROR \fi
\setparticle  
\global\gluonlengthx=\particlelengthx  \global\gluonlengthy=\particlelengthy
\global\gluonbackx=\particlebackx      \global\gluonbacky=\particlebacky      
} 
\gdef\Ngluon{   
\ifcase\LINECONFIGURATION   
\setcoords(600,540,600)(20,620,1220)[0,1050] 
\setunitbox(1600)[tl][r](1600)[bl]
\particleadjusty=195
\or 
\setcoords(-990,-930,-990)(12,615,1215)[0,1050]
\setunitbox(1600)[tr][l](1600)[br]
\particleadjusty=195
\or 
\setcoords(440,390,440)(-10,415,840)[0,850]  
\setunitbox(1250)[tl][r](1250)[bl]
\particleadjustx=0
\particleadjusty=-10
\or 
\setcoords(-820,-770,-820)(-25,400,825)[0,850]  
\particleadjusty=-10  
\setunitbox(1250)[tr][l](1250)[br]
\or \UNIMPERROR  
\or \UNIMPERROR  
\or 
\numupperunits=\unitboxnumberpo
\numlowerunits=\unitboxnumber
\numthirdunits=\unitboxnumberpo
\setcoords(-200,-200,-200)(616,1041,616)[0,850]
\setunitbox(1250)[tl][r](1250)[bl]
\particleadjusty=1238
\particleadjusty=1233
\or 
\numupperunits=\unitboxnumberpo
\numlowerunits=\unitboxnumber
\numthirdunits=\unitboxnumberpo
\setcoords(-200,-200,-200)(620,1045,620)[0,850]
\setunitbox(1250)[tr][l](1250)[br]
\particleadjusty=1245
\else \UNIMPERROR 
\fi
}
\gdef\NEgluon{
\numupperunits=\unitboxnumberpo
\numlowerunits=\unitboxnumber
\numthirdunits=\unitboxnumber
\ifcase\LINECONFIGURATION
\setcoords(900,900,900)(0,900,900)[900,900]
\setunitbox(2200)[tl][tr](401)[b]
\particleadjustx=1100     \particleadjusty=1100
\or 
\setcoords(-180,720,720)(1090,1091,1091)[900,900]
\setunitbox(2200)[br][tr](401)[l]
\particleadjustx=1110     \particleadjusty=1050                      
\else \UNIMPERROR 
\fi
}
\gdef\Egluon{     
\ifcase\LINECONFIGURATION
\setcoords(-210,390,990)(-800,-745,-800)[1050,0]  
\setunitbox(1600)[tr][b](1600)[tl]  
\particleadjustx=130  
\or 
\setcoords(-210,390,990)(800,745,800)[1050,0]  
\setunitbox(1600)[br][t](1600)[bl]
\particleadjustx=130
\or 
\setcoords(-200,225,650)(-625,-575,-625)[850,0]
\setunitbox(1250)[tr][b](1250)[tl]  
\or 
\setcoords(-200,225,650)(625,575,625)[850,0]
\setunitbox(1250)[br][t](1250)[bl]
\or 
\setcoords(-200,430,1060)(-830,-780,-830)[1260,0]
\setunitbox(1660)[tr][b](1660)[tl]
\or 
\setcoords(-200,430,1060)(830,780,830)[1260,0]
\setunitbox(1660)[br][t](1660)[bl]
\or 
\numupperunits=\unitboxnumberpo
\numlowerunits=\unitboxnumber
\numthirdunits=\unitboxnumberpo
\setcoords(440,865,440)(0,50,0)[850,0]
\setunitbox(1250)[tr][b](1250)[tl]
\particleadjustx=1260
\or 
\numupperunits=\unitboxnumberpo
\numlowerunits=\unitboxnumber
\numthirdunits=\unitboxnumberpo
\setcoords(430,855,430)(0,-50,0)[850,0]
\setunitbox(1250)[br][t](1250)[bl]
\particleadjustx=1250
\or 
\setcoords(-160,440,1040)(-600,-550,-600)[1200,0]
\gdef\upperunitbox{\oval(1600,1200)[tr]}
\gdef\thirdunitbox{\oval(1600,1200)[tl]}
\gdef\lowerunitbox{\oval(401,401)[b]}
\else \UNIMPERROR 
\fi
}
\gdef\SEgluon{  
\numupperunits=\unitboxnumberpo
\numlowerunits=\unitboxnumber
\numthirdunits=\unitboxnumber
\ifcase\LINECONFIGURATION
\setcoords(-200,700,700)(-1100,-1100,-1100)[900,-900]
\setunitbox(2200)[tr][br](401)[l]
\particleadjustx=1100     \particleadjusty=-1100  
\or 
\setcoords(890,890,890)(0,-900,-900)[900,-900]
\setunitbox(2200)[bl][br](401)[t]
\particleadjustx=1050     \particleadjusty=-1100 
\else \UNIMPERROR 
\fi
}
\gdef\Sgluon{   
\ifcase\LINECONFIGURATION  
\setcoords(-1000,-940,-1000)(0,-595,-1195)[0,-1050]
\setunitbox(1600)[br][l](1600)[tr]
\particleadjusty=-150
\or 
\setcoords(605,545,605)(-20,-615,-1215)[0,-1050]
\setunitbox(1600)[bl][r](1600)[tl]  
\particleadjusty=-150
\or 
\setcoords(-820,-770,-820)(0,-425,-850)[0,-850]
\setunitbox(1250)[br][l](1250)[tr]  
\or 
\setcoords(440,390,440)(0,-425,-850)[0,-850]
\setunitbox(1250)[bl][r](1250)[tl]  
\or \UNIMPERROR 
\or \UNIMPERROR
\or 
\numupperunits=\unitboxnumberpo
\numlowerunits=\unitboxnumber
\numthirdunits=\unitboxnumberpo
\setcoords(-180,-180,-180)(-635,-1060,-635)[0,-850]
\setunitbox(1250)[br][l](1250)[tr]  
\particleadjusty=-1290
\or 
\numupperunits=\unitboxnumberpo
\numlowerunits=\unitboxnumber
\numthirdunits=\unitboxnumberpo
\setcoords(-180,-180,-180)(-635,-1060,-635)[0,-850]
\setunitbox(1250)[bl][r](1250)[tl]
\particleadjusty=-1290
\else \UNIMPERROR 
\fi
}
\gdef\SWgluon{
\numupperunits=\unitboxnumberpo
\numlowerunits=\unitboxnumber
\numthirdunits=\unitboxnumber
\ifcase\LINECONFIGURATION
\setcoords(-1300,-1300,-1300)(0,-900,-900)[-900,-900]
\setunitbox(2200)[br][bl](401)[t]
\particleadjustx=-1100     \particleadjusty=-1100
\or 
\setcoords(-215,-1115,-1115)(-1107,-1107,-1107)[-900,-900]
\setunitbox(2200)[tl][bl](401)[r]
\particleadjustx=-1120     \particleadjusty=-1120
\else \UNIMPERROR 
\fi
}
\gdef\Wgluon{   
\ifcase\LINECONFIGURATION
\setcoords(-190,-790,-1390)(800,745,800)[-1050,0]
\setunitbox(1600)[bl][t](1600)[br]  
\particleadjustx=-150  
\or 
\setcoords(-190,-790,-1390)(-800,-745,-800)[-1050,0]
\setunitbox(1600)[tl][b](1600)[tr]  
\particleadjustx=-150  
\or 
\setcoords(-200,-625,-1050)(625,575,625)[-850,0]
\setunitbox(1250)[bl][t](1250)[br]  
\or 
\setcoords(-200,-625,-1050)(-625,-575,-625)[-850,0] 
\setunitbox(1250)[tl][b](1250)[tr]  
\or 
\setcoords(-230,-860,-1490)(830,780,830)[-1260,0]
\setunitbox(1660)[bl][t](1660)[br]  
\or 
\setcoords(-230,-860,-1490)(-830,-780,-830)[-1260,0]
\setunitbox(1660)[tl][b](1660)[tr]
\or 
\numupperunits=\unitboxnumberpo
\numlowerunits=\unitboxnumber
\numthirdunits=\unitboxnumberpo
\setcoords(-825,-1250,-825)(0,-50,0)[-850,0]
\setunitbox(1250)[bl][t](1250)[br]  
\particleadjustx=-1250
\or  
\numupperunits=\unitboxnumberpo
\numlowerunits=\unitboxnumber
\numthirdunits=\unitboxnumberpo
\setcoords(-825,-1250,-825)(0,50,0)[-850,0] 
\setunitbox(1250)[tl][b](1250)[tr]  
\particleadjustx=-1250
\else \UNIMPERROR 
\fi
}
\gdef\NWgluon{
\numupperunits=\unitboxnumberpo
\numlowerunits=\unitboxnumber
\numthirdunits=\unitboxnumber
\ifcase\LINECONFIGURATION
\setcoords(-200,-1100,-1100)(1100,1100,1100)[-900,900]
\setunitbox(2200)[bl][tl](401)[r]
\particleadjustx=-1110   \particleadjusty=1100
\or  
\setcoords(-1309,-1309,-1309)(-15,885,885)[-900,900]
\setunitbox(2200)[tr][tl](401)[b]
\particleadjustx=-1120   \particleadjusty=1065
\else \UNIMPERROR 
\fi
}
%
%
%
\gdef\gluonlink{    
\input GLUONLINKS   
\gluonlink}  
\gdef\gluoncap{    
\input GLUONLINKS   
\gluoncap}  

%% file: FTeX/GLUONLINKS.tex
%
%
%
%
%
%
\gdef\gluonlink{
\global\stemlengthx=401 \SETDIR \setadjxy \divide\adjx by -2 \divide\adjy by -2
\ifcase\LDIR  
     \linksetupB \ifodd\LCONFIG\LINKPUT[l]  \else\LINKPUT[r] \fi 
\or  \ifodd\LCONFIG \linksetupBx \LINKPUT[tr]\LINKPUT[l]         
     \else          \linksetupBy \LINKPUT[tr]\LINKPUT[b]     \fi
\or  \linksetupB \ifodd\LCONFIG\LINKPUT[t]  \else\LINKPUT[b] \fi 
\or  \ifodd\LCONFIG \linksetupBy \LINKPUT[br]\LINKPUT[t]         
     \else          \linksetupBx \LINKPUT[br]\LINKPUT[l]     \fi
\or  \linksetupB \ifodd\LCONFIG\LINKPUT[r]  \else\LINKPUT[l] \fi 
\or  \ifodd\LCONFIG \linksetupBx \LINKPUT[bl]\LINKPUT[r]         
     \else          \linksetupBy \LINKPUT[bl]\LINKPUT[t]     \fi
\or  \linksetupB \ifodd\LCONFIG\LINKPUT[b]  \else\LINKPUT[t] \fi 
\or  \ifodd\LCONFIG \linksetupBy \LINKPUT[tl]\LINKPUT[b]         
     \else          \linksetupBx \LINKPUT[tl]\LINKPUT[r]     \fi
\else \UNIMPERROR 
\fi
\ifodd\LDIR\relax 
\else \global\advance\pbackx by \adjx   \global\advance\pbacky by \adjy \fi
\linksetupC
}
%
%
%
\gdef\gluoncap{  
\global\stemlengthx=0  
\ifodd\LDIR\message{NOTE:  Diagonal Gluons are not Capped}\relax
\else
\ifcase\LCONFIG \global\stemlengthx=1000  
\or \global\stemlengthx=1000  
\or \global\stemlengthx=825   
\or \global\stemlengthx=825   
\or \global\stemlengthx=1030  
\or \global\stemlengthx=1030  
\or \global\stemlengthx=0     
\or \global\stemlengthx=0     
\or \global\stemlengthx=800   
\else\UNIMPERROR
\fi 
\ifnum\stemlengthx>400
\global\advance\LDIR by 2 \moduloeight\LDIR   \SETDIR 
\global\advance\LDIR by 6 \moduloeight\LDIR   \setadjxy
\ifodd\LCONFIG \multiply \adjx by -1   \multiply \adjy by -1 \fi
\divide\adjx by 2  \divide\adjy by 2 \linksetupB
\ifcase\LDIR 
     \ifodd\LCONFIG\LINKPUT[tr]  \else\LINKPUT[tl]   \fi   
\or  \relax 
\or  \ifodd\LCONFIG\LINKPUT[br]  \else\LINKPUT[tr]   \fi   
\or  \relax 
\or  \ifodd\LCONFIG\LINKPUT[bl]  \else\LINKPUT[br]   \fi   
\or  \relax 
\or  \ifodd\LCONFIG\LINKPUT[tl]  \else\LINKPUT[bl]   \fi   
\or  \relax 
\else \UNIMPERROR 
\fi
\linksetupD
\else\message{* NOTE:  Attempt to use gluoncap of less that 401 centipoints *}
\fi 
\fi 
}
%
%
\gdef\setadjxy{
\adjx=\stemlengthx   \adjy=\stemlengthx 
\multiply \adjx by \XDIR%
\multiply \adjy by \YDIR%
}
\gdef\advancegluonlength{
\global\advance\particlelengthx by \adjx \global\advance\particlelengthy by\adjy
\global\advance\gluonlengthx by \adjx  \global\advance\gluonlengthy by \adjy
}
\gdef\LINKPUT[#1]{\ifnum\phantomswitch=0\put(\gluonbackx,\gluonbacky)
{\oval(\stemlengthx,\stemlengthx)[#1]}\fi}  
\gdef\linksetupBx{\global\advance\gluonbackx by \adjx}
\gdef\linksetupBy{\global\advance\gluonbacky by \adjy}
\gdef\linksetupB{\linksetupBx\linksetupBy}
\gdef\linksetupC{
\global\advance\pbackx by \adjx  \global\advance\pbacky by \adjy
\global\gluonbackx=\pbackx   \global\gluonbacky=\pbacky
\ifnum\plengthx<0 \multiply\adjx by -1 \fi
\ifnum\plengthy<0 \multiply\adjy by -1 \fi
\advancegluonlength
}
\gdef\linksetupD{
\linksetupC
\SETDIR 
\setadjxy
\divide\adjx by 2  \divide\adjy by 2 \linksetupB
\global\pbackx=\gluonbackx   \global\pbacky=\gluonbacky
\advancegluonlength
\ifnum\phantomswitch=0\put(\pbackx,\pbacky){\line(\XDIR,\YDIR){\stemlength}}\fi
\stemlengthx=\stemlength  
\setadjxy
\global\advance\pbackx by \adjx  \global\advance\pbacky by \adjy
\global\gluonbackx=\pbackx   \global\gluonbacky=\pbacky
\advancegluonlength
}

%% file: FTeX/VERTEX.tex
%
\global\advance\vertexcount by 1 
\newsavebox{\vertexbox}   
\global\newcount\LDIRcount      
\global\newcount\VERTEXNUMBER   
\global\newdimen\VERTEXLINKONE  
\global\newdimen\VERTEXLINKTWO
\global\newdimen\VERTEXLINKTHREE
\global\newdimen\VERTEXLINKFOUR 
\global\newdimen\VERTEXCAPONE   
\global\newdimen\VERTEXCAPTWO
\global\newdimen\VERTEXCAPTHREE
\global\newdimen\VERTEXCAPFOUR  
\global\newdimen\STEMVERTEXONE
\global\newdimen\STEMVERTEXTWO
\global\newdimen\STEMVERTEXTHREE
\global\newdimen\STEMVERTEXFOUR 
\global\newcount\stemlengthcopy
\global\newcount\vertexonex    \global\newcount\vertexoney
\global\newcount\vertextwox    \global\newcount\vertextwoy
\global\newcount\vertexthreex  \global\newcount\vertexthreey
\global\newcount\vertexfourx   \global\newcount\vertexfoury
\global\newcount\vertexmidx    \global\newcount\vertexmidy
\global\newcount\VERTEXLINE  
\global\newcount\FLIPVERTEX    \global\FLIPVERTEX=0
\gdef\flipvertex{\global\FLIPVERTEX=1}  
\newcount\vertadj              \newcount\negvertadj
\gdef\drawvertex#1[#2#3](#4,#5)[#6]{
\global\advance\vertexcount by 1 
\global\LINETYPE=#1           
\global\LINEDIRECTION=#2
\global\VERTEXNUMBER=#3  
\global\vertexonex=#4
\global\vertexoney=#5
\global\unitboxnumber=#6
\global\stemlengthcopy=\stemlength   
%
%
\ifnum\LINETYPE<3 \LINEERROR \fi
\ifnum\LINETYPE=3   
\ifnum\gluoncount=0 \def\gluonlink{\relax} \def\gluoncap{\relax} \fi
  \ifnum\VERTEXNUMBER<3 \UNIMPERROR \fi
  \ifnum\VERTEXNUMBER=3 \THREEPHOTON\fi 
  \ifnum\VERTEXNUMBER=4 \FOURPHOTON \fi 
  \ifnum\VERTEXNUMBER>4 \UNIMPERROR \fi
\fi
\ifnum\LINETYPE=4   
  \ifnum\VERTEXNUMBER<3 \UNIMPERROR \fi
  \ifnum\VERTEXNUMBER=3 \THREEGLUON \fi 
  \ifnum\VERTEXNUMBER=4 \FOURGLUON  \fi 
  \ifnum\VERTEXNUMBER>4 \UNIMPERROR \fi
\fi
\ifnum\LINETYPE>4 \LINEERROR \fi
\clearvertex   
}  
%
\gdef\advtwomodeight#1{
\global\advance\LDIRcount by2\moduloeight\LDIRcount \diagFOURVERT#1[\LDIRcount]}
%
\gdef\THREEPHOTON{
\ifcase\LDIR  
\setvertexA[\S\REG]
\setvertexB[\NW\CURLY](0,0)[2]  \setvertexB[\NE\FLIPPEDCURLY](0,0)[3]
\or  
\setvertexA[\SW\CURLY]
\setvertexB[\N\REG](70,-100)[2] \setvertexB[\E\FLIPPED](0,0)[3]
\or  
\setvertexA[\W0]
\setvertexB[\NE\CURLY](20,0)[2] \setvertexB[\SE\FLIPPEDCURLY](20,0)[3]
\or  
\setvertexA[\NW\CURLY]
\setvertexB[\E\REG](0,0)[2] \setvertexB[\S1](20,100)[3]
\or  
\setvertexA[\N\REG]
\setvertexB[\SE\CURLY](0,0)[2]  \setvertexB[\SW\FLIPPEDCURLY](0,0)[3]
\or  
\setvertexA[\NE\CURLY]
\setvertexB[\S\REG](-20,100)[2] 
\setvertexB[\W\FLIPPED](0,0)[3]
\or  
\setvertexA[\E\REG]
\setvertexB[\SW\CURLY](0,0)[2]  \setvertexB[\NW\FLIPPEDCURLY](0,0)[3]
\or  
\setvertexA[\SE\CURLY]
\setvertexB[\W\REG](0,0)[2]  
\setvertexB[\N\FLIPPED](-40,-100)[3] 
\else \DIRECTERROR
\fi
} 
%
%
\gdef\FOURPHOTON{
\ifnum\LDIR>-1
\global\LDIRcount=\LDIR  \global\advance\LDIRcount by 4 \moduloeight\LDIRcount
\setvertexA[\LDIRcount\REG] \advtwomodeight2 \advtwomodeight3 \advtwomodeight4
\else \UNIMPERROR \fi
\global\FLIPVERTEX=0
}
%
%
\gdef\THREEGLUON{
\vertadj=0   \adjvert  
\ifcase\LDIR  
\setvertexA[\S\CENTRAL]
\setvertexB[\NW\REG](0,0)[2]  \setvertexB[\NE\FLIPPED](0,0)[3]
\or  
\setvertexA[\SW\REG]
\setvertexB[\N\CURLY](-170,442)[2] \setvertexB[\E 3](420,-183)[3]
\setvertexC(442,442)[bl]
\or  
\setvertexA[\W6]
\setvertexB[\NE\REG](0,0)[2] \setvertexB[\SE\FLIPPED](0,0)[3]
\or  
\setvertexA[\NW\REG]
\setvertexB[\E\CURLY](420,183)[2]  \setvertexB[\S3](-183,-442)[3]
\setvertexC(442,-442)[tl]
\or  
\setvertexA[\N\CENTRAL]
\setvertexB[\SE\REG](0,0)[2]  \setvertexB[\SW\FLIPPED](0,0)[3]
\or  
\setvertexA[\NE\REG]
\setvertexB[\S\CURLY](170,-442)[2]
\setvertexB[\W\FLIPPEDCURLY](-420,183)[3]
\setvertexC(-442,-442)[tr]
\or  
\setvertexA[\E\CENTRAL]
\setvertexB[\SW\REG](0,0)[2]  \setvertexB[\NW\FLIPPED](0,0)[3]
\or  
\setvertexA[\SE\REG]
\setvertexB[\W\CURLY](-420,-183)[2]
\setvertexB[\N\FLIPPEDCURLY](170,442)[3]
\setvertexC(-442,442)[br]
\else \DIRECTERROR
\fi
} 
%
%
%
%
\gdef\FOURGLUON{
\ifodd\LDIR\vertadj=0 \else  \vertadj=412 \fi    \adjvert
\ifcase\LDIR  
\setvertexA[\S\CURLY]         \MIDADJUST(\negvertadj,\vertadj)
\WFOURVERT2    \NFOURVERT3    \EFOURVERT4
\or \FOURPHOTON  
\or  
\setvertexA[\W\CURLY]         \MIDADJUST(\vertadj,\vertadj)
\NFOURVERT2  \EFOURVERT3  \SFOURVERT4
\or \FOURPHOTON    
\or  
\setvertexA[\N\CURLY]         \MIDADJUST(\vertadj,\negvertadj)
\EFOURVERT2   \SFOURVERT3   \WFOURVERT4
\or \FOURPHOTON    
\or  
\setvertexA[\E\CURLY]         \MIDADJUST(\negvertadj,\negvertadj)
\SFOURVERT2  \WFOURVERT3  \NFOURVERT4
\or \FOURPHOTON    
\else \DIRECTERROR
\fi
\global\FLIPVERTEX=0   
} 
%
%
\gdef\MIDADJUST(#1,#2){  
\ifnum\FLIPVERTEX=0
\global\advance\vertexmidx by #1   \global\advance\vertexmidy by #2
\setvertexC(0,\vertadj)[bl]        \setvertexC(0,\negvertadj)[tr]
\setvertexC(\vertadj,0)[tl]        \setvertexC(\negvertadj,0)[br]
\else  
\global\particleadjustx=#1  \global\particleadjusty=#2  
\ifnum\LDIR=0 \global\multiply\particleadjustx by -1 \fi
\ifnum\LDIR=2 \global\multiply\particleadjusty by -1 \fi
\ifnum\LDIR=4 \global\multiply\particleadjustx by -1 \fi
\ifnum\LDIR=6 \global\multiply\particleadjusty by -1 \fi
\global\advance\vertexmidx by \particleadjustx  
\global\advance\vertexmidy by \particleadjusty
\setvertexC(0,\negvertadj)[tl]        \setvertexC(0,\vertadj)[br]
\setvertexC(\negvertadj,0)[tr]        \setvertexC(\vertadj,0)[bl]
\fi
}
\gdef\NFOURVERT#1{
  \ifnum\FLIPVERTEX=0 \setvertexB[\N\CURLY](\negvertadj,\vertadj)[#1]
  \else \setvertexB[\N\FLIPPEDCURLY](\vertadj,\vertadj)[#1] \fi}
\gdef\SFOURVERT#1{
  \ifnum\FLIPVERTEX=0 \setvertexB[\S\CURLY](\vertadj,\negvertadj)[#1]
  \else \setvertexB[\S\FLIPPEDCURLY](\negvertadj,\negvertadj)[#1] \fi}
\gdef\EFOURVERT#1{
  \ifnum\FLIPVERTEX=0 \setvertexB[\E\CURLY](\vertadj,\vertadj)[#1]
  \else \setvertexB[\E\FLIPPEDCURLY](\vertadj,\negvertadj)[#1] \fi}
\gdef\WFOURVERT#1{
  \ifnum\FLIPVERTEX=0 \setvertexB[\W\CURLY](\negvertadj,\negvertadj)[#1] 
  \else \setvertexB[\W\FLIPPEDCURLY](\negvertadj,\vertadj)[#1] \fi}
\gdef\diagFOURVERT#1[#2]{\setvertexB[#2\FLIPVERTEX](0,0)[#1]} 
%
%
%
\gdef\setvertexA[#1#2]{
\global\savebox\vertexbox(0,0){
\begin{picture}(0,0)
\global\adjx=#2
\ifnum\FLIPVERTEX=1 \global\advance\adjx by 1  
      \ifnum\VERTEXNUMBER=3 \global\FLIPVERTEX=0 \fi  
  \fi  
\ifdim\STEMVERTEXONE=1pt\backstemmed \fi
\drawline\LINETYPE[#1\adjx](0,0)[\unitboxnumber]
\global\stemlength=\stemlengthcopy  
\ifdim\VERTEXLINKONE=1pt\gluonlink \fi 
\ifdim\VERTEXCAPONE=1pt\gluoncap \fi 
\global\vertexmidx=\particlebackx  \global\vertexmidy=\particlebacky  
\end{picture}
}
\global\multiply\vertexmidx by -1  \global\multiply\vertexmidy by -1  
\global\advance\vertexmidx by \vertexonex  
\global\advance\vertexmidy by \vertexoney
\ifdim\STEMVERTEXONE=1pt\backstemmed \fi
\drawline\LINETYPE[#1\adjx](\vertexmidx,\vertexmidy)[\unitboxnumber]
\global\stemlength=\stemlengthcopy  
\ifdim\VERTEXLINKONE=1pt\gluonlink \fi 
\ifdim\VERTEXCAPONE=1pt\gluoncap \fi 
} 
\gdef\setvertexB[#1#2](#3,#4)[#5]{
\global\adjx=\vertexmidx   \global\adjy=\vertexmidy
\global\advance\adjx by #3   \global\advance\adjy by #4
\VERTEXLINE=#5
\ifcase\VERTEXLINE\UNIMPERROR  
\or \UNIMPERROR  
\or \ifdim\STEMVERTEXTWO=1pt\backstemmed \fi  
\or \ifdim\STEMVERTEXTHREE=1pt\backstemmed\fi 
\or \ifdim\STEMVERTEXFOUR=1pt\backstemmed\fi  
\else \UNIMPERROR                             
\fi
\drawline\LINETYPE[#1#2](\adjx,\adjy)[\unitboxnumber]
\global\stemlength=\stemlengthcopy  
\ifcase\VERTEXLINE\UNIMPERROR  
\or \UNIMPERROR  
\or \ifdim\VERTEXLINKTWO=1pt\gluonlink \fi 
    \ifdim\VERTEXCAPTWO=1pt\gluoncap \fi 
    \global\vertextwox=\particlebackx  \global\vertextwoy=\particlebacky  
\or \ifdim\VERTEXLINKTHREE=1pt\gluonlink\fi 
    \ifdim\VERTEXCAPTHREE=1pt\gluoncap\fi 
    \global\vertexthreex=\particlebackx  \global\vertexthreey=\particlebacky  
\or \ifdim\VERTEXLINKFOUR=1pt\gluonlink\fi 
    \ifdim\VERTEXCAPFOUR=1pt\gluoncap\fi 
    \global\vertexfourx=\particlebackx  \global\vertexfoury=\particlebacky  
\else \UNIMPERROR  
\fi
}
\gdef\setvertexC(#1,#2)[#3#4]{
\global\adjx=\vertexmidx   \global\adjy=\vertexmidy
\global\advance\adjx by #1   \global\advance\adjy by #2
\absstemlength=1250  
\ifnum \VERTEXNUMBER=4 \absstemlength=\vertadj\double\absstemlength \fi
\ifnum\phantomswitch=0 
   \put(\adjx,\adjy) {\oval(\absstemlength,\absstemlength)[#3#4]}\fi
}
\gdef\adjvert{\negvertadj=\vertadj  \multiply\negvertadj by -1}
%
%
\gdef\vertexlink#1{
\global\VERTEXLINE=#1
\ifcase\VERTEXLINE\UNIMPERROR
\or \global\VERTEXLINKONE=1pt    \or \global\VERTEXLINKTWO=1pt
\or \global\VERTEXLINKTHREE=1pt  \or \global\VERTEXLINKFOUR=1pt  
\else\UNIMPERROR\fi}
\gdef\vertexlinks{
\global\VERTEXLINKONE=1pt     \global\VERTEXLINKTWO=1pt
\global\VERTEXLINKTHREE=1pt  \global\VERTEXLINKFOUR=1pt  }
%
%
%
\gdef\vertexcap#1{
\global\VERTEXLINE=#1
\ifcase\VERTEXLINE\UNIMPERROR
\or \global\VERTEXCAPONE=1pt    \or \global\VERTEXCAPTWO=1pt
\or \global\VERTEXCAPTHREE=1pt  \or \global\VERTEXCAPFOUR=1pt  
\else\UNIMPERROR\fi}
\gdef\vertexcaps{
\global\VERTEXCAPONE=1pt     \global\VERTEXCAPTWO=1pt
\global\VERTEXCAPTHREE=1pt  \global\VERTEXCAPFOUR=1pt  }
%
%
%
\gdef\stemvertex#1{
\global\VERTEXLINE=#1
\ifcase\VERTEXLINE\UNIMPERROR
\or \global\STEMVERTEXONE=1pt    \or \global\STEMVERTEXTWO=1pt  
\or \global\STEMVERTEXTHREE=1pt  \or \global\STEMVERTEXFOUR=1pt  
\else\UNIMPERROR\fi}
\gdef\stemvertices{
\global\STEMVERTEXONE=1pt     \global\STEMVERTEXTWO=1pt  
\global\STEMVERTEXTHREE=1pt  \global\STEMVERTEXFOUR=1pt  }
\gdef\clearvertex{
\global\STEMVERTEXONE=0pt    \global\STEMVERTEXTWO=0pt  
\global\STEMVERTEXTHREE=0pt  \global\STEMVERTEXFOUR=0pt 
\global\VERTEXLINKONE=0pt    \global\VERTEXLINKTWO=0pt  
\global\VERTEXLINKTHREE=0pt  \global\VERTEXLINKFOUR=0pt  
\global\VERTEXCAPONE=0pt    \global\VERTEXCAPTWO=0pt  
\global\VERTEXCAPTHREE=0pt  \global\VERTEXCAPFOUR=0pt  
\global\stemlength=275}  
\global\stemlengthcopy=\stemlength
\clearvertex  
\global\stemlength=\stemlengthcopy

%% file: FTeX/LOOPS.tex
\global\newcount\loopfrontx    \global\newcount\loopfronty
\global\newcount\loopbackx    \global\newcount\loopbacky
\global\newcount\loopmidx    \global\newcount\loopmidy
\global\newdimen\CENTRALLOOP
\gdef\drawloop#1[#2#3](#4,#5){
\global\CENTRALLOOP=0pt  
\global\LINETYPE=#1  
\ifnum\LTYPE=\gluon\relax\else\UNIMPERROR\LTYPE=1\message{Reverting to Gluons}
\fi
\global\LINEDIRECTION=#2  
\global\fourthlineadjx=#3 
\ifnum\fourthlineadjx=0 
  \global\CENTRALLOOP=1pt  
  \global\fourthlineadjx=8
  \global\LDIR=0
\fi   
\global\fourthlineadjy=\fourthlineadjx  
\global\advance\fourthlineadjy by -4
\global\loopfrontx=#4   \global\loopfronty=#5
\ifdim\CENTRALLOOP=1pt
  \global\advance\loopfrontx by -2413  \global\advance\loopfronty by -425
\fi                          
\global\unitboxnumber=1  
\ifnum\LINETYPE=\photon \unitboxnumber=2 \fi
\checkdir
\drawline\LINETYPE[\LDIR\LCONFIG](\loopfrontx,\loopfronty)[\unitboxnumber]
\DRAWLOOP
\ifnum\fourthlineadjy>-1 
\global\loopmidx=\loopfrontx   \global\loopmidy=\loopfronty
\global\advance\loopmidx by \loopbackx  \global\advance\loopmidy by \loopbacky
\divide\loopmidx  by 2 \divide\loopmidy by 2  
\ifdim\CENTRALLOOP=1pt
  \global\advance\loopfrontx by 200    \global\advance\loopfronty by 425
  \global\advance\loopbackx by -200    \global\advance\loopbacky by -425
\fi
\fi 
}
\gdef\DRAWLOOP{
\global\advance\fourthlineadjx by -1
\ifnum\fourthlineadjx=0\relax  
\else
\ifnum\fourthlineadjx=\fourthlineadjy 
   \global\loopbackx=\pbackx   \global\loopbacky=\pbacky
\fi
\global\advance\LDIR by 1
\moduloeight\LDIR
\checkdir
\drawline\LINETYPE[\LDIR\LCONFIG](\pbackx,\pbacky)[\unitboxnumber]
\fi 
\ifnum\fourthlineadjx>1 \DRAWLOOP  \fi  
}
\gdef\checkdir{
\ifnum\LTYPE=\gluon
\ifodd\LDIR \global\LCONFIG=0 \else \global\LCONFIG=2 \fi
\fi 
}

%% file: TitlePage.tex
\headnote{ 
    \hspace*{\fill}{\small\sf UNITUE-THEP-99/10, \hskip .5cm IU-NTC-99/06, 
                    \hskip .5cm FAU-TP3-99/05} \\[-6pt]
    \hspace*{\fill}{\small\sf http://xxx.lanl.gov/abs/nucl-th/9909082}
                   }

\title{Current Conservation in the Covariant\\ 
       Quark-Diquark Model of the Nucleon}

\author{M.~Oettel\inst{1} \and M.~A.~Pichowsky\inst{2} \and
  L.~von~Smekal\inst{3} }

\institute{Universit\"at T\"ubingen, Institut f\"ur Theoretische Physik,
           Auf der Morgenstelle 14, 72076 T\"ubingen, Germany 
            \and  
	   Nuclear Theory Center, Indiana University, Bloomington IN
            47405, USA, and \\
           Department of Physics \& Supercomputer Computations
           Research Institute, Florida State University,\\
           Tallahassee, FL 32306-4130, USA \and 
           Universit\"at Erlangen-N\"urnberg, 
           Institut f\"ur Theoretische Physik III,
           Staudtstr.~7, 91058 Erlangen, Germany}

\date{(30 September 1999, revised 2 February 2000)}
\def\makeheadbox{}

\abstract{
The description of baryons as fully relativistic bound states of quark and
glue reduces to an effective Bethe-Salpeter equation with quark-exchange
interaction when irreducible 3-quark interactions are neglected and
separable 2-quark (diquark) correlations are assumed. 
This covariant quark-diquark model of baryons is studied with
the inclusion of the quark substructure of the diquark correlations. 
In order to maintain electromagnetic current conservation it is then
necessary to go beyond the impulse approximation. A conserved current is
obtained by including the coupling of the photon to the exchanged quark
and direct ``seagull'' couplings to the diquark structure.
Adopting a simple dynamical model of constituent quarks and
exploring various parametrisations of scalar diquark correlations,
the nucleon Bethe-Salpeter equation is solved and the proton and neutron
electromagnetic form factors are calculated numerically. 
The resulting magnetic moments are still about 50\% too small, the
improvements necessary to remedy this are discussed. The results obtained
in this framework provide an excellent description of the electric form
factors (and charge radii) of the proton, up to a photon momentum transfer
of 3.5GeV$^2$, and the neutron. 
}
\maketitle


%% file: Sec1.tex
\section{Introduction}

To the precision accessible to current measurements, the proton is the
only known hadron that is stable under the effect of all interactions.
Protons are thus ideal for use in beams or targets for scattering
experiments designed to explore the fundamental dynamics of the strong
interaction. From this, it follows that more is known about the proton and
its nearly stable isospin partner, the neutron, than any other hadron.
There is an abundance of observable properties of the nucleon, from 
elastic scattering form factors to electromagnetic form factors and parton
distributions, which are being measured with increasing precision at
various accelerator facilities around the world. 
In particular, the ratio of the electric and the magnetic form factor of the
proton is subject to current measurements at TJNAF, and current
experiments at MAMI~\cite{MAMI} and NIKHEF~\cite{NIKHEF} show great
promise that soon we will have a precise knowledge of the electromagnetic
properties of neutron as well. Hence, the development of an accurate and
tractable covariant framework for the nucleon in terms of the underlying
quarks and gluons is clearly desirable. 

Models of the nucleon and other baryons are numerous and have had varying
degrees of success as they are usually designed to describe particular
properties of baryons. Some of the frameworks that have been
employed are non-rela\-tivis\-tic~\cite{Gel64,Kar68,Fai68} and 
relativistic~\cite{Fey71} quark potential models, bag
models~\cite{Cho74,Has78}, skyrmion models~\cite{Sky61,Adk83,Schw89,Hol93} or
the chiral soliton of the Nambu-Jona-Lasinio (NJL)
model~\cite{Alk96a,Chr96}.     
The relativistic bound state problem of 3-quark Faddeev type was studied
extensively within the NJL model~\cite{Rei90,Buc92,Ish93,Hua94,Buc95,Ish95}
and its non-local generalization, the global color model~\cite{Cah89,Bur89}.
In addition, some complementary aspects of these models have been 
combined,  {\it e.g.}, the chiral bag model~\cite{Cho75,Rho94} and 
a hybrid model that implements the NJL-soliton picture of baryons within
the quark-diquark Bethe-\-Salpe\-ter (BS) framework \cite{Zue97}. 

The present study is concerned with the further development of a description of
baryons (and the nucleon, in particular) as bound states of quarks and
gluons in a fully-covariant quantum field theoretic framework based on the 
Dyson-Schwinger equations of QCD. Such a framework has already reached a high
level of sophistication for mesons.  
In these studies, the quark-antiquark scattering kernel is modeled as a
confined, non-perturbative gluon exchange. Such a gluon exchange can be
provided by solutions to the Dyson-Schwinger equations for the propagators
of QCD in the covariant gauge~\cite{Sme97,Sme98}. In phenomenological studies
the gross features of such solutions are mimicked by the use of a
phenomenological quark-antiquark kernel. With this kernel, the
dressed-quark propagator is 
obtained as the solution of its (quenched) Dyson-Schwinger equation (DSE),
and the same kernel is then employed in the quark-antiquark Bethe-Salpeter
equation (BSE)\footnote{
We warn the reader not to confuse the meaning of this acronym with another one 
adopted recently. See, {\it e.g.}, Ref.~\cite{Lancet}.}
from which one obtains the masses of the meson bound states
and their BS amplitudes. Once the dressed-quark propagators and meson BS
amplitudes have been obtained, observables can be calculated in a  
straight-forward manner. The success of this approach has been demonstrated
in many phenomenological applications, such as 
electromagnetic form factors~\cite{Bur96}, decay widths, 
$\pi$-$\pi$ scattering~\cite{Rob96}, vector-meson
electroproduction~\cite{Pic97}, to name a few.
Summaries of this Dyson-\-Schwinger/\-Bethe-\-Salpeter description of
mesons may be found in Refs.~\cite{Miranski,Rob94,Tan97}.   

The significantly more complicated framework required for 
an analogous description of baryons based on the quantum field theoretic
description of bound states in 3-quark correlations has meant that baryons
have received far less attention than mesons.
Consequently, the utility of such a description for the baryons
is considerably less understood, even on the phenomenological level. 
(For example, the ramifications of various truncation schemes,
required in any quantum field theoretic treatment of bound states, which
have been explored extensively in the meson sector are completely unknown
in the baryon sector.) The developments in this direction employ a
picture based on separable 2-quark correlations, {\it i.e.}, diquarks,
interacting with the 3rd quark which allows a treatment of the
relativistic bound state in a manner analogous to the 2-body BS 
problem~\cite{Kus97,Hel97b,Oet98}. Although these studies employ 
simplified model assumptions for the quark propagators and diquark
correlators, they de\-mon\-strate the utility of the approach in general, and
these simplified model assumptions can be replaced by more realistic ones in
the future as more is understood about the underlying dynamics of quarks and
gluons.  
 
In the present article we generalize this framework for the 3-quark bound
state problem in quantum field theory to accommodate the quark-substructure
of the separable diquark correlations. This necessary extension has
important implications on the electromagnetic properties of the nucleons. We
explicitly construct a conserved current for their electromagnetic couplings,
and we verify analytically that it yields the correct charges of both
nucleons.  
To achieve this we employ Ward and Ward-Takahashi identities for the 
quark correlations which arise from electromagnetic gauge invariance.  
In order to avoid unnecessary complications,
our present study considers only simple constituent-quark and diquark
propagators. However, the generalization to include dressed
propagators, most importantly to account for confined nature of quarks and
diquarks, requires only minor modifications.

The organization of the article is as follows.
In Sec.~\ref{dq_corrs}, some general properties of the two-quark correlations
(or diquarks) are discussed. The importance of constructing
a diquark correlator that is antisymmetric under quark exchange is
emphasized. The parametrisations of the Lorentz-scalar isoscalar diquark
correlations which are employed in the numerical calculations of the subsequent
sections are introduced to reflect these properties in deliberately simple
ways. The framework is sufficiently general to accommodate the results for
the diquark correlators as they become available from studies of the underlying
dynamics of quark and gluon correlations in QCD. 
Of course, diquark correlations other than the scalar diquark will be
important for a more complete description of the nucleon. Certainly necessary
is the inclusion of other channels, such as the axialvector diquark, when the
description is extended to the octet and the decuplet baryons. 
Nonetheless, for the purposes of the present study, only scalar 
diquarks are considered; the generalization to include other diquarks is
straight-forward~\cite{Oet98}. In Sec.~\ref{QDBSE}, the nucleon BS equation
and the quark-exchange kernel are introduced. One particular feature of this
kernel is that it necessarily depends on the total momentum of the nucleon 
bound state $P_n$. This conclusion is based on general arguments, such as
the exchange-antisymmetry of the diquark correlations. It is important to
obtain the correct normalizations and charges of the nucleon bound states. 
In particular, to ensure current conservation requires a considerable
extension beyond the impulse approximation in the calculations of
electromagnetic form factors for baryons.\footnote{The 
$P_n$-dependence of the exchange-kernel violates one of the
necessary conditions for current conservation in the {\em impulse}
approximation. See Sec.~\ref{TECO}.} The numerical solutions for the
nucleon BS amplitudes obtained herein preserve the invariance 
of observables under a re-routing of the relative momentum, a requirement 
that follows from the translational invariance of the relativistic bound
state problem.
In Sec.~\ref{SecNucNorm}, the normalization condition for
the nucleon BS amplitude is derived and the nucleon electromagnetic current
is obtained in Sec.~\ref{TECO}. In Sec.~\ref{WIaS}, the ``seagull''
contributions to the electromagnetic current, necessary for current
conservation, are derived from the Ward and Ward-Takahashi
identities. In Sec.~\ref{ElmFFs}, expressions for the electromagnetic
form factors of the nucleon are derived, the numerical calculations are
described in Sec.~\ref{NC}, and the results for the form factors are
presented and discussed in Sec.~\ref{Res}. The conclusions of this study are
provided in Sec.~\ref{Conclusions}.   
Several appendices have also been included with further
details which may provide the interested reader with addition insight into
the framework.

%% file: Sec2.tex
\section{Diquark Correlations}
\label{dq_corrs}

To obtain a solution of the 3-particle Faddeev equations in quantum field
theory requires a truncation of the interaction kernel. 
A widely-employed truncation scheme is to neglect contributions to
the Faddeev kernel which arise from irreducible 3-quark interactions.
This allows one to rewrite the Dyson series for the 3-quark Green function
as a coupled system of equations. The first  being the BS equation for the
2-quark scattering amplitude and the other being the Faddeev equation,
which describes the coupling of the 3rd quark to these 2-quark correlations. 
As a result, the nature of the gluonic interactions of quarks enters only in
the BSE of the 2-quark subsystem, {\it i.e.}, in the quark-quark
interaction kernel. The solution of the full inhomogeneous BS problem for
this 2-quark system is simplified by assuming separable contributions
(explained below), hereafter referred to as diquarks, to account for the
relevant 2-quark correlations within the hadronic bound state. 
Herein, a ``diquark correlation'' thus refers to the use of a separable
4-quark function in the $\bar{3}$ representation of the $SU(3)$ color group.  
The utility of such diquark correlations for a description of baryon bound
states is a central element of the present approach.

While in the NJL model the 2-quark scattering amplitude has the property
of being separable, in general this is an additional assumption 
useful to simplify the 3-quark bound state problem in quantum field
theory.   
An example for separable contributions would be (a finite sum of) 
isolated poles at timelike total momenta $P^2$ of the diquark.
Such poles allow the use of homogeneous BSEs to obtain the respective
amplitudes from the gluonic interaction kernel of the quarks, and thus to
calculate these separable contributions to the 2-quark scattering
amplitude.

However, the validity of using a diquark correlator to parametrise the 
2-quark correlations phenomenologically, does not rely on the existence of
asymptotic diquark states. Rather, the diquark correlator may be devoid of
singularities for timelike momenta, which may be interpreted as one possible
realization of diquark confinement. In principle, one may appeal to models
employing a general, separable diquark correlator which need not have any
simple analytic structure, in which case no particle interpretation for the
diquark would be possible. The implementation of this model of 
confined diquark is straight-forward and does not introduce
significant changes to the framework. The use of diquark correlations in this
capacity is quite general and does not necessarily imply the existence 
diquarks, which have not been observed experimentally. 
 
The absence of asymptotic-diquark states may be explained in a number
of ways. Although, it has been observed that solutions of the BSE in ladder
approximation yield asymptotic color-$\bar{3}$ diquark states
\cite{Pra89}, when terms {\em beyond ladder approximation} are maintained,
the diquarks cease to be bound~\cite{Ben96}. That is, the addition of terms
beyond the ladder approximation to the BS kernel, in a way which preserves
Goldstone's theorem at every order, has a minimal impact on
solutions for the color-singlet meson channels. In contrast, such terms have a
significant impact on the  color-$\bar{3}$ diquark channels. 
In Ref.~\cite{Ben96}, it was demonstrated that these
contributions to BS kernel beyond ladder approximation are predominantly
repulsive in the color-$\bar 3$ diquark channel. It was furthermore
demonstrated in a simple confining quark model that the strength of these
repulsive contributions suffices to move the diquark poles from
the timelike $P^2$-axis and far into the complex $P^2$-plane.
While the particular, confining quark model is in conflict with locality,
the same effect was later verified within the NJL model~\cite{Hel97a}. 
This suggests that this mechanism for diquark confinement, which
eliminates the possibility of producing asymptotic diquark states,
might hold independent of the particular realization of quark confinement.

In a local quantum field theory, on the other hand, colored
asymptotic states do exist, for the elementary fields as well as possible
colored composites such as the diquarks, but not in the physical subspace of
the indefinite metric space of covariant gauge theories.   
The analytic structure of correlation functions, the holomorphic envelope of
extended permuted tubes in coordinate space, is much the same in this
description as in quantum field theory with a positive definite inner product
(Hilbert) space. In particular, 2-point correlations in momentum space 
are generally analytic functions in the cut complex $P^2$-plane with the cut
along the timelike real axis. Confinement is interpreted as the
observation that both absorptive as well as anomalous thresholds in hadronic
amplitudes are due only to other hadronic states~\cite{Oeh95}. However, the
implementation of this algebraic notion of confinement
seems much harder to realize in phenomenological applications.

For the present purposes of developing a general framework for the
description of baryon bound states, the question as to whether one should
or should not model the diquark correlations in terms of functions with or
without singularities for total timelike momentum $P^2$ is irrelevant.
However, we reiterate that the present framework is able to accommodate
both of these descriptions of confinement with only straight-forward
modifications.  

The goal of the present study is to assess the utility of
describing baryons as bound states of quark and diquark correlations, in a
framework in which the diquark correlations are assumed to be 
{\em separable}. (The term separable refers to the property that a 4-point 
Green function $G(p,q;P)$ be independent of the scalar $qp$, where $q$
and $p$ are the relative momenta of the two incoming and outgoing particles,
respectively, and $P$ is the total momentum.)
To provide a simple demonstration of the general formalism, and its
application to the calculation of nucleon form factors, we assume that the
diquark correlator corresponds to a single scalar-diquark pole at $P^2 =
m_s^2$ which is both separable and constituent-like. In this description, the
(color-singlet) baryon thus is a bound state of a color-$3$ quark and a
color-$\bar{3}$ (scalar) diquark correlation.

\input pictex/Fig1

Assuming identical quarks, consider the 4-point quark Green function in
coordinate space given by 
\begin{eqnarray} 
G_{\alpha\beta\gamma\delta}(x_1,x_2,x_3,x_4) &=&  \label{Gen4q-G} \\   
 && \hskip -1cm \langle 
T(q_\gamma(x_3) q_\alpha(x_1) \bar q_\beta(x_2) \bar q_\delta(x_4)) \rangle
\; , \nn
\end{eqnarray}
where $\alpha, \beta, \gamma$, and $\delta$ denote the Dirac indices of
the quarks and $T$ denotes time-ordering of the quark fields
$q_{\alpha}(x)$. 
The assumption of a separable diquark correlator, 
corresponding to the diagram shown in Fig.~\ref{dqpole}, can be written in
momentum space as
$G^{\hbox{\tiny sep}}_{\alpha \beta \gamma \delta}(p,q,P)$ where 
\begin{eqnarray} 
G_{\alpha\gamma , \beta\delta}^{\hbox{\tiny sep}}(p,q,P) \, &:=&  
 e^{-iPY}  \, \int d^4\!X\, d^4\!y\, d^4\!z \,\,  e^{iqz}
e^{-ipy} \nn \\ 
&& \hskip 1.5cm  e^{iPX} G_{\alpha\beta\gamma\delta}^{\hbox{\tiny sep}} (x_1,x_2,x_3,x_4) \nn
 \\ 
&=& D(P) 
\; \chi_{\gamma\alpha}(p,P) \bar \chi_{\beta\delta}(q,P) \;
\; ,  \label{dq_pole_ms} 
\end{eqnarray}
where $D(P)$ is the diquark propagator, 
$X = \sigma x_1 + (1\!-\!\sigma) x_3 $,  $ Y =
(1\!-\!\sigma') x_2 + \sigma'  x_4 $, the total momentum partitioning of
the outgoing and incoming quark pairs are given respectively 
by $ \sigma $ and $ \sigma' $ both in $ [0,1]$, and $y = x_1 - x_3$, $z = x_2
- x_4$. 

As described above, the separable form of the diquark correlation of
Eq.~(\ref{dq_pole_ms}) does not necessarily entail the existence of an
asymptotic diquark state.  
The framework developed herein makes no restrictions on the particular
choice of the diquark propagator $D(P)$. Technically, model
correlations which mimic confinement through the absence of timelike poles
are easy to implement as shown in Ref.~\cite{Hel97b}.  

Nonetheless, for the purpose of demonstration, here we employ the simplest
form for this propagator corresponding to a simple (scalar) diquark bound
state in the 2-quark Green function $G$ of Eq.~(\ref{Gen4q-G}).  
The appearance of such an asymptotic diquark state
requires the diquark propagator to have a simple pole at $P^2 = m_s^2$,
where $m_s$ is the mass of the diquark bound state, {\it i.e.},   
\begin{eqnarray}
D(P) = \frac{i}{P^2 - m_s^2 + i \epsilon} \; .  \label{pole_dqcorr}
\end{eqnarray}
Then $\chi(p,P)$ and its adjoint $\bar\chi(p,P) = \gamma_0
\chi^\dagger(p,P) \gamma_0 $ are the BS wave functions of the (scalar) diquark
bound state which are defined as the matrix elements of two quark fields or two
antiquark fields between the bound state and the vacuum,  
respectively. Further details concerning these definitions are given in 
Appendix~\ref{AppDqDetails}. 

It is convenient to define the {\em truncated} diquark BS amplitudes
(sometimes referred to as BS vertex functions) $\widetilde\chi$ and 
$\widetilde{\bar\chi}$ from the BS wave functions in
Eq.~(\ref{dq_pole_ms}) by amputating the external quark propagators,
\begin{eqnarray}
\widetilde\chi (p,P) &:=&   S^{-1}((1\!-\!\sigma)P+p) 
\; \chi(p,P) \; {S^{-1}}^T\!(\sigma P-p) \, , \\
\widetilde{\bar\chi} (p,P)  &:=&  {S^{-1}}^T\!(\sigma P-p)  
\; \bar\chi (p,P) \;  S^{-1}((1\!-\!\sigma) P+p) \, . 
\end{eqnarray} 
The convention employed here is to use the same symbols for both the BS
wave functions and the truncated BS amplitudes, with the latter denoted by 
a tilde.  
An important observation made in Appendix~\ref{AppDqDetails}, which is of
use in the following discussions, is that the for identical quarks,
antisymmetry under quark exchange constrains the diquark BS
amplitudes $\widetilde\chi$ to satisfy:
\begin{eqnarray} 
\widetilde\chi(p, P) \, = \, \left. - \widetilde\chi^T(-p,P)\right|_{\sigma
\leftrightarrow (1-\sigma)}  \! .  
\label{dq_asym} 
\end{eqnarray} 
Note that $\sigma $ and $(1\!-\!\sigma)$ have been interchanged here.

In principle, at this point one could go ahead and specify the form of the
kernel for the quark-quark BSE in ladder approximation, and obtain diquark BS
amplitudes in much the same manner in which solutions for the mesons are
obtained. 
However, as discussed above, the appearance of stable bound state solutions
might be an artifact of the ladder approximation rather than
the true nature of the quark-quark scattering amplitude.
Therefore, for our present purposes various simple model
parametrisations for diquark BS amplitudes are explored, rather than
using a particular solution of the diquark homogeneous BSE.
The motivation is to explore the general aspects and implications of using
a separable diquark correlation for the description of the nucleon bound
state.   

For separable contributions of the pole type~(\ref{pole_dqcorr}), but
possibly complex mass (with Re$(m_s^2) > 0$), one readily obtains standard  
BS normalization conditions to fix the overall strength of the quark-quark
coupling to the diquark for a given parametrisation of the diquark
structure. These are obtained from the inhomogeneous quark-quark BSE for the
Green function of Eq.~(\ref{Gen4q-G}) employing pole dominance for $P^2$
sufficiently close to $m_s^2$.\footnote{For other separable contributions,
{\it e.g.}, of the form of non-trivial entire functions for which one
necessarily has a singularity at $|P^2| \to \infty $, the 
use of the inhomogeneous BSE to derive normalization conditions
relies on the existence of full solutions to (\protect\ref{dqIBSE}) of the
separable type.} To sketch their derivation consider the
inhomogeneous BSE which is of the general form, 
\begin{eqnarray} 
G(p,q,P) \, = \, \left( {G^{(0)}}^{-1} (p,q,P) - K(p,q,P) \right)^{-1} \; .
\label{dqIBSE}
\end{eqnarray}
Here $G^{(0)}$ denotes the antisymmetric Green function for the
disconnected propagation of two identical quarks. The definition of its 
inverse ${G^{(0)}}^{-1}$ and a brief discussion of how to derive it, may
be found in Appendix~\ref{AppDqDetails}.  
With the simplifying assumption that the quark-quark interaction kernel
$K$ be independent on the total diquark momentum $P$; that is, 
$K(p,q,P) \equiv K(p,q)$  (which is satisfied in the ladder approximation for
example), the derivative of $G$ with respect to the total momentum $P$, gives
the relation 
\begin{eqnarray} 
 &&  - P^\mu\frac{\partial}{\partial P^\mu} \, G(p,q,P) \, = \,
 \int \frac{d^4k}{(2\pi)^4}\frac{d^4k'}{(2\pi)^4} \\ 
 && \hskip .6cm 
  G(p,k,P) \left(
P^\mu\frac{\partial}{\partial P^\mu} \, {G^{(0)}}^{-1}\!(k,k',P) \right)
 G(k',q,P) \nn \; .
\end{eqnarray}    
Upon substitution of $G(p,q,P)$ as given by
Eqs.~(\ref{dq_pole_ms}) and (\ref{pole_dqcorr}), and equating the residues of
the most singular terms, one obtains the non-linear constraint for the
normalization of the diquark BS amplitudes $\chi$ and $\bar\chi$,
\begin{eqnarray}
1 & \, \stackrel{!}{=} \, &  \frac{-i}{4 m_s^2} \int
\frac{d^4p}{(2\pi)^4}  \label{dq_norm} \\ 
&&  \left\{ 
\hbox{tr} \left( S^T(p_\beta) \widetilde{\bar\chi}(p,P) \left(
P\frac{\partial}{\partial P} S(p_\alpha)\right) \widetilde\chi(p,P) \right)
\right.  \nn\\
&& \hskip .5cm + \,  \left.
\hbox{tr} \left(\widetilde{\bar\chi}(p,P) S(p_\alpha)  \widetilde\chi(p,P)
\left( P\frac{\partial}{\partial P} S^T(p_\beta)\right) \right)
\right\} \; , \nn
\end{eqnarray} 
where $p_\alpha = p + (1\!-\!\sigma)P$ and $ p_\beta = -p + \sigma P$, {\it
i.e.}, $P= p_\alpha + p_\beta$ and $p = \sigma p_\alpha - (1\!-\!\sigma)
p_\beta $. 
The scalar diquark contribution, relevant to the present study of
the nucleon bound state, is color-$\bar{3}$ and isosinglet.
Lorentz covariance requires its Dirac structure to be the sum of four
independent contributions, each proportional to a function of two
independent momenta.
For simplicity only a single term is maintained here, which has the
following structure: 
\begin{eqnarray} 
\widetilde\chi(p,P) =  \gamma_5C \,  \frac{1}{N_s} 
	\tilde P(p^2,pP) \; , 
\label{sdq_amp_def}
\end{eqnarray}
where $C$ is the charge conjugation matrix ($C = i\gamma_2 \gamma_0$ in the
standard representation).   
The normalization constant $N_s$ is fixed from the condition given in
Eq.~(\ref{dq_norm}) for a given form of $\tilde P(p^2,pP)$. 

It may seem reasonable to neglect the dependence of this invariant
function on the scalar $pP$;
such a simplification would yield the leading moment of an
expansion of the angular dependence in terms of orthogonal polynomials,
which has been shown to provide the dominant contribution to the meson
bound state amplitudes in many circumstances.  
However, in the present case the antisymmetry of the diquark
wave function, {\it c.f.}, Eqs. (\ref{dq_asym}), entails that   
\begin{eqnarray} 
\tilde P(p^2, pP) \, =\, \left. 
\tilde P(p^2, - pP) \right|_{\sigma
\leftrightarrow (1-\sigma)} \; . \label{sym_P}
\end{eqnarray}
For $ \sigma \not= 1/2 $ and thus for $\bar p := p
\big|_{\sigma \leftrightarrow (1-\sigma)} \not= p$, it is not
possible to neglect the $pP$ dependence in the amplitude
without violating the quark-exchange antisymmetry. 
To maintain the correct quark-exchange antisymmetry, we assume
instead that the amplitude depends on both scalars, $p^2$ and $pP$
in a specific way.  In particular, we assume the diquark BS amplitude is
given by a function that depends on the scalar 
\begin{eqnarray} 
  x &:=&  p_\alpha p_\beta - \sigma (1-\sigma) \, m_s^2 
     = - (1 - 2\sigma) pP - p^2  \nn \\
 &=&  (1 - 2\sigma) \bar pP - \bar p^2 \label{x_def}
\end{eqnarray} 
with $\bar p  =  (1-\sigma) p_\alpha - \sigma p_\beta $ and 
$p_{\{\alpha , \beta\}}$ as given above. 
For equal momentum partitioning between the quarks in the
diquark correlation $\sigma = 1/2$, the scalar $x$ reduces to the negative
square of the relative momentum, $x = -p^2$. 
The two scalars that may be constructed from the available momenta $p$ and
$P$ (noting that $P^2 = m^2_{s}$ is fixed) which have definite symmetries
under quark exchange are given by the two independent combinations
$p_\alpha p_\beta$ (which is essentially the same as above $x$) and 
$p_\alpha^2 - p_\beta^2 $. 
The latter may only appear in odd powers which are associated with
higher moments of the BS amplitude.  Hence, these are neglected by setting 
\begin{eqnarray} 
\tilde P(p^2, pP) \equiv P(x)  
\end{eqnarray}  
which can be shown to satisfy the antisymmetry constraint given by
Eq.~(\ref{sym_P}) $\forall \sigma \in [0,1]$.
 Finally, the diquark BS normalization $N_s $, as obtained from
Eq.~(\ref{dq_norm}), is given by
\begin{eqnarray}
 N_s^2  \, &=& \,  \frac{-i}{4 m_s^2} \int
\frac{d^4p}{(2\pi)^4} \; P^2(x) 
\label{Ns} \\ 
&& \hskip 1cm   P\frac{\partial}{\partial P} \;  
\hbox{tr} \big[  S(p+(1\!-\!\sigma)P)  S(-p+\sigma P)  \big] \; . \nn
\end{eqnarray}

The numerical results presented in the following sections explore the
ramifications of several Ans{\"a}tze for $P(x)$,
\begin{eqnarray}
P_{n\mbox{\tiny-P}}(x)  =  \left(\frac{\gamma_n}{x + \gamma_n} \right)^n
\hskip -.1cm    \mbox{or} \;
 P_{\hbox{\tiny EXP}}(x)  =  \exp \, \{ - x/\gamma_{\hbox{\tiny EXP}} \}
, \label{P_amps}
\end{eqnarray} 
where the integer $n= 1,2 ...$ and corresponds to monopole,
dipole,... diquark BS amplitudes.  
Their widths $\gamma_n$, $\gamma_{\hbox{\tiny EXP}}$ are determined 
from the nucleon BSE by varying them until the diquark normalization given
by Eq.~(\ref{Ns}) and coupling strength $g_s^2$ necessary to produce the 
correct nucleon bound-state mass are equal.  
This is carried out numerically and described in detail in Sec.~\ref{QDBSE}.

For completeness, various Gaussian forms that peak for values of $x = x_0
\ge 0 $, 
\begin{eqnarray}
P_{\hbox{\tiny GAU}}(x) \, = \, \exp \, \{ -
 (x-x_0)^2 /\gamma_{\hbox{\tiny GAU}}^2   \} \;
 ,  \label{P_gau}
\end{eqnarray} 
are also explored.
Such forms with finite $x_0$ were suggested for diquark amplitudes as
a result of a variational calculation of an approximate diquark BSE in
Ref.~\cite{Pra88} and have been used in the nucleon calculations of 
Ref.~\cite{Bur89}. 
Therein, a fit to the Gaussian form given by Ref.~\cite{Pra88}
(and Eq.~(\ref{P_gau}) above) was employed with a width of 
$\gamma_{\hbox{\tiny GAU}} \simeq 0.11$GeV$^2$ and
$x_0/\gamma_{\hbox{\tiny GAU}}\simeq 1.7$. 
From a calculation within the present framework, which is described in
Sec.~\ref{QDBSE}, we observe that the necessary value for
$\gamma_{\hbox{\tiny GAU}}$ to obtain a reasonable nucleon mass is about
an order of magnitude smaller than the value given in Ref.~\cite{Bur89}.
Furthermore in Sec.~\ref{ElmFFs}, we observe that the effect of a finite
$x_0$ on the electric form factor of the neutron rules out the use of a
Gaussian form with $x_0 \not = 0$.

Some of the parametrisations used for diquark correlations in previous
quark-diquark model studies of the nucleon~\cite{Kus97,Hel97b,Oet98},  
correspond to neglecting the substructure of diquarks entirely; 
this may be reproduced in the present framework by setting  
$\tilde P(p^2,pP)=1$. 
By neglecting the diquark substructure, the diquark BS normalizations such as
$N_s$ here are not well-defined. The strengths of the
quark-diquark couplings are undetermined and chosen as free
adjustable parameters of the model (one for each diquark channel
maintained). In the present, more general approach, 
these couplings are determined by the normalizations of the respective 
diquark amplitudes. At present this is the scalar diquark normalization 
$N_s$ alone which in turn determines the coupling strength $g_s=1/N_s$
between the quark and diquark which binds the nucleon.

The aim of our present study is to generalize the notion of diquark
correlations by going beyond the use of a point-like diquark and include a
diquark substructure in a form similar to that of mesons.
In the most general calculation of the three-body bound state, 
the strength of the quark-diquark coupling is not at one's disposal; it
arises from the elementary interactions between the quarks. 
By using the diquark BS normalization condition of Eq.~(\ref{Ns}), we
can assess whether the quark-diquark picture of the nucleon is still 
able to provide a reasonable description of the nucleon bound state if the
coupling strength is obtained from the diquark BSE rather than forcing the
quark and diquark to bind by adjusting their couplings freely. 
Whether the quark-diquark coupling is sufficiently strong to produce bound
baryons will eventually be determined by the strength of the 
quark-quark interaction kernel that leads to the diquark BS amplitude.

Finally, the use of a diquark BS amplitude with a {\em finite} width improves
on the previous calculations in yet another respect.
Without the substructure of the diquark, an additional ultraviolet
regularization had to be introduced in the exchange kernel of the BSE for the
nucleon in Refs.~\cite{Kus97,Hel97b,Oet98}. 
In the present study, the finite-sized substructure of the diquark
leads to a nucleon BSE which is completely regular in the ultraviolet 
in a natural way. 

In this section, we have discussed some general features of the
diquark amplitude employed in our present study. The important observations
are the implications of its antisymmetry under quark exchange which
constrains the functional dependence on the quark momenta, and the
derivation of the BS normalization condition,  Eq.~(\ref{Ns}), to fix the
quark-diquark coupling strength.
By taking these into account, we will find that the precise form of
the Ansatz for the diquark BS amplitude $P(x)$ has little qualitative
influence on the resulting nucleon amplitudes as long as $P(x)$ falls off
by at least one power of $x$ for large $x$ corresponding to a large spacelike
relative momentum.

%% file: pictex/Fig1.tex
%
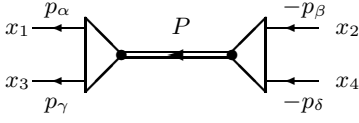
\begin{figure}[t]
\vskip 1cm
\begin{picture}(16000,6000)(-12500,-3000)
\thicklines
\drawline\fermion[\NW\REG](0,0)[2000]
\drawline\fermion[\S\REG](\pbackx,\pbacky)[\fermionlengthy]
\drawline\fermion[\S\REG](\pbackx,\pbacky)[\fermionlength]
\drawline\fermion[\NE\REG](\pbackx,\pbacky)[2000]
\thinlines
\drawline\fermion[\W\REG](\pfrontx,-1000)[2000]
\drawarrow[\W\ATBASE](\pmidx,\pmidy)
\drawline\fermion[\W\REG](\pfrontx,1000)[2000]
\drawarrow[\W\ATBASE](\pmidx,\pmidy)
\put(-100,0){\circle*{400}}
\drawline\fermion[\E\REG](0,100)[2000]
\drawline\fermion[\E\REG](0,-100)[2000]
\thicklines
\drawarrow[\W\ATBASE](2000,0)
\put(-12000,4000){\framebox{$G_{\alpha\beta\gamma\delta}^{\hbox{\tiny sep}}$}}
\put(1800,800){$P$}
\put(-3000,-2000){$p_\gamma$}
\put(-3000,1600){$p_\alpha$}
\put(-4500,-1200){$x_3$}
\put(-4500,800){$x_1$}
\drawline\fermion[\NE\REG](4000,0)[2000]
\drawline\fermion[\S\REG](\pbackx,\pbacky)[\fermionlengthy]
\drawline\fermion[\S\REG](\pbackx,\pbacky)[\fermionlength]
\drawline\fermion[\NW\REG](\pbackx,\pbacky)[2000]
\thinlines
\drawline\fermion[\E\REG](\pfrontx,-1000)[2000]
\drawarrow[\W\ATBASE](\pmidx,\pmidy)
\drawline\fermion[\E\REG](\pfrontx,1000)[2000]
\drawarrow[\W\ATBASE](\pmidx,\pmidy)
\put(4100,0){\circle*{400}}
\thinlines
\drawline\fermion[\W\REG](4000,100)[2000]
\drawline\fermion[\W\REG](4000,-100)[2000]
\thicklines
\put(6000,-2000){$-p_\delta$}
\put(6000,1600){$-p_\beta$}
\put(8000,-1200){$x_4$}
\put(8000,800){$x_2$}
\end{picture}
\label{dqpole}
\caption{The diquark pole in the 4-quark Green function.}
\end{figure}

%% file: Sec3.tex
\section{The Quark-Diquark Bethe-Salpeter Equation of the Nucleon}
\label{QDBSE}

By neglecting irreducible 3-quark interactions in the
kernel of the Faddeev equation giving the 6-point Green function that 
describes the fully-interacting propagation of 3 quarks, the Dyson series
for this 6-point Green function reduces to a coupled set of two-body
Bethe-Salpeter equations, see for example, Ref.~\cite{Hua94}.
As discussed in the previous section, for the purpose of demonstrating the
framework developed herein, we choose to employ constituent quark and
diquark propagators.  However, the framework itself is much more general.  
The assumptions under which it is developed require only that
the irreducible 3-quark interactions are neglected in the nucleon Faddeev
kernel and that the diquark correlations be well-parametrised by a sum of
separable terms (which may or may not have poles associated with asymptotic
diquark states).

Maintaining only the (flavor-singlet, color-$\bar 3$) scalar diquark
channel in these correlations, corresponding to Eqs. (\ref{dq_pole_ms}) 
and (\ref{pole_dqcorr}), the appearance of a stable nucleon bound state
coincides with the development of a pole in the Green function
$G_{\alpha\beta}$ describing the fully-interacting (spin-${1}/{2}$) quark
and (scalar) diquark correlations which is of the form,
\begin{eqnarray}
G^{\hbox{\tiny pole}}_{\alpha\beta} (p,k,P_n)  &=&  \label{nuc_pole_cont}  \\
&& \left(\psi
(p,P_n)  \frac{ i ( \fslash P_n + M_n ) }{P_n^2 - M_n^2 + i \epsilon} 
\bar\psi(-k,P_n) \right)_{\alpha\beta}   . \nn 
\end{eqnarray}
Here $k$ and $p$ are the relative momenta between the quark and diquark,
incoming and outgoing, respectively, $P_n$ is the four-momentum of
the nucleon with mass $M_n$, $\alpha$ and $\beta$ denote the Dirac indices
of the quark, and the nucleon BS wave function $\psi(p,P_n)$ is related to
it's adjoint according to
\begin{eqnarray}
\bar \psi(p,P_n) \,  = \,  \gamma_0 \psi^\dagger(-p,P_n) \gamma_0
            \,  = \,  C \psi^T(p,-P_n) C^{-1} \, . 
\end{eqnarray}
The truncated nucleon BS amplitude $\widetilde\psi$ is defined as 
\begin{eqnarray}
\psi(p,P_n) =  D((1\!-\!\eta)P_n - p) S(\eta P_n + p) \widetilde\psi(p,P_n)
\; , 
\end{eqnarray}
where $\eta \in [0,1]$ is the fraction of the nucleon momentum $P_n$
carried by the quark.  
The resulting homogeneous Bethe-Salpeter equation for the nucleon bound
state reads, 
\begin{eqnarray}
\widetilde\psi_{\alpha\beta}(p,P_n)  =  \int \frac{d^4k}{(2\pi)^4} \;
K_{\alpha\gamma}(p,k;P_n) \, \psi_{\gamma\beta}(k,P_n) 
\; , \label{hom_nuc_BSE}  
\end{eqnarray}
where $K_{\alpha\beta}(p,k;P_n)$ is the kernel of the nucleon BSE
which represents the exchange of one of the quarks within the diquark 
with the spectator quark. Maintaining the quark-exchange antisymmetry 
of the diquark correlations, this kernel provides the full antisymmetry 
within the nucleon ({\it i.e.}, Pauli's principle) in the quark-diquark
model \cite{Rei90}.  
\input{pictex/Fig3}
The exchange kernel is shown in Fig.~\ref{kernel} and given by  
\begin{eqnarray}
K_{\alpha\beta}(p,k;P_n)\, &=& \label{xker} \\ 
 && \hskip -1cm \,- \frac{1}{2} \, 
 \widetilde \chi_{\alpha\gamma}(p_1,q+p_\alpha) 
\, S^T_{\gamma\delta}(q) \,
\widetilde{\bar\chi}^T_{\delta\beta}(p_2,q+p_\beta) \nn \\ 
&=& \,  \frac{1}{2 N_s^2}\,  P(x_1) P(x_2)  \;
 S_{\alpha\beta}(q) \; , \label{x_kern_par}  
\end{eqnarray}
where the factor -1/2 arises from the flavor coupling between the quark and
diquark and $p_1$ and $p_2$ are the relative momenta of the quarks within
the incoming and outgoing diquark, respectively, such that
\begin{eqnarray}
              p_1 = \sigma p_\alpha - (1\!-\!\sigma) q \; , 
\quad \hbox{and} \quad p_2 = (1\!-\!\sigma') q - \sigma' p_\beta \; . 
\end{eqnarray}
The respective momentum partitionings are $\sigma$ and $\sigma'$
and need not be equal. 
The total momenta of the incoming and the outgoing diquark are $q+p_\alpha$
and $q+p_\beta$. 
The momentum $q$ of the exchanged quark 
is expressed in terms of the total nucleon momentum $P_n$ and relative
momenta $k$ and $p$ as 
\begin{eqnarray}
q \, =\, ( 1\!-\!2\eta) P_n - p -k \; . \label{ex_qk_mom}
\end{eqnarray} 
Using the definitions of $q$ and the quark momenta $p_\alpha = \eta P_n
+p$ and $ p_\beta = \eta P_n + k$, the relative momenta in the diquark
BS amplitudes can be expressed as
\begin{eqnarray}
p_1 \, & =&\,  (\sigma \eta - (1\!-\!\sigma)(1\!-\!2\eta)) P_n +
(1\!-\!\sigma) k + p  
\; , \label{p_1} \\
p_2 \, & =&\,  ((1\!-\!\sigma') (1\!-\!2\eta)  - \sigma'\eta) P_n - k -
(1\!-\!\sigma')p 
\; . 
\end{eqnarray} 
The corresponding arguments of the quark-exchange-anti\-sym\-metric diquark
BS amplitudes follow readily from their definition in Eq.~(\ref{x_def}),
\begin{eqnarray}  
x_1 \, &=& \, - p_1^2 - (1\!-\!2\sigma) ((1\!-\!\eta) p_1 P_n 
- p_1 k)
\; , \label{x_1}\\ 
x_2 \, &=& \, - p_2^2 + (1\!-\!2\sigma') ((1\!-\!\eta) p_2 P_n 
- p_2 p) 
\; . \label{x_2}
\end{eqnarray}  

Before proceeding, it is worth summarizing some of the important aspects
of this framework: 
\begin{enumerate}
\item The dependence of the exchange kernel on the total
nucleon bound-state momentum $P_n$ is {\em crucial} in order to obtain the
correct relationship between the electromagnetic charges of the proton and
neutron bound states and the normalizations of their BS amplitudes. 
\item The momentum $q$ of the exchanged quark is {\em independent} of the
nucleon momentum $P_n$ {\em only for} the particular value of momentum
partitioning $\eta = 1/2$.
\item The relative momenta $p_1$ and $p_2$ of the quarks within the diquarks
are only independent of the total momentum $P_n$ of the nucleon for the
particular choice: 
\begin{eqnarray}
\sigma = \sigma' = \frac{1-2\eta}{1-\eta} \; . \label{dq_mom_part}
\end{eqnarray}
The {\em symmetric} arguments of $x_{1}$ and $x_2$ in the diquark BS
amplitudes are independent of the total momentum $P_n$ only if, in
addition to the above criteria, $\sigma = \sigma' = 1/2$ as well, and
hence $\eta = 1/3$.
In fact, this conclusion can be further generalized: 
Any exchange-symmetric argument of the diquark amplitude can differ
from the definition of Eq.~(\ref{x_def}) only by a term that is
proportional to the square of the total diquark momentum. 
From this, it is possible to show that the diquark BS amplitudes can be
independent of $P_n$ only if $\eta = 1/3$. This is shown in
Appendix~\ref{SuppNBSE}. 
\end{enumerate}   

Unlike the ladder-approximate kernels commonly employed in
phenomenological studies of the meson BSE, the quark-exchange kernels
of the BSEs for baryon bound states, such as the one depicted in
Fig.~\ref{kernel}, {\em necessarily} depend on the total momentum of the
baryon $P_n$ for {\em all} values of $\eta \in [0,1]$. 
This has important implications on the normalization of the bound-state BS
amplitudes as well as on the calculation of the electromagnetic charges of
the bound states.  In particular, considerable extensions of the framework
beyond the impulse approximation are required in the calculation of
electromagnetic form factors in order to ensure that the electromagnetic
current is conserved for baryons.  This issue is explored in detail in
Sec.~\ref{TECO}.

The form of the pole contribution arising from the nucleon  bound state to
the quark-diquark 4-point correlation function in 
Eq.~(\ref{nuc_pole_cont}) constrains the nucleon BS amplitude to obey the
identities: 
\begin{eqnarray}
\widetilde\psi(p ,P_n) \Lambda^+(P_n) &=& \widetilde\psi(p ,P_n)\; , \\
\Lambda^+(P_n) \widetilde{\bar\psi}(p,P_n) &=&
\widetilde{\bar\psi}(p,P_n) \; , 
\end{eqnarray}
where $\Lambda^+(P_n) \, =\, (\fslash P_n + M_n)/2M_n $. 
It follows from this that the most general Lorentz-covariant
form of the nucleon BS amplitude can be parametrised by
\begin{eqnarray}
\widetilde\psi(p ,P_n)\, &=& \label{psiDec} \\
&& \hskip -1.5cm  S_1(p,P_n)\, \Lambda^+(P_n) \, +\,
S_2(p,P_n) \,  \Xi(p,P_n)\, \Lambda^+(P_n) \; , \nn \\
\widetilde{\bar\psi}(p,P_n) \, &=& \label{psibarDec} \\
&&\hskip -1.5cm  S_1(-p,P_n)\,  \Lambda^+(P_n) \, +\, 
S_2(-p,P_n)\,  \Lambda^+(P_n)\,  \Xi(-p,P_n) \; ,  \nn
\end{eqnarray}
where $\Xi(p,P_n) = (\fslash p - p P_n/M_n)/M_n $, and $S_1(p,P_n)$ and
$S_2(p,P_n)$ are Lorentz-invariant functions of $p$ and $P_n$. 
This provides a separation of the positive and negative energy
components of the nucleon BS amplitude. 
In Appendix~\ref{SuppNBSE}, some of the consequences of this decomposition
are explored.  In particular, it allows one to rewrite the
homogeneous BSE of Eq.~(\ref{hom_nuc_BSE}) in a compact manner, in terms
of a 2-vector $S^T(p,P_n) := (S_1(p,P_n),\, S_2(p,P_n))$ as
\begin{eqnarray}
S(p,P_n) &=& \label{EucSBSE} \\
&& \hskip -1.2cm  \frac{1}{2N_s^2}  \int \frac{d^4k}{(2\pi)^4} \; P(x_1) P(x_2)
\,  D(k_{s}) \,  T(p,k,P_n)  \, S(k,P_n) \; , \nn
\end{eqnarray}
where $k_s = (1\!-\!\eta)P_n-k\, $ and  $T(p,k,P_n)$
is a $2\times 2$ matrix in which each of the four elements is
given by a trace over the Dirac indices of the quark
propagator $S(k_q)$ (with $k_q = \eta P_n + k$), 
the propagator $S(q)$ of the exchanged quark and
a particular combination of Dirac structures derived from the 
decomposition of the nucleon amplitude in Eq.~(\ref{psiDec}), see
Appendix~\ref{SuppNBSE}.

\begin{figure}[t]
\leftline{\hskip -.1cm 
\epsfig{file=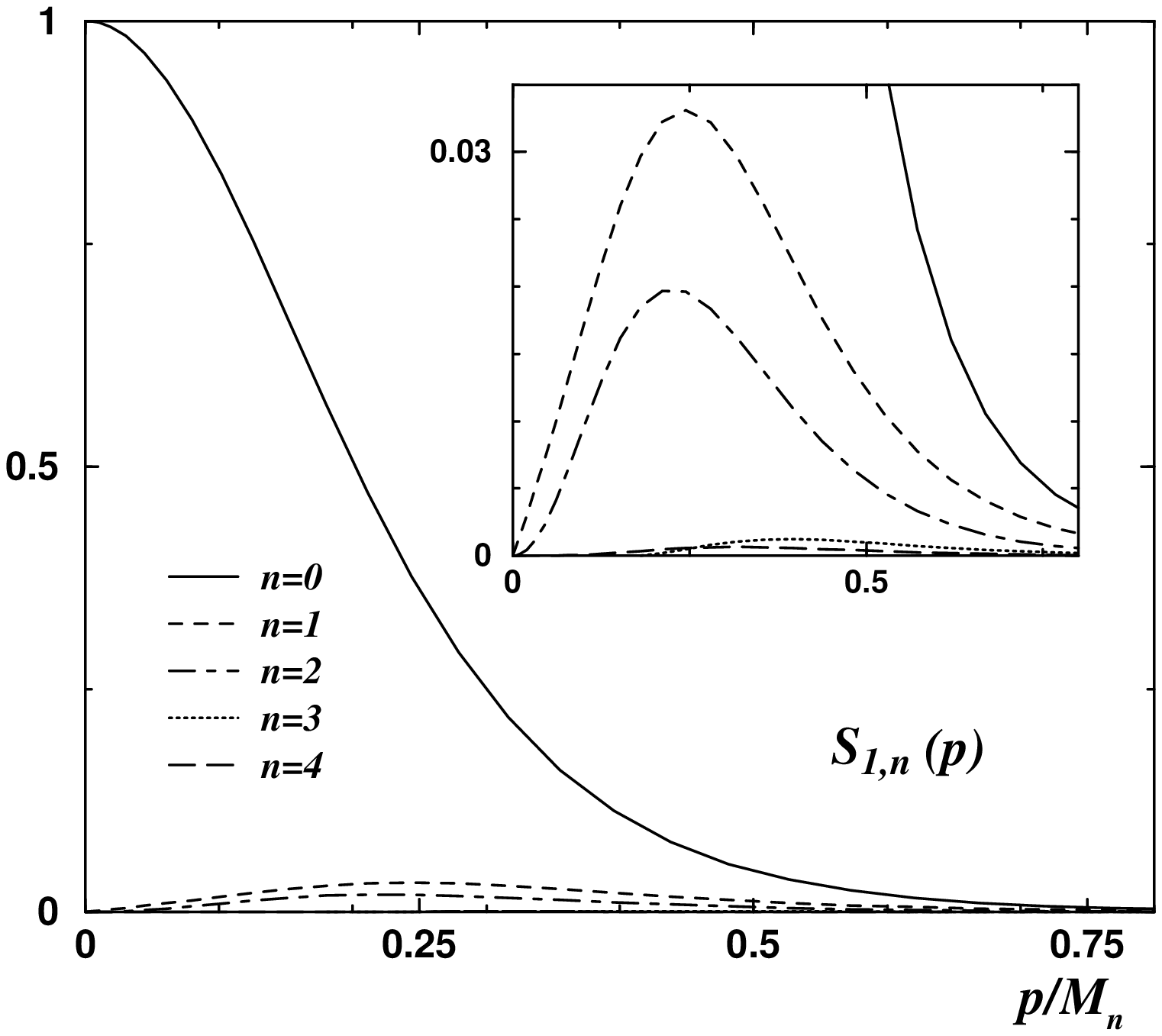,width=0.49\linewidth} 
\hskip .2cm \epsfig{file=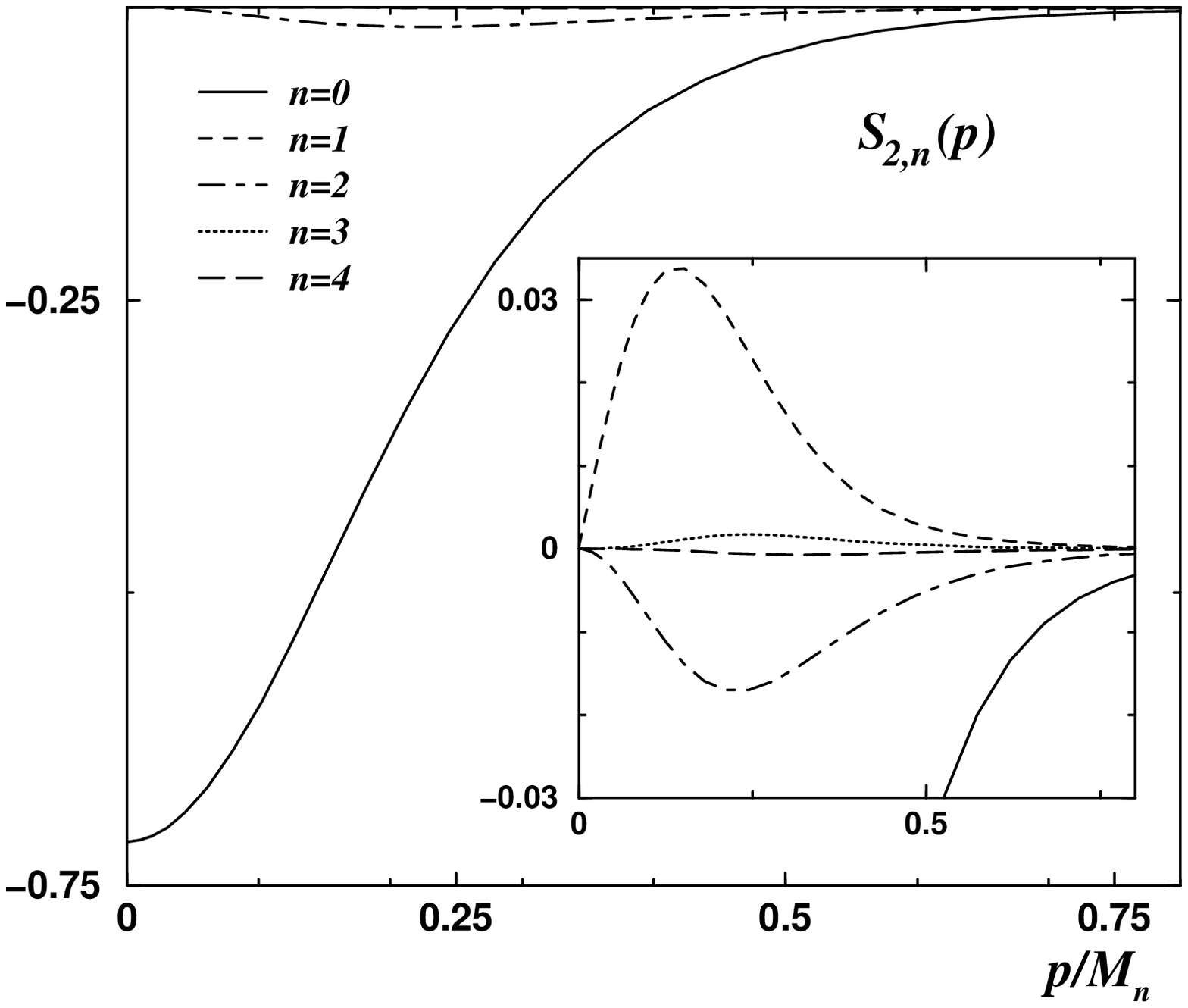,width=0.49\linewidth} }
\caption{The first 5 moments of the nucleon BS amplitudes $S_1$ (left)
and $S_2$ (right).}
\label{S1_S2}
\end{figure}

Upon carrying out these traces, the resulting BSE is
transformed into the Euclidean metric by introducing 4-dimensional polar
coordinates corresponding to the rest frame of the nucleon 
according to the following prescriptions (where ``$\to $'' denotes the formal
transition from the Minkowski to the Euclidean metric):
\begin{eqnarray}
&&p^2 \, \to \, -p^2\; , \quad P_n^2 \to M_n^2 \; , \quad 
 p P_n \, \to \, i M_n  p\,  y \; ,\label{WickRot}   \\
&& \mbox{and} \quad S(p,P_n) \, \to \, S(p,y)  \; . \nn 
\end{eqnarray}   
Written in terms of these variables, the nucleon BS amplitude is a
function of the square of the relative momentum $p^2$ and the cosine 
$y \in [-1,+1]$ of the azimuthal angle between the four-vectors $p$ and 
$P_n$. The dependence of the nucleon BS amplitudes $S(p,y)$ on the 
angular variable $y$ is approximated by an expansion to the order $N$ in
terms of the complete set of Chebyshev polynomials $T_n(y)$,
\begin{eqnarray}
S(p,y) \, &\simeq& \, \sum_{n=0}^{N-1} (-i)^n \, S_n(p) T_n(y)
\label{ChebyS} \\
 S_n(p) &=& i^n \, \frac{2}{N} \sum_{k=1}^N S(p,y_k) T_n(y_k) \; ,
 \label{ChebyM} \\
\hbox{where} \; y_k &=& \cos\left( \frac{\pi (k- 1/2)}{N} \right)  
\nn \end{eqnarray}
are the zeros of the Chebyshev polynomial $T_N(y)$ of degree $N$ in $y$. 
Here, we employ Chebyshev polynomials of the first kind with a 
convenient (albeit non-standard) normalization for the zeroth Chebyshev
moment from setting $T_0 = 1/\sqrt{2}$. 
The explicit factor of $(-i)^n$ is introduced into Eq.~(\ref{ChebyS}) to
ensure that the moments $S_n(p)$ are real functions of positive $p \equiv
\sqrt{p^2}$ for all $n$. 
The BSE for these moments of the nucleon BS amplitude is then
\begin{eqnarray}
S_m(p) &=& \label{Eq:38}\\
&& \, - \frac{1}{2N_s^2} \, \int \frac{k^3 dk}{(4\pi)^2} \,
 \sum_{n=0}^{N-1} \,  i^{m-n} \; T_{mn}(p,k) \, S_n(k) \; , \nn
\end{eqnarray}
where the ($2N\times 2N$) matrix $T_{mn}(p,k)$ is obtained from $T(p,k,P_n)$
by expanding in terms of Chebyshev polynomials both amplitudes $S$ in the
nucleon BSE of Eq.~(\ref{EucSBSE}); that is, summing over the 
$y_k$ on both sides of Eq.~(\ref{EucSBSE}) according to
Eq.~(\ref{ChebyM}) and using Eq.~(\ref{ChebyS}) for the BS amplitude on the
right-hand side. 
The definition of the matrix $T_{mn}$ appearing in Eq.~(\ref{Eq:38})
furthermore includes the explicit 
diquark propagator and BS amplitudes $P(x_1)$ and $P(x_2)$
of the integrand in Eq.~(\ref{EucSBSE}) as well as the integrations over all
angular variables. Its exact form is provided in Appendix~\ref{SuppNBSE}.     

\begin{figure}[t]
\centerline{\epsfig{file=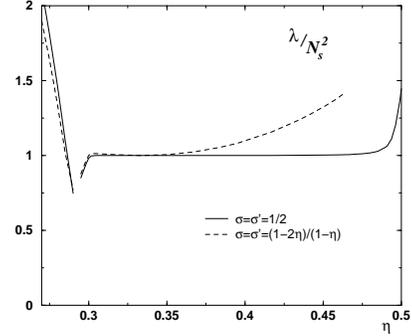,width=0.6\linewidth}}
\caption{Dependence of $\lambda/N_s^2$ on 
$\eta $ for fixed $\sigma = \sigma' = 1/2$ (solid) and with $\sigma $,
$\sigma'$  from Eq.~(\protect\ref{dq_mom_part}) (dashed). The width
$\gamma_2$ is fixed to  yield $\lambda/N_s^2 = 1 $ at $\eta = 1/3$.} 
\label{eta_fig}
\end{figure}

In the calculations presented herein, we restrict ourselves to the use of
free propagators for constituent quark and diquark, with masses 
$m_q$ and $m_s$ respectively, as the simplest model to parametrise
the quark and diquark correlations within the nucleon. Measuring 
all momenta in units of the nucleon bound-state mass $M_n$
leaves $m_q/M_n$ and $m_s/M_n$ as the only free parameters.
Using for the scalar diquark amplitude $P(x)$ the forms of
Eqs.~(\ref{P_amps}) or (\ref{P_gau}) with fixed widths $\gamma$, the
homogeneous BSE for the nucleon in Eq.~(\ref{hom_nuc_BSE}) is converted 
into an eigenvalue equation of the form,
\begin{eqnarray}
     \lambda \,\widetilde\psi(p,P_n)  \, =\,     \int
     \frac{d^4k}{(2\pi)^4} \;   N_s^2 K(p,k,P_n) \, \psi(p,P_n) \; ,
\end{eqnarray}
with the additional constraint that $\lambda = N_s^2$ (note that $N_s^2 K$ is
independent of $N_s$). This equation is solved iteratively and the largest
eigenvalue $\lambda$ is found which corresponds to the nucleon ground-state. 
To implement the constraint, we calculate the diquark normalization $N_s^2$ 
from Eq.~(\ref{Ns}) and compare it to the eigenvalue $\lambda$. This
procedure is repeated with a new value for the width $\gamma$ of the diquark
BS amplitude until the eigenvalue obtained from the BSE agrees with the
normalization integral in Eq.~(\ref{Ns}); that is, until the constraint 
$\lambda = N_s^2$ is satisfied. 

A typical solution of the nucleon BSE is shown in Fig.~\ref{S1_S2}.
Five orders were retained in the Chebyshev expansion.   
The figure depicts the relative importance of the first four Chebyshev
moments of the nucleon amplitudes $S_1(p,P_n)$ and $ S_2(p,P_n)$. 
The respective results for their fifth moments are too small ($\le 10^{-4}$) 
to be distinguished from zero on the scale of Fig.~\ref{S1_S2}. 
This provides an indication for the high accuracy obtained from the Chebyshev
expansion to this order. This particular solution, which will be shown to
provide a good description of the electric form factors in Sec.~\ref{ElmFFs},
was obtained for the dipole form of the diquark amplitude, {\it i.e.}, from
Eq.~(\ref{P_amps}) with $n=2$, using $m_q = 0.62 M_n$ and $m_s = 0.7 M_n$
(with $\sigma = \sigma' = 1/2 $, $\eta = 1/3$).  
The value for the corresponding diquark width $\gamma_2$, necessary to yield
$\lambda = N_s^2 $ (to an accuracy of $10^{-3}$), resulted thereby to be
$\gamma_2 = (0.294\, M_n)^2$. 

The dependence of the BS eigenvalue on the momentum partitioning parameter
$\eta$ is shown in Fig.~\ref{eta_fig}. The complex domains of the constituent
quark and diquark momentum variables, $k_q^2$ and 
$k_s^2$ respectively, that are sampled by the integration over the relative
momentum $k$ in the nucleon BSE are devoid of poles for  
$ 1 - m_s/M_n < \eta < m_q /M_n$. The pole in the momentum $q^2$ of the 
exchanged quark of the kernel in Eq.~(\ref{x_kern_par}) is avoided by
imposing the further bound $\eta > (1- m_q/M_n)/2$ on the momentum partitioning
parameter. Therefore, with the present choice of constituent masses it is for
values of $\eta$ in the range $ 0.3 < \eta < 0.6$, for which our numerical
methods can yield stable results.

In principle, the integrations necessary to solve the nucleon BSE should
always be real and never lead to an imaginary result (since $m_q + m_s
>M_n$). More refined numerical methods would be necessary, however, 
when integrations were to be performed in presence of 
the poles in the constituent propagators that occur within the integration
domain for values of $\eta$ outside the above limits. 

\begin{figure}[t]
\leftline{\hskip -.1cm 
\epsfig{file=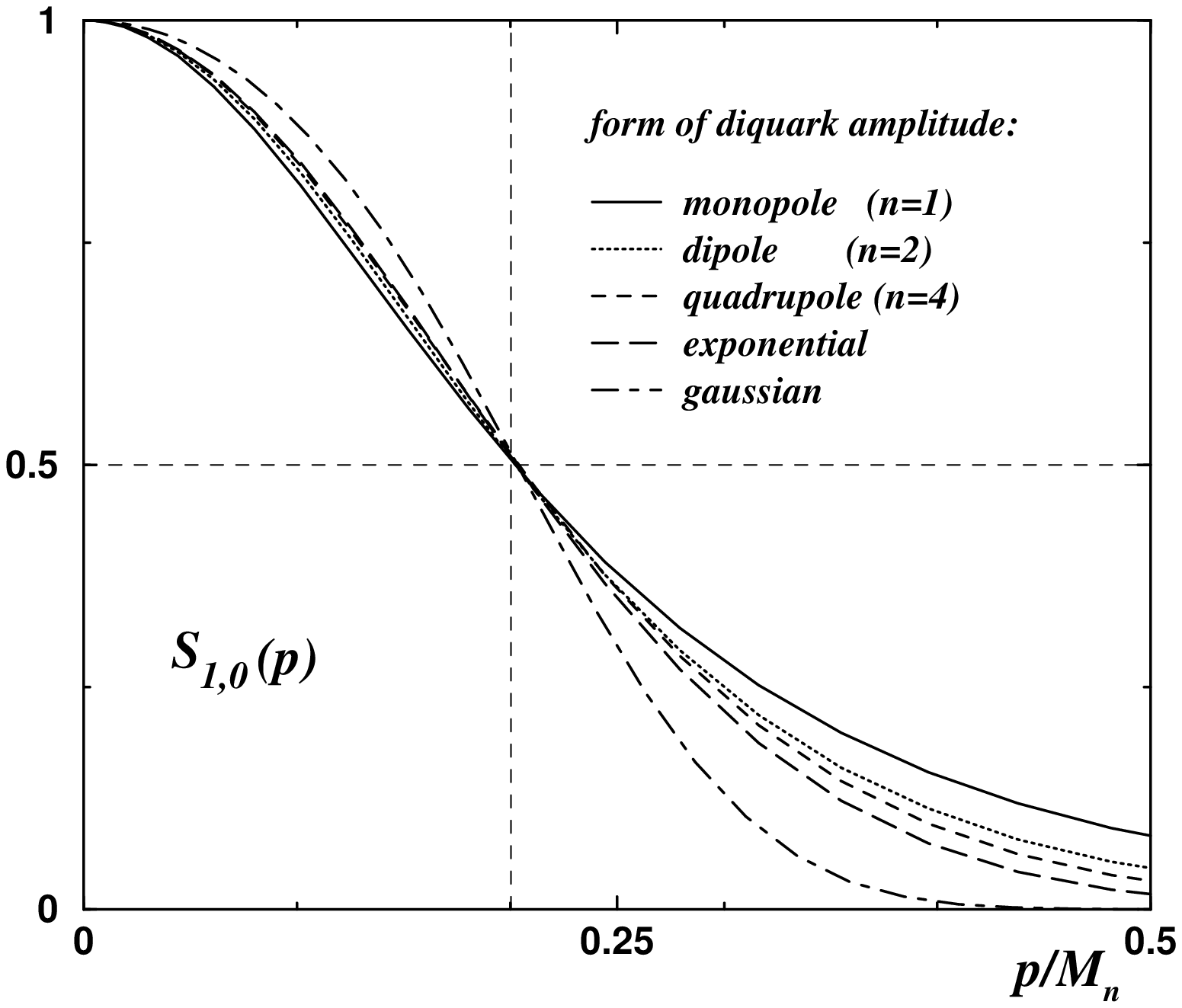,width=0.49\linewidth} 
\hskip .2cm \epsfig{file=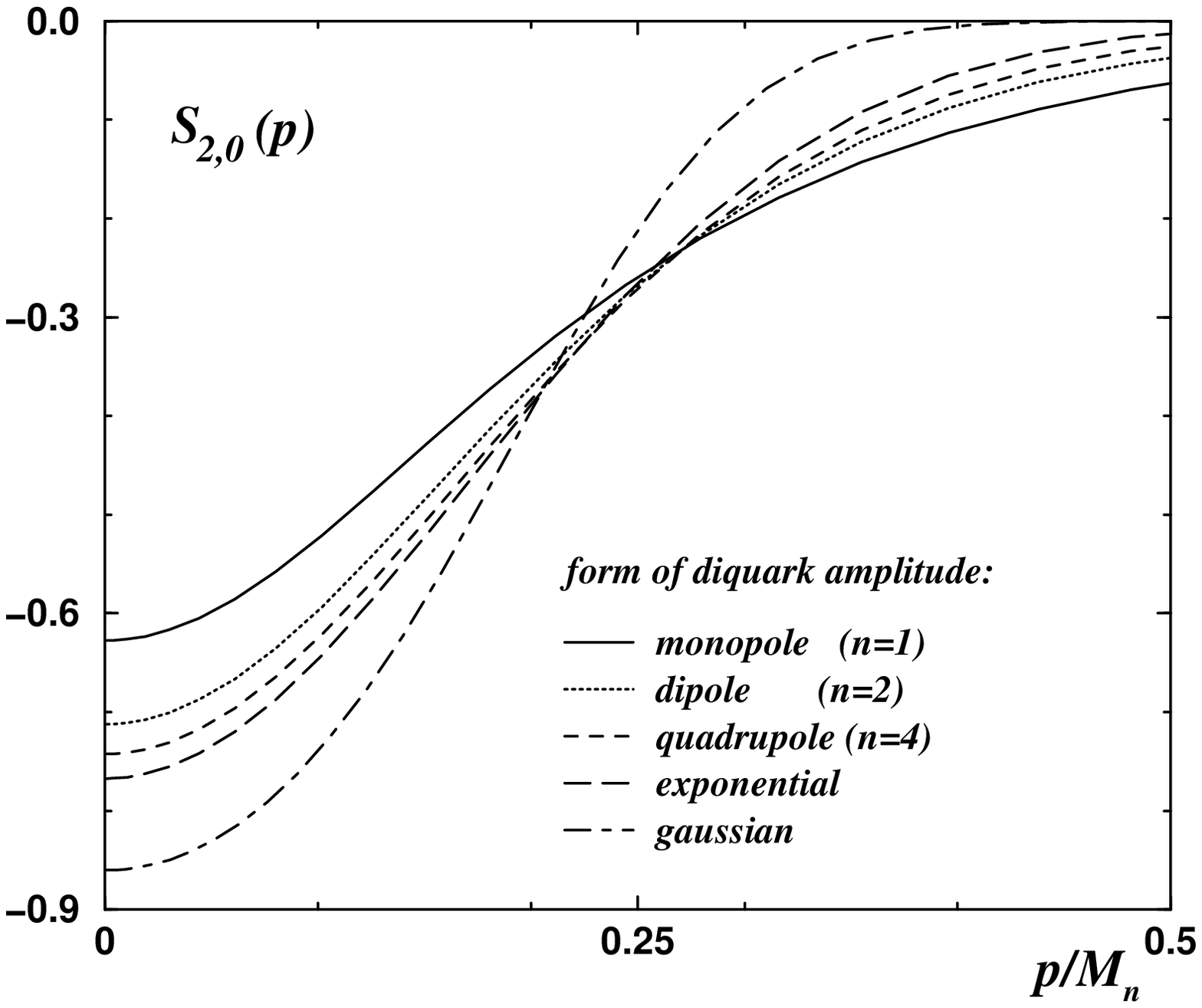,width=0.49\linewidth} }
\caption{The leading moments of the nucleon BS amplitudes $S_1$ (left)
and $S_2$ (right) for the diquark amplitudes 
(\protect\ref{P_amps},\protect\ref{P_gau}) with $m_q$ adjusted to yield 
$S_{1,0}(p)\vert_{p = 0.2M_n} = 1/2$.}
\label{S1b_S2b}
\end{figure}

The momentum partitioning within the nucleon $\eta$ is not an observable.
Hence, for any value of $\eta$ for which our numerical results are stable,
we expect the eigenvalue of the nucleon BSE $\lambda$ (which implicitly 
determines the nucleon mass) to be independent of $\eta$.  
In Fig.~\ref{eta_fig}, the dependence on $\eta$ of the nucleon BSE
eigenvalue $\lambda$ is explored using a fixed value for $\sigma = \sigma'
= 1/2$ (solid curve) and using the values of $\sigma$ and $\sigma'$ from
Eq.~(\protect\ref{dq_mom_part}) (dashed curve). 

In the first case, the arguments of the diquark amplitudes simplify to 
$x_i = - p_i^2$ with $p_1 = -(1\!-\!3\eta)P_n/2 + p + k/2$
and $p_2 = (1\!-\!3\eta)P_n/2 - k - p/2 $. 
This implies that the real parts of $x_i$ are guaranteed to be positive only
for  $\eta = 1/3$.  For values of $\eta\not=1/3$, a negative contribution
arises from the timelike nucleon bound-state momentum $P_n$.  
If the $n$-pole forms for the diquark amplitudes are employed, this results
in the appearance of artificial poles whenever $(1\!-\!3\eta)^2 M_n^2/4  \ge
\gamma_n $.  The value of $\gamma_2 = (0.294\, M_n)^2$ as it results
here, would entail the appearance of a pole for $\eta \ge 0.53$. This
explains why our numerical procedure becomes unstable as $\eta$ approaches the
value $1/2$ in this case.

No such timelike contribution to the $x_i$ arises in the second
case shown in Fig.~\ref{eta_fig}. 
Here, for $\sigma = \sigma' = (1\!-\!2\eta)/(1\!-\!\eta)$ 
the relative momenta $p_i$ are independent of $P_n$, and there are no terms
$\propto M_n^2 $ in $x_i$; see, for example, Eqs.~(\ref{x_1},\ref{x_2}). 
However, as $\eta \to 1/2$, we find that $\sigma , \sigma' \to 0$ and the
diquark normalization integrals are then affected by singularities.

\begin{table}[b]
\begin{center}

\begin{tabular}{lcccc}
\noalign{\leftline{Fixed width $S_1(p)\vert_{p = 0.2M_n} = 1/2$,
see Fig.~\protect\ref{S1b_S2b}:}}
 $P(x)$ & $m_s\; [M_n]$ & $m_q\; [M_n]$  &  $\sqrt{\gamma} \; [M_n]$ & 
                                                     $ g_s = 1/N_s $ \\ \hline 
  $n= 1$ &  0.7  & 0.685  & 0.162 &    117.1\\[-2pt]
  $n= 2$ &  0.7  & 0.620  & 0.294 &    91.80\\[-2pt]
  $n= 4$ &  0.7  & 0.605  & 0.458 &    85.47\\[-2pt]
  $n= 6$ &  0.7  & 0.600  & 0.574 &    84.37\\[-2pt]
  $n= 8$ &  0.7  & 0.598  & 0.671 &    83.76\\[-2pt]
   EXP   &  0.7  & 0.593  & 0.246 &    82.16\\[-2pt]
   GAU   &  0.7  & 0.572  & 0.238 &    71.47\\  
\noalign{\leftline{Fixed masses, see Fig.~\ref{S1a_S2a}:}} 
  $n= 1$ &  0.7  & 0.62   & 0.113 &    155.8\\[-2pt]
  $n= 2$ &  0.7  & 0.62   & 0.294 &    91.80\\[-2pt]
  $n= 4$ &  0.7  & 0.62   & 0.495 &    81.08\\[-2pt]
  $n= 6$ &  0.7  & 0.62   & 0.637 &    78.61\\[-2pt]
   EXP   &  0.7  & 0.62   & 0.283 &    74.71
\end{tabular}

\end{center}
\caption{Summary of parameters used for the various diquark amplitudes 
 $P(x)$, {\it c.f.},
 Eqs.~(\protect\ref{P_amps},\protect\ref{P_gau}). \label{FWHM_table}}  
\end{table}

In the restricted range allowed to $\eta$, the results for the nucleon BSE
obtained herein are found to be independent of $\eta$ to very good 
accuracy when the equal momentum partitioning between the quark in the
diquarks is used, $\sigma = \sigma' = 1/2 $. 
In this case, the diquark normalization in Eq.~(\ref{Ns}) yields a fixed
value for $N^2_s$ and any $\eta$ dependence of the ratio
$\lambda/N_s^2$ has to arise entirely from the nucleon BSE. The $\eta$
independence of $\lambda$ thus demonstrates that the solutions to the nucleon
BSE are under good control. 
 
The more considerable $\eta$-dependence observed for $\sigma = \sigma' =
(1\!-\!2\eta)/(1\!-\!\eta)$ arises from the dependence of $N_s^2$ on
$\sigma$, $\sigma'$.  Here, the limitations of the model assumptions 
for the {\em diquark} BS amplitudes are manifest.  The dependence of the
diquark BS amplitudes on the diquark bound-state mass, higher Chebyshev
moments (which have been neglected) and the Lorentz structures
(which we have neglected), are all responsible for this observed $\sigma$
dependence.  Such a dependence would not occur, had the diquark
amplitudes employed herein been calculated from the diquark BSE.

For the numerical calculations of electromagnetic form factors presented
in Sec.~\ref{TECO}, we therefore choose to restrict the model to 
$\sigma = \sigma' = 1/2 $ and vary $\eta$.   
From the observed independence of the nucleon BSE solutions to $\eta$ 
(when $\sigma = \sigma' = 1/2$), one expects to find that calculated
nucleon observables, such as the electromagnetic form factors, will display
a similar independence of $\eta$ as well.

\begin{figure}[t]
\leftline{\hskip -.1cm 
\epsfig{file=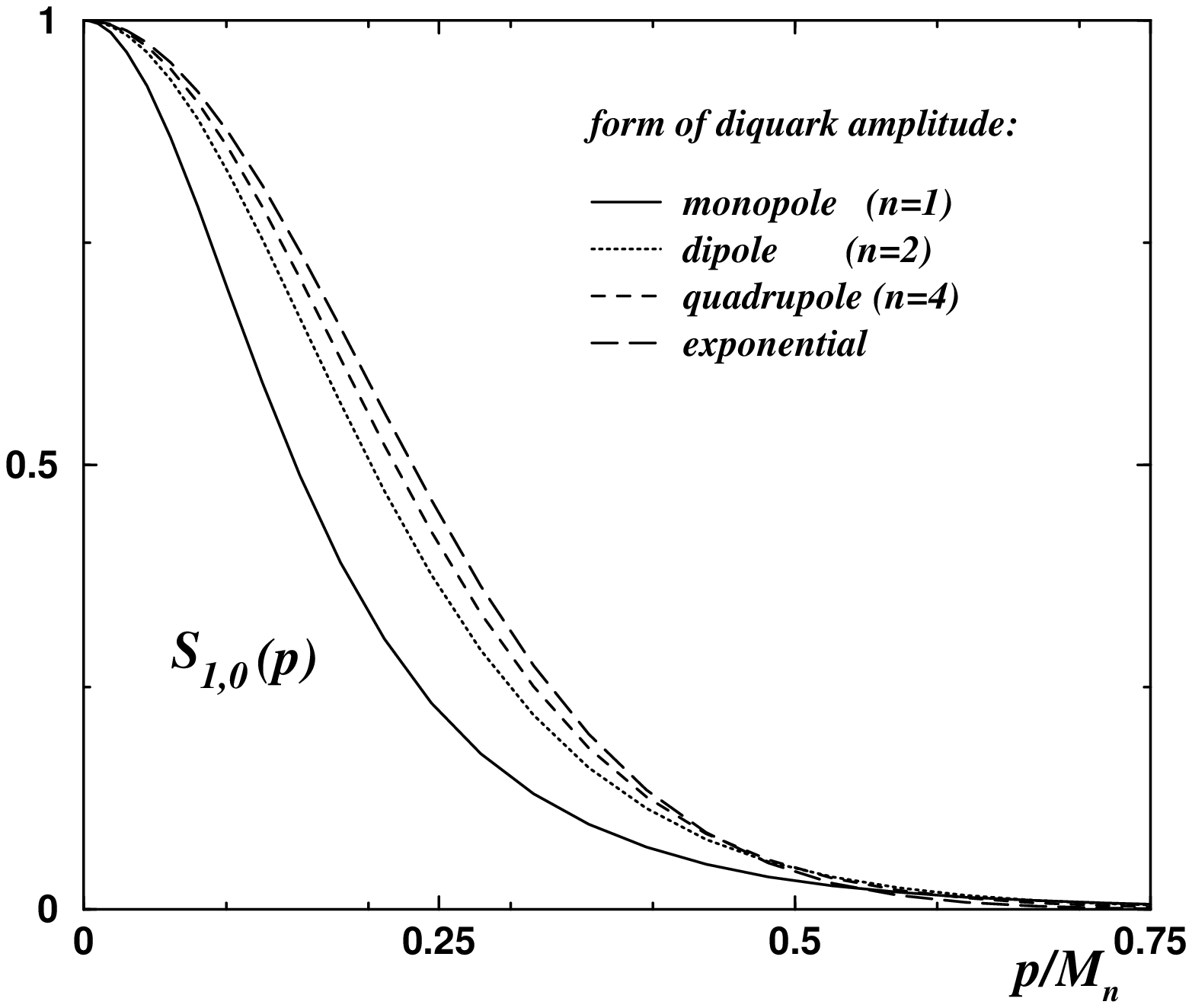,width=0.49\linewidth} 
\hskip .2cm \epsfig{file=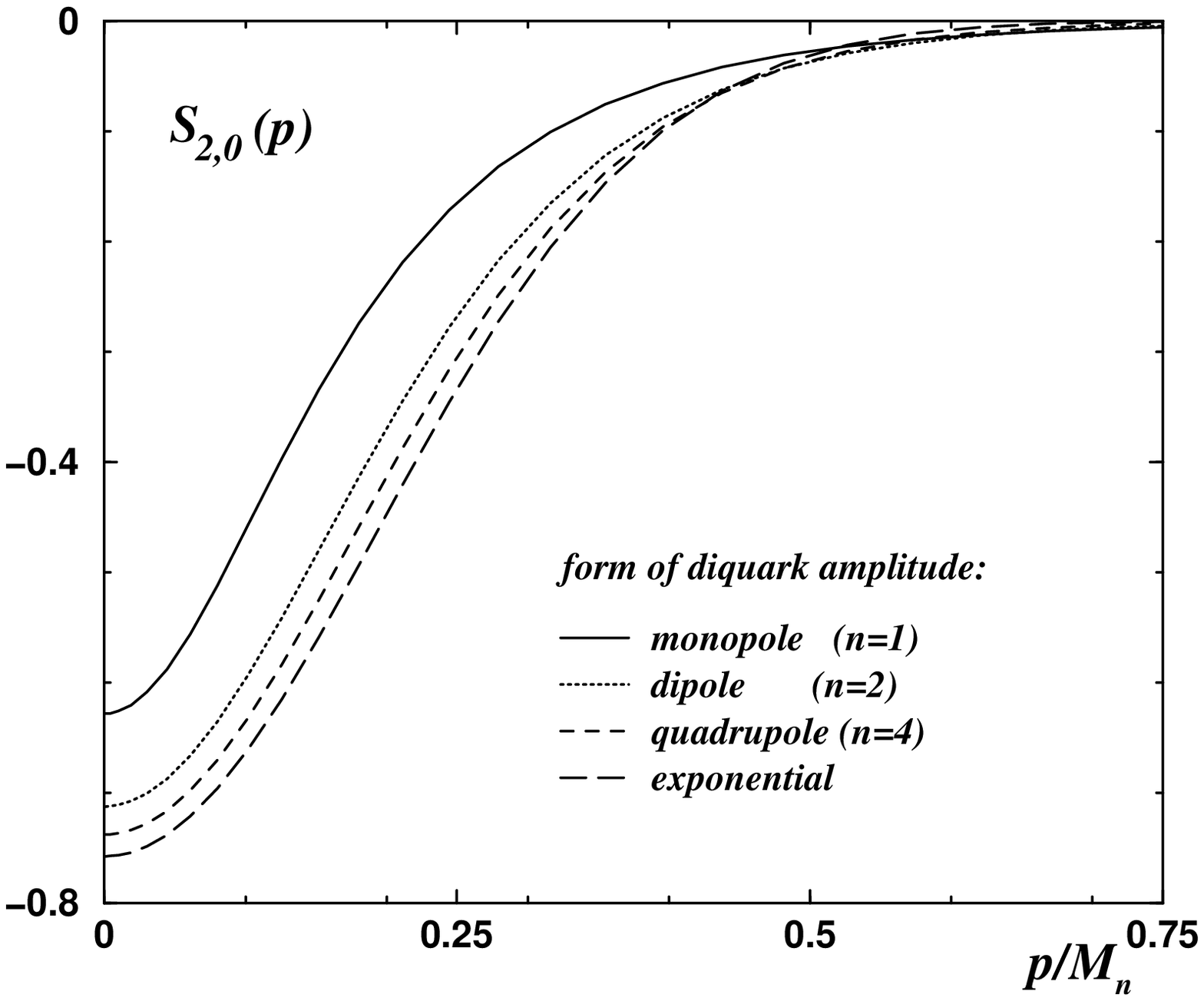,width=0.49\linewidth} }
\caption{The moments $S_{1,0}$ (left) and $S_{2,0}$ (right) of the nucleon
amplitudes for the diquark amplitudes of
Eqs.~(\protect\ref{P_amps},\protect\ref{P_gau})  
with fixed masses, $m_s = 0.7 M_n$, $m_q = 0.62 M_n$.} 
\label{S1a_S2a}
\end{figure}

In Figure~\ref{S1b_S2b} the zeroth Chebyshev moments $S_{1,0}(p)$
and $S_{2,0}(p)$ obtained from a numerical solution of the nucleon BSE
with $\sigma = \sigma' =1/2 $ and $\eta = 1/3$ are shown, for various
diquark amplitudes of the forms given in Eqs.~(\ref{P_amps}) and
Eq.~(\ref{P_gau}) with $x_0 =0$.  
To provide a comparison between these different forms of diquark BS
amplitude,  we have chosen to keep the diquark mass fixed and vary the
quark mass until a solution of the nucleon BSE was found with the property
that 
\begin{eqnarray}  S_{1,0}(p)\vert_{p = 0.2M_n}
\stackrel{!}{=} 1/2 \; .\label{FWHMC} 
\end{eqnarray} 
The resulting values for $m_q$ along with the corresponding diquark widths
and couplings $g_s \equiv 1/N_s$ are summarized in Table~\ref{FWHM_table}.  

We observe that the mass of the quark required to satisfy the condition in
Eq.~(\ref{FWHMC}) tends to smaller values for
higher $n$-pole diquark amplitudes. 
On the other hand, for fixed constituent quark (and diquark) mass
the higher $n$-pole diquark amplitudes lead to wider nucleon amplitudes.
This is shown in Fig.~\ref{S1a_S2a} and the corresponding diquark widths
and couplings are given in the lower part of Table~\ref{FWHM_table}. 
The mass values hereby correspond to the results shown in the previous
Figure~\ref{S1b_S2b} for $n= 2$.

\begin{figure}[t]
\leftline{\hskip -.1cm 
\epsfig{file=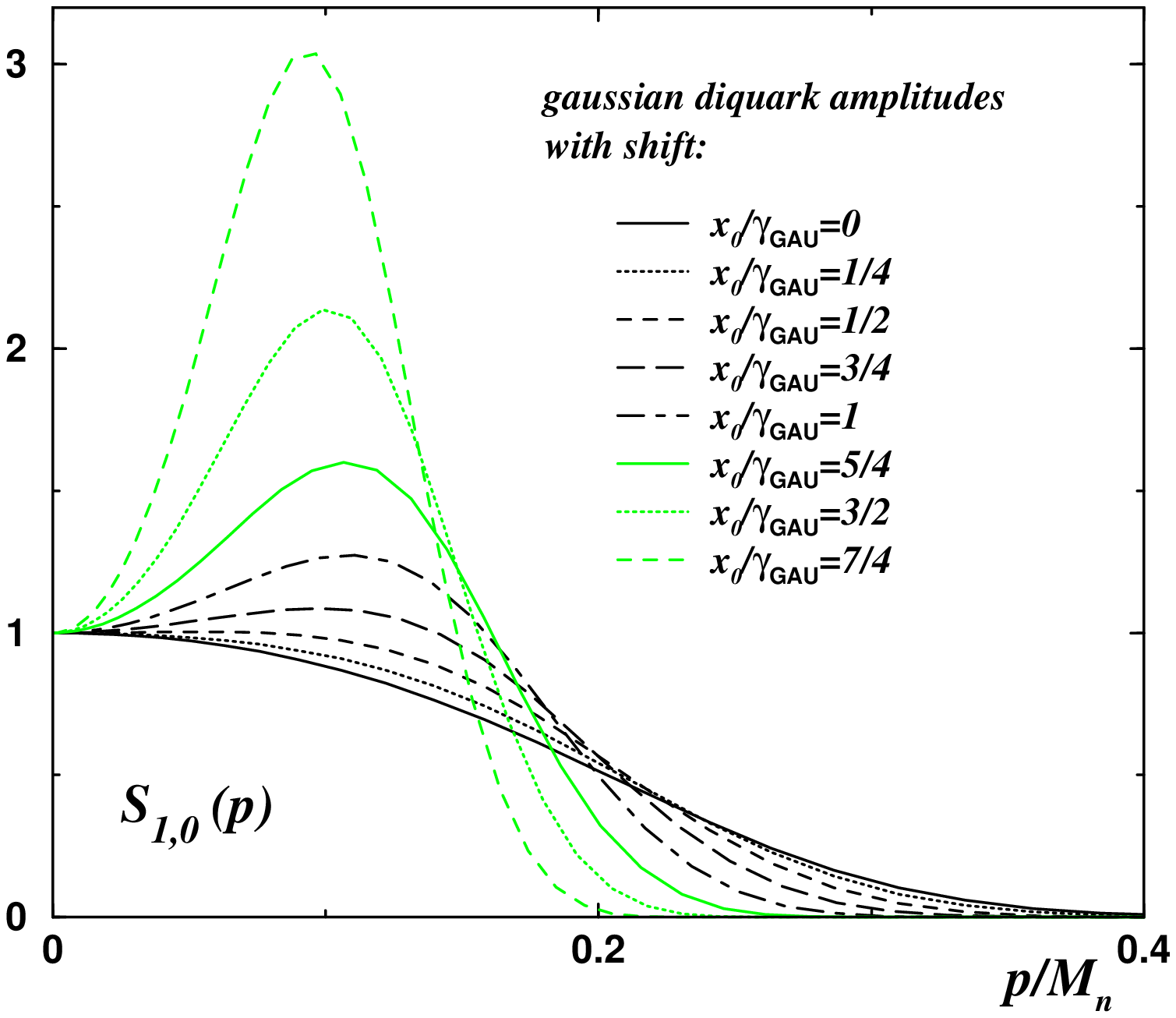,width=0.49\linewidth} 
\hskip .2cm \epsfig{file=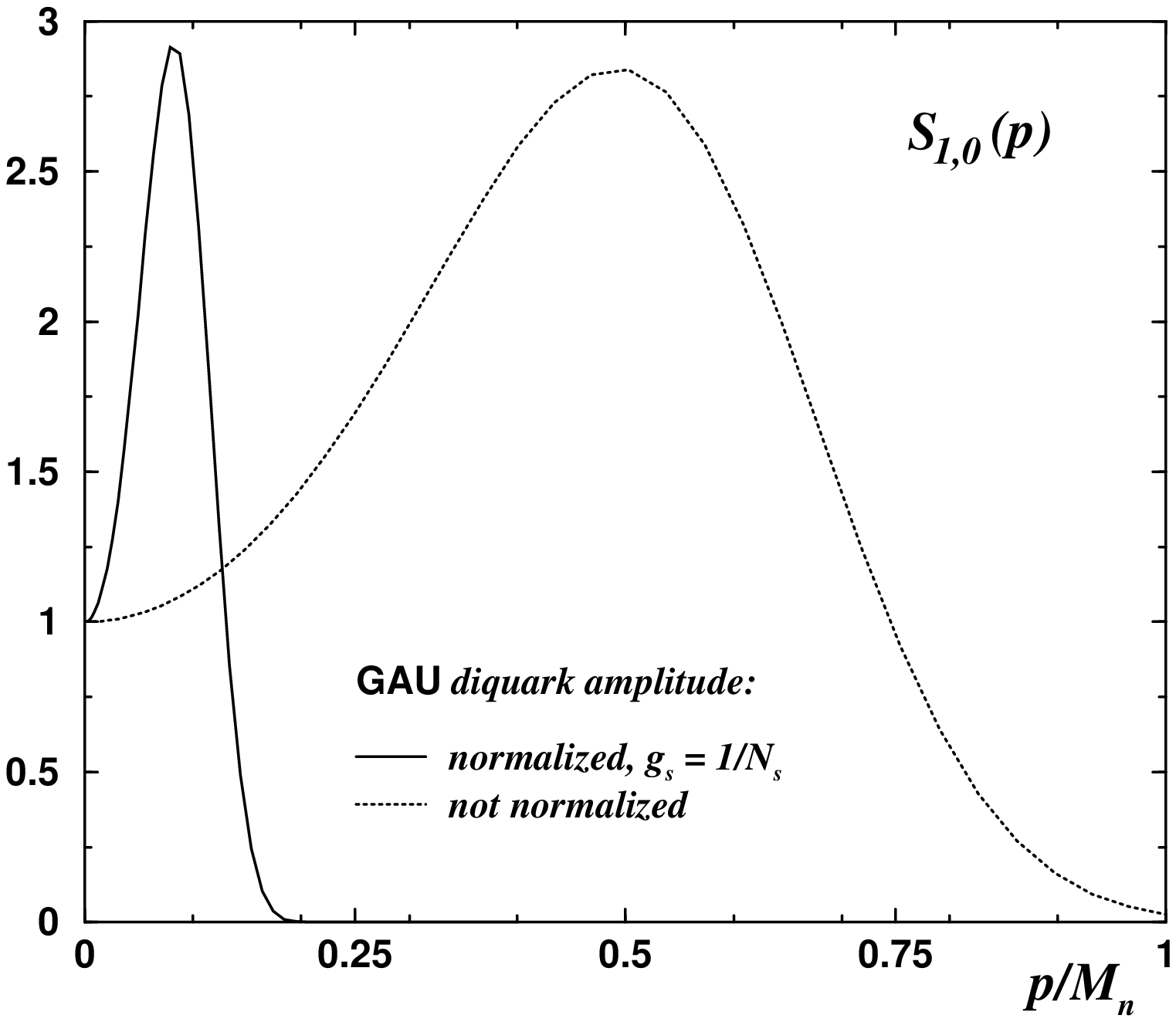,width=0.49\linewidth} }
\caption{The moment $S_{1,0}$ for the Gaussian diquark amplitudes of
Eq.~(\protect\ref{P_gau}) with various shifts $x_0/\gamma_{\mbox{\tiny GAU}}$
(left), see Table~\protect\ref{Sgau_table}, and a comparison (right) of our
result for $x_0/\gamma_{\mbox{\tiny GAU}} = 19/11$ (with $m_s = 0.7 M_n$,
$m_q = 0.555 M_n$) with a result we obtain from using
$\gamma_{\mbox{\tiny GAU}} = 0.11 $GeV$^2$ as in Ref.~\protect\cite{Bur89}.}  
\label{S1a_S1gau}
\end{figure}

The qualitative effect of shifting the maximum in the Gaussian forms in 
Eq.~(\ref{P_gau}) for the diquark amplitudes by a finite amount $x_0$, is
shown in Fig.~\ref{S1a_S1gau}. 
The curve for $x_0 = 0$ resembles the Gaussian result given in
Fig.~\ref{S1b_S2b} with masses $m_s = 0.7 M_n$ and  $m_q = 0.572 M_n$,   
which are kept fixed in the results for finite shifts $x_0$. 
The corresponding widths and normalizations of these forms for the diquark
amplitudes are given in Table~\ref{Sgau_table}. 
While the widths $\gamma$ and the couplings $g_s$ are not free parameters
in our approach, the additional parameter  
$x_0/\gamma_{\hbox{\tiny GAU}}$ introduced into the form of the diquark BS
amplitude is free to vary. 
In contrast to the Gaussian form of the model diquark BS amplitude
employed in Ref.~\cite{Bur89} with $x_0/\gamma_{\hbox{\tiny GAU}} =
0.19/0.11 \simeq 1.73$, by implementing the diquark normalization condition, 
we find that the width must be about 30 times smaller to provide a sufficient
interaction strength $g_s = 1/N_s^2$ necessary to bind the nucleon.

To provide for a closer comparison with the results of Ref.~\cite{Bur89},
we have also solved the nucleon BSE using the values of parameters
employed in that study.  That is, we have solved the nucleon BSE using
$P_{\hbox{\tiny GAU}}(x)$ with the parameter $x_0/\gamma_{\hbox{\tiny
GAU}} = 0.19/0.11$  and using the constituent quark and diquark masses
$m_q = 0.555 M_n$  and $m_s = 0.7 M_n$, respectively. 
(Note that we use the nucleon mass as the intrinsic momentum scale of our
framework, whereas the momentum scale employed in Ref.~\cite{Bur89} was
1~GeV.)
The value employed in Ref.~\cite{Bur89} for the diquark mass 
$m_s = $ 0.568~GeV implies that $M_n =$ 811~MeV in order to compare to 
our calculations (with $m_s = 0.7 M_n$), and our value for the quark mass
($m_q = 0.555 M_n$) thus corresponds to $m_q = $ 0.45~GeV.
With these para\-meters, we obtain $\gamma_{\hbox{\tiny GAU}} =  (0.216 \,
M_n)^2 =$ $ 0.308 \, 10^{-2} \mbox{GeV}^2$, 
which may be compared to $\gamma_{\hbox{\tiny GAU}}
=  (0.409\, M_n)^2 = 0.11\, \mbox{GeV}^2$ used in Ref.~\cite{Bur89}.

\begin{wraptable}[15]{l}{6cm}
\begin{tabular}{ccc}
\noalign{\leftline{$m_s = 0.7 M_n$, $m_q = 0.572 M_n$:}}
  $x_0/\gamma_{\hbox{\tiny GAU}} $  &  $\sqrt{\gamma_{\hbox{\tiny GAU}}}\; [M_n]$
    &  $ g_s = 1/N_s $ \\ \hline   
    0    & 0.238 &    71.47\\[-2pt]
   1/4   & 0.209 &    69.09\\[-2pt]
   1/2   & 0.181 &    72.64\\[-2pt]
   3/4   & 0.154 &    81.96\\[-2pt]
    1    & 0.130 &    97.75\\[-2pt]
   5/4   & 0.109 &    121.3\\[-2pt]
   3/2   & 0.091 &    154.3\\[-2pt]
   7/4   & 0.077 &    198.5 
\end{tabular}
\caption{Width and normalizations of Gaussian amplitudes, 
 Eq.~(\protect\ref{P_gau}) for various shifts $x_0$, see
Fig.~\protect\ref{S1a_S1gau} (right). \label{Sgau_table}}
\end{wraptable}
\noindent 
The same results for the nucleon ampli\-tu\-des as shown above are used for the
calculations of the electromagnetic form factors of the proton and the
neutron in Sec.~\ref{ElmFFs}. 
We will show that when $n$-pole diquark BS amplitudes with $n=2$ and $n=4$
are employed, one obtains results for the electromagnetic form factors
that are in good agreement with the phenomenological dipole fit for the
electric proton form factor, while higher powers or exponential forms tend
to produce form factors which decrease too fast with increasing momentum
transfer $Q^2$.  
We also show that use of a Gaussian form for the diquark BS amplitude
with $x_0 \not = 0$ leads to neutron electric form factor 
that has a {\em qualitatively} different behavior than that obtained by
using the other model forms of diquark BS amplitudes. 
Furthermore, this form leads to nodes in the neutron electric form factor
for small $Q^2$, a feature for which there is {\em no} experimental
evidence.

%% file: pictex/Fig3.tex
%
\begin{figure}[t]
\vskip 1cm
\begin{picture}(16000,6000)(-14500,-3000)
\thicklines
\put(0,0){\circle{2000}}
\put(-300,-200){$\widetilde\chi$} 
\put(0,4000){\circle{2000}}
\put(-300,3800){$\widetilde{\bar\chi}$} 
\drawline\fermion[\N\REG](0,1000)[2000]
\drawarrow[\N\ATBASE](\pmidx,\pmidy)
\drawline\fermion[\NW\REG](-1000,0)[5500]
\drawarrow[\NW\ATBASE](\pmidx,\pmidy)
\drawline\fermion[\W\REG](3000,4000)[2000]
\drawarrow[\W\ATBASE](\pmidx,\pmidy)
\thinlines
\drawline\fermion[\W\REG](3000,100)[2000]
\drawline\fermion[\W\REG](3000,-100)[2000]
\drawline\fermion[\SW\REG](-1000,4100)[2440]
\drawline\fermion[\SW\REG](\pbackx,\pbacky)[1000]
\drawline\fermion[\SW\REG](\pbackx,\pbacky)[2000]
\drawline\fermion[\SW\REG](-1000,3900)[2300]
\drawline\fermion[\SW\REG](\pbackx,\pbacky)[1000]
\drawline\fermion[\SW\REG](\pbackx,\pbacky)[2000]
\thicklines
\drawarrow[\SW\ATBASE](\pmidx,1000)
\drawarrow[\W\ATBASE](2000,0)
\put(-14000,4000){\framebox{$K_{\alpha\beta}$}}
\thinlines
\drawline\fermion[\W\REG](8000,2000)[1000]
\drawarrow[\W\ATBASE](\pbackx,\pbacky)
\put(7200,2400){$P_n$}
\put(3600,3800){$\beta$}
\put(-6400,3800){$\alpha$}
\put(200,2000){$q$}
\put(-6000,-800){$q+p_\beta$}
\put(2000,4600){$p_\beta$}
\put(2000,-800){$q+p_\alpha$}
\put(-5000,4600){$p_\alpha$}
\end{picture}
\vskip -.4cm
\centerline{$\scriptstyle 
p\, = \,(1\!-\!\eta) p_\alpha - \eta (q+p_\beta) 
\; , \quad k\, =\,  (1\!-\!\eta) p_\beta - \eta (q+p_\alpha) $.}

\label{kernel}
\caption{The quark-exchange kernel of the nucleon BSE.}
\vspace{.5cm}
\end{figure}

%% file: Sec4.tex
\section{Normalization of Nucleon Bethe-Salpeter Amplitude} 
\label{SecNucNorm}

In order to reproduce the correct electromagnetic charges of the physical 
asymptotic states, it is required that their Bethe-Salpeter (or Faddeev)
amplitudes be normalized according to the normalization conditions obtained
from the fully-interacting Green functions of the elementary fields. 
These normalization conditions are 
derived from  {\em inhomogeneous} BS (or Faddeev) equations.   
In the present framework, the quark-diquark 4-point Green function
is given by (suppressing Dirac indices)
\begin{eqnarray} 
G(p,q,P) \, = \, \left( {G^{(0)}}^{-1} (p,q,P) - K(p,q,P) \right)^{-1} \; .
\end{eqnarray}
Here, $K(p,q,P)$ is the quark-exchange kernel of Eq.~(\ref{x_kern_par})
depicted in Fig.~\ref{kernel}, and $G^{(0)}(p,q,P)$ is the disconnected 
contribution to this Green function, given by 
\begin{eqnarray}
G^{(0)}_{\alpha\beta}(p,k,P) \, := \, (2\pi)^4 \delta^4(p-k) \,
S_{\alpha\beta}(k_q) \, D(k_s) \delta_{\eta \eta'} \; ,
\end{eqnarray}
with $k_q =\eta P + k$ and $k_s = (1\!-\!\eta)P - k$.
Taking the derivative of $G(p,k,P)$ with respect to the total momentum $P$
and then examining the leading contributions in the limit that 
$P^2 \rightarrow M_n^2$ which arise from the nucleon pole contribution
from Eq.~(\ref{nuc_pole_cont}), one finds 
\begin{eqnarray} 
 M_n \, \Lambda^+(P_n)  &\stackrel{!}{=}& 
 \; i \int \frac{d^4p}{(2\pi)^4}
\frac{d^4k}{(2\pi)^4}  \label{gen_nuc_norm} \\
 && \hskip -1.2cm \bar\psi(-p,P_n)  \left(
P_\mu\frac{\partial}{\partial P_\mu} G^{-1}(p,k,P) \right)_{P = P_n}  
\! \psi(k,P_n) \; . \nn  
\end{eqnarray} 
The most important difference in normalizing the amplitudes in the
Bethe-Salpeter framework developed here and a genuine {two-body}
BSE, is the dependence of the BS kernel on the total  momentum of the
bound-state $P$. In phenomenological studies of 2-body bound states 
BS kernels, such as ladder-approximate ones for example, are most commonly
employed in a form which does not depend on the bound-state momentum. 
However, in the present approach, the original 3-body nature of the
nucleon bound state {\em requires} the exchange kernel of the reduced 
BS equation for quark and diquark to depend on the total momentum $P_n$ of
the nucleon.   
In particular, when $\eta \not= 1/2$, the propagator for the
exchanged quark in the kernel depends on the total momentum $P_n$ and for
$\eta  \not= 1/3$, the diquark BS amplitudes depend on $P_n$.
This added complication, which is easily
avoided by ladder-approximate studies of meson bound states is unavoidable
for studies of baryons. 
Thus, the normalization condition for the nucleon will always contain
contributions from the derivative of the kernel.

Following the above procedure, one obtains a normalization condition of
the form
\begin{eqnarray}
1 \, &\stackrel{!}{=}&  \label{nuc_norm} \\
&& \hskip -.1cm  \eta N_q + (1-\eta) N_D + (1-2\eta) N_X + (1-3\eta)
N_P \; , \nn
\end{eqnarray} 
where 
\begin{eqnarray}
N_q \, &=& \, - \frac{P^\mu}{2M_n} i \int \frac{d^4k}{(2\pi)^4} 
\;  D(k_s)
 \label{NqDef} \\
&& 
\hskip 2cm  
\tr \bigg[ \widetilde{\bar\psi}(-k,P) \left(\frac{\partial}{\partial
k_q^\mu} S(k_q) \right) \widetilde \psi(k,P) \bigg] \; ,\nn \\
N_D \, &=&  \, - \frac{P^\mu}{2M_n} i \int \frac{d^4k}{(2\pi)^4} 
\;  \left( \frac{\partial}{\partial k_s^\mu} D(k_s) \right)
 \label{NDDef} \\
&&\hskip 3cm 
   \tr \bigg[ \widetilde{\bar\psi}(-k,P) S(k_q) \widetilde
\psi(k,P)\bigg] \; ,\nn \\
N_X \, & = & \, - \frac{P^\mu}{2M_n} i \int
\frac{d^4p}{(2\pi)^4}\frac{d^4k}{(2\pi)^4} 
\;  \frac{1}{2N_s^2} \,  P(-p^2_1)
P(-p^2_2)  
\label{NXDef} \\ 
&& \hskip 2cm
\tr\bigg[ \bar\psi(-p,P) \left(\frac{\partial}{\partial
q^\mu} S(q) \right) \psi(k,P) \bigg]   \; , \nn  \\
N_P \, & = &  \, - \frac{P^\mu}{2M_n} i \int
\frac{d^4p}{(2\pi)^4}\frac{d^4k}{(2\pi)^4}\;  
  \frac{1}{2N_s^2} 
\label{NPDef} \\
&&
\hskip .5cm \bigg(  \,  p_{1\mu}  \, P'(-p^2_1) P(-p^2_2) 
\, -  \,  p_{2\mu} \, P(-p^2_1) P'(-p^2_2) \, \bigg)  \nn\\
&& \hskip 3cm 
\tr\bigg[ \bar\psi(-p,P)  S(q) \psi(k,P) \bigg]  \; , \nn 
\end{eqnarray}
with $ p_1 = -(1\!-\!3\eta) P/2 + p + k/2  \, , \;
p_2 =  (1\!-\!3\eta) P/2 - p/2 - k \, , \; \mbox{for} \; \sigma\! =\!\sigma'
\!=\!1/2 $.

It is clear from Eq.~(\ref{nuc_norm}) that either the exchanged quark or the
presence of a diquark substructure, or in general both, provide
non-vanishing contributions to the nucleon BS normalization.  
Maintaining these terms is critical for the correct
determination of the electromagnetic charges of baryons since the nucleon 
normalization and electromagnetic form factors are intimately related by
the differential Ward identities. 

In the following section, we show that in calculations of the electromagnetic
form factors of the nucleon, use of the usual impulse
approximation~\cite{Man55} (which includes only contributions arising from
the coupling of the photon to the quark and diquark, that are related to
the terms $N_q$ and $N_D$ in the normalization condition), by itself is
insufficient to 
guarantee electromagnetic current and charge conservation of the nucleon. 

One of the additional contributions required for the proper calculation of
electromagnetic form factors that goes beyond the usual 
impulse approximation is the coupling of the photon to the exchanged-quark
in the kernel from Eq.~(\ref{xker}).    
We will show that this term provides a crucial
contribution to the nucleon electromagnetic form factors and helps maintain
electromagnetic current conservation of the nucleon.
It is interesting to note that the contribution of this term is important
even in the special case that $\eta = 1/2$ in which the term $N_X$ does not
contribute to the normalization condition of Eq.~(\ref{nuc_norm}).  

We will show in the next section that in the presence of a diquark with
a non-trivial substructure, additional, direct couplings of the photon to
this substructure are required to maintain current conservation.
Similar contributions have previously been found important 
in several other contexts \cite{Oth89,Wan96}. These couplings are commonly
referred to as ``sea\-gulls''.

We conclude this section by summarizing that the respective 
contributions to the nucleon normalization condition of Eq.~(\ref{nuc_norm})
given in Eqs. (\ref{NqDef}) to (\ref{NDDef}) correspond to the usual impulse
approximate contributions arising from the constituent quark and diquark,
plus the contribution from the exchange-quark in the BS kernel, and the
seagull contributions which arise from the quark-substructure in the diquark
correlations.

%% file: Sec5.tex
\section{The Electromagnetic Current in Impulse Approximation and Beyond} 
\label{TECO}

The electromagnetic current operator $J^{\hbox{\tiny em}}_\mu(x)$ in impulse
approximation is determined by the disconnected contributions from
the electromagnetic couplings of the spectator quark ($J^\mu_{q}$) and the
scalar diquark ($J^\mu_{D}$) which in the Mandelstam formalism are calculated
from the following momentum space kernels~\cite{Man55}, 
\begin{eqnarray}  
  J^\mu_{q}(p,P';k,P) \, &=& \, (2\pi)^4 \delta^4(p-k-\hat\eta Q) \\
&&\hskip -1cm  q_{q} \,  \Gamma^\mu_{q} (\eta P' + p, \eta P +
  k)  \;  D^{-1}(\hat\eta P - k) \; , \nn \\
  J^\mu_{D}(p,P';k,P) \, &=& \, (2\pi)^4 \delta^4(p-k+\eta Q)  \\
&&\hskip -1cm q_{D}  \, \Gamma^\mu_{D} (\hat\eta P' - p, \hat\eta
  P - k)  \;  S^{-1}(\eta P + k) \; , \nn
\end{eqnarray} 
with $Q = P' - P$ and $\eta + \hat\eta = 1$. The charge of
the spectator quark in the nucleon is denoted $q_{q}$, the charge of the 
scalar diquark is $q_{D}$, and $q_{q} + q_{D} = $ 1 and 0 for the
proton and neutron, respectively. 
The Ward-Takahashi identities for the quark and diquark electromagnetic
vertices are
\begin{eqnarray}  
  Q_\mu \Gamma^\mu_{q}(p+Q,p) \, &=& \, iS^{-1}(p+Q) \, - \, iS^{-1}(p)
  \label{qk_WTI_nuc} \; , \\
  Q_\mu \Gamma^\mu_{D}(p+Q,p) \, &=& \, iD^{-1}(p+Q) \, - \, iD^{-1}(p)  \; .
\label{dq_WTI_nuc}
\end{eqnarray} 
From these one can immediately write down the nucleon matrix elements for
the divergences of the corresponding Mandelstam currents between initial
and final nucleon states with momentum and spin $P, \, s$ and $P', \, s'$
respectively, as 
\begin{eqnarray}  
\langle P',s' | \partial_\mu J^\mu_{q}(0) | P,s \rangle  &=&
 \,    q_{q}  \int \frac{d^4k}{(2\pi)^4} \, \Big\{ \label{dsqc} \\
&& \hskip -2.5cm 
 \bar u(P',s')\, \widetilde{\bar\psi}(-(k+\hat\eta Q), P')  \psi(k,P) \,
 u(P,s) \nn\\
&& \hskip -2cm 
-\,  \bar u(P',s') \, \bar\psi(-(k+\hat\eta
Q), P')  \widetilde{\psi}(k,P) \,  u(P,s)  \Big\}  , \nn \\ 
\langle P',s' | \partial_\mu J^\mu_{D}(0) | P,s \rangle  &=&
 \,      q_{D} \int \frac{d^4k}{(2\pi)^4} \,  \Big\{  \label{dsdc} \\
&& \hskip -2.5cm
  \bar u(P',s') \,  \widetilde{\bar\psi}(-(k -\eta Q), P')  \psi(k,P) \,
 u(P,s) \nn\\
&& \hskip -2cm 
- \, \bar u(P',s') \,  \bar\psi(-(k-\eta
Q), P')  \widetilde{\psi}(k,P)  \, u(P,s) \Big\} \; . \nn
\end{eqnarray} 
Here, we insert the BSEs for the 
amplitudes $\widetilde\psi$ and
$\widetilde{\bar\psi}$ which can be written in the
compact form,
\begin{eqnarray} 
     \widetilde\psi(p,P)  &=& \int \frac{d^4k}{(2\pi)^4} \, K(p,k,P)
     \, \psi(p,P) \; , \label{BSEpsi} \\
     \widetilde{\bar\psi}(-p,P)  &=& \int \frac{d^4k}{(2\pi)^4} \,
     \bar\psi(-k,P)\,  K(p,k,P) \; . \label{BSEpsibar}
\end{eqnarray} 
After shifting the integration momentum by $p+\hat\eta Q \to p$
in the second terms of Eqs.~(\ref{dsqc}) and (\ref{dsdc}), one obtains, 
\begin{eqnarray}  
 \langle P',s' | \partial_\mu J^\mu_q(0) | P,s \rangle \, &=& \,  q_q  
\int \frac{d^4p}{(2\pi)^4}  \frac{d^4k}{(2\pi)^4}  \, \Big\{ \label{mac1} \\
&& \hskip -3.5cm  
   \bar u(P',s') \, \bar\psi(-p, P') K(p, k+\hat\eta Q, P')
\psi(k,P)  \, u(P,s) \nn \\ 
&& \hskip -3.2cm 
- \, \bar u(P',s')\,   \bar\psi(-p, P')  K(p-\hat\eta Q, k, 
P) \psi(k,P)  \, u(P,s) \Big\}  \; , \nn \\
 \langle P',s' | \partial_\mu J^\mu_D(0) | P,s \rangle \, &=& \,  q_D  
\int \frac{d^4p}{(2\pi)^4}  \frac{d^4k}{(2\pi)^4} \, \Big\{ \label{mac2} \\
&& \hskip -3.5cm  
   \bar u(P',s') \,  \bar\psi(-p,
P') K(p, k-\eta Q, P')  \psi(k,P) \, u(P,s)  \nn\\ 
&& \hskip -3.2cm 
-\, \bar u(P',s') \, \bar\psi(-p, P')  K(p+\eta Q,k,P) \psi(k,P)  \,
u(P,s) \Big\} \; .  \nn 
\end{eqnarray} 
Examination of the terms in Eqs.~(\ref{mac1}) and (\ref{mac2}) reveals
that their sum gives rise to a conserved current only if $K(p,k,P) $  $\equiv
K(p-k) $.  That is, if the nucleon BS 
kernel is independent of the total nucleon bound-state momentum $P$ and, in
addition, it only depends on the {\em difference} of the relative momenta
$p-k$. 
These criteria are satisfied in studies of meson bound states within the
ladder approximation, see for example Ref.~\cite{Tan97}.
However, we observe that even in absence of an explicit
dependence on the total nucleon momentum $P$, the exchange kernel of the
BSE, as obtained from the nucleon Faddeev equation, {\em necessarily}
depends on the sum of the relative momenta $p+k$ and not their difference. 
It follows that even with approximating the exchange kernel by a  
$P$-independent one, which corresponds to neglecting diquark substructure
together with using $\eta = 1/2$ as in Refs.~\cite{Hel97b,Oet98}, the
electromagnetic current of the nucleon is not conserved in the impulse
approximation and it is already necessary to include an additional coupling
of the photon to the quark-exchange kernel.   

It is interesting to note that similar photon-kernel couplings are
required in other systems as well.  For example, while the nucleon-nucleon
scattering kernels due to meson exchanges depend only on the difference of
the relative momenta ({\it i.e.}, $K \equiv K(p-k)$), the isospin dependence of
a charged meson that is exchanged between the two nucleons requires 
one to introduce additional photon couplings to the exchanged meson to
maintain current conservation, in this case the charged pion.
Such contributions play an important role in determining the
electromagnetic form factors of the deuteron \cite{Gro87}. For the importance
of meson-exchange currents in few-nucleon systems, see also the recent
review, Ref.~\cite{Car98}.

The coupling of the exchange quark (with electromagnetic charge $q_X$) 
in the kernel of Eq.~(\ref{xker}) gives rise to the additional
contribution $J_X$ to the nucleon current:
\begin{eqnarray}
&& \hskip -.9cm J^\mu_X(p,P';k,P)\, = \,\hbox{\hfill}  \, 
- q_X \, \frac{1}{2} \label{J_exchange} \\ 
&&\hskip -.5cm   \widetilde \chi(p_1,\hat\eta P-k) 
\, S^T(q) \, {\Gamma^\mu_q}^T(q',q) \, S^T(q') \,
\widetilde{\bar\chi}^T(p_2',\hat\eta P' -p) \; , \nn \\
\hbox{with} && 
q \, = \,  \hat\eta P - \eta P' - p - k \; , 
\; \; q' \, = \, q + Q \; ,\nn \\
&& \hskip -.9cm 
p_1 \,=\,  \sigma (\eta P' + p) - \hat\sigma q  \; ,  \; 
\; p_2' \, = \,  - \sigma' (\eta P+k) + \hat\sigma' q' \; ,  \nn \\
\hbox{and~} && \sigma  + \hat\sigma =  \sigma'  + \hat\sigma'  =  1 \; . \nn
\end{eqnarray} 
From the Ward-Takahashi identity for the quark-photon vertex in
Eq.~(\ref{qk_WTI_nuc}), one finds that the divergence of the
exchanged-quark contribution to the nucleon electromagnetic current is 
\begin{eqnarray}
Q_\mu J^\mu_X(p,P';k,P)\, &=& \label{QJ_X} \\ 
&&\hskip -2.5cm \,- i q_X \, \frac{1}{2} \, \biggl( 
 \widetilde \chi(p_1,\hat\eta P-k) 
\, S^T(q) \, \widetilde{\bar\chi}^T(p_2',\hat\eta P' -p)   \nn \\
&&\hskip -1cm -  \,
 \widetilde \chi(p_1,\hat\eta P-k) 
\,  S^T(q') \,\widetilde{\bar\chi}^T(p_2',\hat\eta P' -p) \, \biggr) \; . \nn
\end{eqnarray} 
To provide a comparison with the quark and diquark electromagnetic
currents given above for the quark-exchange kernel of Eq.~(\ref{xker})
written using the same momentum conventions as used 
in Eq.~(\ref{J_exchange}), we rewrite the divergences of the Mandelstam
currents given in Eqs.~(\ref{mac1}) and (\ref{mac2}) as 
\begin{eqnarray}
Q_\mu J^\mu_q(p,P';k,P)\, &=& \label{e5.13} \\ 
&&\hskip -3cm \,- i q_q \, \frac{1}{2} \, \biggl( 
 \widetilde \chi(p_1,\hat\eta P-k) 
\, S^T(q) \, \widetilde{\bar\chi}^T(p_2'-Q,\hat\eta P' -p)  \nn \\
&&\hskip -2cm -  \, 
 \widetilde \chi(p_1-Q,\hat\eta P-k) 
\,  S^T(q') \,\widetilde{\bar\chi}^T(p_2',\hat\eta P' -p) \, \biggr) \; , \nn \\
Q_\mu J^\mu_D(p,P';k,P)\, &=& \label{e5.14} \\ 
&&\hskip -3cm \,- i q_D \, \frac{1}{2} \, \biggl( 
 \widetilde \chi(p_1-\hat\sigma Q,\hat\eta P-k +Q) 
\, S^T(q') \, \widetilde{\bar\chi}^T(p_2',\hat\eta P' -p)  \nn \\
&&\hskip -2.5cm 
-  \,  \widetilde \chi(p_1,\hat\eta P-k) 
\,  S^T(q) \,\widetilde{\bar\chi}^T(p_2' - \hat\sigma' Q ,\hat\eta P' -p -Q)
\, \biggr) \; . \nn 
\end{eqnarray} 
Since $ q_q \, - \, q_D \, + \, q_X \, = \, 0 $,
one thus finds that, in the case of a  point-like diquark BS amplitude, 
({\it i.e.}, neglecting any momentum dependence in the diquark BS amplitudes
$\widetilde\chi$ and $\widetilde{\bar\chi}$), the sum of the three
currents given above now yields a conserved electromagnetic current for
the nucleon; that is,
\begin{eqnarray} 
Q_\mu  \, \biggl( J^\mu_q \, + \, J^\mu_D \, + \, J^\mu_X \biggr) \,
= \, 0 \; . 
\end{eqnarray} 
For this cancellation all three contributions are crucial.
In particular, the contributions from the photon coupling to the exchanged
quark $J^{\mu}_{X}$, as well as the impulse approximate contributions 
$J^\mu_q \, + \, J^\mu_D$ must all be included. 
For the quark-diquark model of baryons, these three contributions to the
current correspond to those given in Ref. \cite{Bla99a} for the general
structure of the {\em one-particle contributions} to the current of a
3-particle Faddeev bound state (when the interactions are due to  
a separable 2-particle scattering kernel).\footnote{Here, {\em one-particle}
refers to a contribution that arises from a one-particle irreducible {\em
3-point vertex} for the photon coupling.} In the NJL model the one-particle
contributions yield the complete current of the 3-particle bound
state~\cite{Ish95}.   

However, if the substructure of the diquark BS amplitudes is taken
into account and the diquark BS amplitude is dependent on any momentum, 
the cancellation of the longitudinal pieces in the one-particle
contributions, Eqs.~(\ref{QJ_X}) to~(\ref{e5.14}), is destroyed. Additional 
photon couplings which are not of the one-particle type become necessary.
The violations to current conservation from the one-particle contributions 
can be displayed in the present framework in a way which will become useful in
following sections. Rearranging the six terms from the
Eqs.~(\ref{QJ_X}),~(\ref{e5.13}) and (\ref{e5.14}) in such a way as to
factor out two terms,  
\begin{eqnarray} 
S_1(p,P';k,P) \, &:=&  \, - i q_q \, \widetilde \chi(p_1-Q,\hat\eta P-k) 
\, + \\ 
&& \hskip -2cm i q_D \,  \widetilde \chi(p_1-\hat\sigma Q,\hat\eta
P-k +Q)  \, - 
\, i q_X  \, \widetilde\chi(p_1,\hat\eta P-k) \; ,\nn \\[4pt] 
S_2(p,P';k,P) \, &:=& \, - i q_q \, \widetilde{\bar\chi}(p_2'-Q,\hat\eta P'
-p) \, + \\
&& \hskip -2cm   i q_D \,  \widetilde{\bar\chi}(p_2'-\hat\sigma'
Q,\hat\eta P'-p -Q) 
\, -  \,i q_X  \,\widetilde{\bar\chi}(p_2',\hat\eta P' -p) \;  , \nn
\end{eqnarray} 
one obtains 
\begin{eqnarray} 
 Q_\mu  \, \biggl( J^\mu_q \, + \, J^\mu_D \, + \, J^\mu_X \biggr) \,
&=& \\
&& \hskip -2.5cm
 - \frac{1}{2} \biggl( S_1(p,P';k,P)  S^T(q')
\,\widetilde{\bar\chi}^T(p_2',\hat\eta P' -p) \nn\\
&& \hskip -2cm - \,  \widetilde \chi(p_1,\hat\eta P-k) 
\, S^T(q) \, S_2^T(p,P';k,P) \, \biggr) \; .  \nn
\end{eqnarray}
It will be demonstrated in the subsection below that these contributions are
exactly canceled by the so-called ``seagull'' contributions which arise
from one-particle irreducible {\em 4-point couplings} of the two quarks, the
diquark and photon.  Such terms must be included whenever the substructure
of a diquark bound state (in form of a momentum-dependent diquark BS
amplitude) is included in the description of the nucleon. 
Analogous seagull contributions were previously found necessary  
in $\gamma$-meson-baryon-baryon couplings to satisfy the corresponding
Ward-Takahashi identities \cite{Oth89,Wan96}. 

Note that terms analogous to the explicit one-particle contributions to the
bound state currents presented in this section can also be obtained by
employing a ``generalized'' impulse approximation for the 3-particle Faddeev
amplitudes. This procedure was recently adopted in an exploratory study of the
electromagnetic nucleon form factors \cite{Blo99}. In this study five
distinct (one-particle) contributions to the form factors arose which were
calculated using parameterizations of a simplified nucleon Faddeev
amplitude. From the separable-kernel Faddeev equation one readily verifies,
however, that only three of these five contributions are independent. These
three have the exact same topology as the one-particle contributions
presented above. Starting from the generalized impulse approximation, the
relative weights of these contributions differ, however, from those needed
for current conservation. The latter can be systematically derived from a
gauging technique~\cite{Ish98}. The discrepancy in the weights is due to an
overcounting of the (generalized) impulse approximation which can therefore
not lead to a conserved current~\cite{Bla99b}. This problem is independent of
the necessity for the additional seagull contributions which persists when
non-pointlike diquarks are used. We will now address these contributions.

%% file: Sec6.tex
\subsection{Ward Identities and Seagulls} \label{WIaS}

The Ward-Takahashi identity for the quark-photon vertex,
Eq.~(\ref{qk_WTI_nuc}), follows from the equal-time commutation relation for
the electromagnetic quark-current operator $j_\mu(x)$ with the quark field
(with charge $q_q$), 
\begin{eqnarray} 
&&\ [j^0(x), q(y) ] \, \delta(x_0 - y_0) \, = \, - q_{q} \, q(x) \,
\delta^4 (x-y) \; , \nonumber\\        
&&\ [j^0(x), \bar q(y) ] \, \delta(x_0 - y_0) \, = \,  q_{q} \, \bar q(x)
\, \delta^4 (x-y) \; . \label{charge_con}
\end{eqnarray}  
Formal problems with equal-time commutation relations for interacting fields
can be avoided by replacing the canonical formalism with a Lagrangian
formulation based on relativistic causality rather than to single out a sharp
timelike surface \cite{Pei52}. 
However, as an operational device for the derivation of Ward identities,
the equal-time commutation relations of Eqs.~(\ref{charge_con}) will
nevertheless give the correct result.

Consider the 5-point Green function that describes the photon coupling to 
four quarks $G^\mu_{\alpha\gamma ,\beta\delta}$.
Using the notation that $q_{q\alpha}$, $q_{q\beta}$, $q_{q\gamma}$, and
$q_{q\delta}$ denote the charges of the quark fields with Dirac indices
denoted by $\alpha$, $\beta$, $\gamma$ and $\delta$, respectively, the
Ward identity is given by  
\begin{eqnarray}
&& \hskip -.2cm 
\partial_\mu^{z} \, \langle T\big( q_\gamma(x_3)  q_\alpha(x_1) \bar
q_\beta(x_2) \bar q_\delta(x_4) \, j^\mu(z) \big) \rangle \, = 
\label{wti5pt} \\
&& \hskip .1cm  
- \, \big( \, q_{q\alpha} \, \delta^4(x_1 - z)  \,  +\,  q_{q\gamma} \, \delta^4(x_3 - z)  - \, q_{q\beta}
\, \delta^4(x_2 - z) \,   \nn  \\ 
&& \hskip 1cm  -\,  q_{q\delta}  \, \delta^4(x_4-z) \, \big) \;
\langle T\big( q_\gamma(x_3) q_\alpha(x_1) \bar q_\beta(x_2) \bar
q_\delta(x_4)  \big) \rangle \; .  \nn
 \end{eqnarray} 
The 4-point function on the right-hand side has the diquark pole
contribution given in Eq.~(\ref{dq_pole_ms}) and depicted in
Fig.~\ref{dqpole}. 
The Fourier transformation of the left-hand side of Eq.~(\ref{wti5pt})
allows one to define,
\begin{eqnarray}
G^\mu_{\alpha\gamma , \beta\delta}(p,P';k,P) \, &:=& \\ 
&& \hskip -2cm 
\int d^4\!x_1\, d^4\!x_2\, d^4\!x_3  \, d^4\!x_4 \,\;  e^{ip_\alpha x_1}\,
e^{ip_\beta x_2} \, e^{ip_\gamma x_3} \,  e^{ip_\delta x_4} \nn \\
&& \hskip -.6cm \langle T\big( q_\gamma(x_3) q_\alpha(x_1) \bar q_\beta(x_2)
\bar q_\delta(x_4) \, j^\mu(0)  \big) \rangle \; . \nn
\end{eqnarray} 
Here, $p = \sigma p_\gamma - \hat\sigma p_\alpha$,  $k = \sigma' p_\beta -
\hat\sigma' p_\delta$ and $P' = P + Q $ as before. 
It is straight forward to verify the following Ward identity for this
5-point Green function from the pole contribution to the 4-quark Green
function, given in Eq.~(\ref{dq_pole_ms}), which determines the dominant
contribution when the diquark momenta are close to the diquark pole at
$P^2 = {P'}^2 = m_s^2$.
\begin{eqnarray} 
i Q_\mu G^\mu_{\alpha\gamma , \beta\delta}(p,P';k,P) \, &:=& \\
&& \hskip -2.5cm 
q_{q\alpha} \, \frac{i}{P^2 -m_s^2 + i\epsilon} \, \chi_{\gamma\alpha}(p +
\hat\sigma Q, P)  \, \bar\chi_{\beta\delta}(k,P) \nn\\
&& \hskip -3cm
 + \, q_{q\gamma} \, \frac{i}{P^2 -m_s^2 + i\epsilon} \,
\chi_{\gamma\alpha}(p -\sigma Q, P)  \, \bar\chi_{\beta\delta}(k,P) \nn \\
&& \hskip -3cm
 - \, q_{q\beta} \, \frac{i}{{P'}^2 -m_s^2 + i\epsilon} \,
\chi_{\gamma\alpha}(p,P') \, \bar\chi_{\beta\delta}(k-\sigma' Q,P') \nn\\
&& \hskip -3cm
- \, q_{q\delta} \, \frac{i}{{P'}^2 -m_s^2 + i\epsilon} \,
\chi_{\gamma\alpha}(p,P') \, \bar\chi_{\beta\delta}(k+\hat\sigma'Q,P') \; .\nn 
\end{eqnarray} 
To explicitly demonstrate that this does indeed give the additional
contributions necessary to current conservation of the BSE solution for
the nucleon, one needs to consider the irreducible 4-point coupling of the
photon to the quarks and the diquark derived from the following definition:
\begin{eqnarray} 
\big( S(p_\gamma) \,  M^\mu(p_\gamma,p_\alpha,P_d) \,  S^T(p_\alpha)
\big)_{\gamma\alpha} \,  D(P_d) \, & := & \\
&& 
\hskip -5.5cm  Z^{-1} \, \int \frac{d^4k}{(2\pi)^4}  \, G^\mu_{\alpha\gamma ,
\beta\delta}(p,P_d+Q;k,P_d) \, \widetilde\chi_{\delta\beta}(k,P_d) \; , \nn
\end{eqnarray} 
with $p_\alpha = - p + \sigma (P_d +Q)  $, $p_\gamma = p + \hat\sigma (P_d
+Q)  $, and 
\begin{eqnarray} 
Z \, := \, \, \int \frac{d^4k}{(2\pi)^4}  \, \tr\big[
\bar\chi(k,P_d) \,  \widetilde\chi(k,P_d)\, \big] \; . 
\end{eqnarray}
The Ward identity for the 5-point Green function then entails, 
\begin{eqnarray}
iQ_\mu M^\mu(p_\gamma,p_\alpha,P_d) \, & = & \\
&& \hskip -2cm  q_{q\alpha}  \,
\widetilde\chi(p + \hat\sigma Q, P_d ) \, S^T(p_\alpha - Q)  \, S^{-1\,
T}(p_\alpha) \nn \\
&& \hskip -1.6cm + \,   q_{q\beta}  \, S^{-1}(p_\gamma) \, S(p_\gamma -Q)  \,
\widetilde\chi(p - \sigma Q , P_d ) 
\nn\\ 
&& \hskip -1.6cm - \,  \Delta_\Phi(Q^2) \; \widetilde\chi(p, P_d + Q )  \,
\frac{P_d^2 - m_{s}^2}{(P_d+Q)^2  - m_s^2 + i \epsilon} \; ,  \nn 
\end{eqnarray}
with
\begin{eqnarray}
Q  \, = \,  p_\gamma + p_\alpha - P_d \; , \;\;
p \, = \, \sigma p_\gamma - \hat\sigma p_\alpha \; , \nn
\end{eqnarray} 
($\sigma + \hat\sigma= 1$) and $\Delta_\Phi(Q^2)$ is defined by
\begin{eqnarray}
\Delta_\Phi(Q^2) &:=& \,  Z^{-1} \, \int \frac{d^4k}{(2\pi)^4}  \, \Big\{ \\ 
&& \hskip -1cm  
 q_{q\beta}\,  \tr\big[ S^T(-k +\hat\sigma' P_d +Q) \times \nn \\
&& \hskip -.3cm
\widetilde{\bar\chi}(k-\sigma' Q,P_d+Q) \, S(k + \sigma'P_d) \,
\widetilde\chi(k,P_d) \,  \big] \nn \\  
&& \hskip -1cm 
+ \,q_{q\delta} \, 
\tr\big[ S^T(-k +\hat\sigma' P_d ) \times \nn\\ 
&& \hskip -.3cm
\widetilde{\bar\chi}(k+\hat\sigma'
Q,P_d+Q) \, S(k+\sigma'P_d+Q) \,\widetilde\chi(k,P_d) \big] \Big\}   \; . \nn
\end{eqnarray}
In the limit $Q\to 0$, this is normalized in such a way as to yield the
charge of the scalar diquark; that is, 
$\Delta_\Phi(0) = q_{q\beta} + q_{q\delta} \equiv  q_{\Phi} $.  

In a more detailed and complete calculation, the coupling of the diquark
to the photon would itself have to be done within a Mandelstam formalism. 
To achieve this, the Ward identity of Eq.~(\ref{qk_WTI_nuc}) would be used
in the Mandelstam formalism to construct the Ward identity for the quark
substructure of the diquark, thereby replacing the naive Ward identity of
Eq.~(\ref{dq_WTI_nuc}) with a more accurate identity which accurately
depicts the quark substructure of the diquark. 
This added complication can be worked out in a straight-forward manner by
introducing a few additional technical details. 
However, the basic principle of such couplings, as derived from Ward
identities, can be seen from the following simplifying assumption, which
will be used in the following sections. 
Assume that $\Delta_\Phi(Q^2)$ is independent of the photon momentum,
such that $\Delta_\Phi(Q^2) = \Delta_{\Phi}(0) = q_{\Phi}$.   
This assumption is sufficient in
order to obtain the correct charges for the nucleon bound state. 
From this starting point, the electromagnetic diquark form factor can be
easily included in a minor extension of the framework and follows simply
from  the inclusion of a dependence on the photon momentum $Q^2$ of
$\Delta_\Phi(Q^2)$.

By including the effect of only the charge of the diquark, that is setting
$\Delta_\Phi(Q^2) \equiv q_\Phi$ for all photon momenta $Q$, the divergence
of the amplitude $M^\mu$ is written as
\begin{eqnarray}
Q_\mu M^\mu(p_\alpha,p_\beta,P_d) &=&  \\
&& \hskip -1.5cm  Q^\mu M^{legs}_\mu(p_\alpha,p_\beta,P_d) 
\, + \,    Q^\mu M^{sg}_\mu(p_\alpha,p_\beta,P_d) \; , \nn 
\end{eqnarray}
where $M^{legs}$ contains the couplings of the photon to the amputated quark
and diquark legs according to their respective Ward identities,
\begin{eqnarray}
iQ^\mu M^{legs}_\mu(p_\alpha,p_\beta,P_d) \, & = &  \\
&& \hskip -2cm \phantom{+} \, q_{q\alpha} \left(
S^{-1}(p_\alpha) -  S^{-1}(p_\alpha - Q) \right) \times \nn\\
 &&  S(p_\alpha - Q) 
\widetilde\chi(p - \hat\sigma Q , P_d ) \nn\\ 
&& \hskip -2cm + \,  q_{q\beta} \widetilde\chi(p + \sigma Q , P_d )  
S^T(p_\beta - Q) \times \nn\\
&&  \left( S^{-1\, T}(p_\beta) -  S^{-1\, T} (p_\beta -
Q)  \right) \nn\\
&& \hskip -2cm - \,  q_\Phi \widetilde\chi(p,   
P_d + Q )  D(P_d+Q)  \nn\\
&&   \left(D^{-1}(P_d)  - D^{-1}(P_d+Q) \right) \; ,  \nn 
\end{eqnarray} 
where the term $M^{sg}$ describes the one-particle irreducible seagull
couplings and its divergence is given by  
\begin{eqnarray}
iQ^\mu M^{sg}_\mu(p_\alpha,p_\beta,P_d) \, & = & \label{sgWI}\\
&& \hskip -2cm q_{q\alpha} \, \widetilde\chi(p - \hat\sigma
Q , P_d )  \,  + \,   q_{q\beta} \, \widetilde\chi(p + \sigma Q , P_d ) \nn\\
&& \hskip 1cm  
- \,  q_\Phi \widetilde\chi(p, P_d + Q )  \; . \nn
\end{eqnarray} 

These seagull couplings are exactly what is needed to arrive at a conserved
electromagnetic current for the nucleon. 
Upon substitution of the charges of the spectator and exchanged quark and
the scalar diquark, $q_{q\alpha} = q_{q}$, $q_{q\beta} = q_X$ and $q_\Phi
= q_D$, respectively, one finds  
\begin{eqnarray} 
S_1(p,P';k,P) \, = \, Q^\mu M^{sg}_\mu(\eta P'+p,q+Q,\eta P - k) \, ,
\end{eqnarray}
with $ Q = P' - P $. 
A solution to this Ward identity, with transverse terms added so as to keep
the limit $Q\to 0$ regular, which follows from a standard 
construction, {\it c.f.}, Refs.~\cite{Oth89,Wan96}, is provided by
\begin{eqnarray}
 i M^{sg}_\mu(p_\alpha,p_\beta,P_d) \, & = & \label{sgM}  \\ 
&& \hskip -3cm \phantom{+}
 q_{q} \frac{(2p_\alpha -
 Q)_\mu }{p_\alpha^2 - (p_\alpha - Q)^2} 
\bigg( \widetilde{\chi}(p_\alpha -
 Q,p_\beta,P_d) -  \widetilde{\chi}(p_\alpha ,p_\beta ,P_d) \bigg) \nn \\
&&  \hskip -3cm  
 + q_X \frac{(2p_\beta - Q)_\mu }{p_\beta^2 - (p_\beta -
 Q)^2} \bigg( \widetilde{\chi}(p_\alpha ,p_\beta -Q,P_d) -
 \widetilde{\chi}(p_\alpha ,p_\beta ,P_d) \bigg) \nn \\ 
&& \hskip -3cm  - q_D \frac{(2P_d + Q)_\mu }{(P_d+Q)^2 - P_d^2} \bigg(
 \widetilde{\chi}(p_\alpha ,p_\beta ,P_d+Q) - 
 \widetilde{\chi}(p_\alpha ,p_\beta ,P_d) \bigg) \nn 
\end{eqnarray}
with $Q = p_\alpha + p_\beta -P_d$. 
Analogously, for the other seagull, one finds
\begin{eqnarray}
 i \bar M^{sg}_\mu(p_\alpha,p_\beta,P_d) \, & = & \label{sgMbar}\\ 
&& \hskip -3cm \phantom{+} 
q_{q} \frac{(2p_\alpha - Q)_\mu }{p_\alpha^2 - (p_\alpha - Q)^2} 
\bigg( \widetilde{\bar\chi}(p_\alpha -
 Q,p_\beta,P_d) -  \widetilde{\bar\chi}(p_\alpha ,p_\beta ,P_d) \bigg) \nn \\
&& \hskip -3cm  +  q_X \frac{(2p_\beta - Q)_\mu }{p_\beta^2 - (p_\beta -
 Q)^2} \bigg( \widetilde{\bar\chi}(p_\alpha ,p_\beta -Q,P_d) -
 \widetilde{\bar\chi}(p_\alpha ,p_\beta ,P_d) \bigg) \nn \\ 
&&\hskip -3cm   
-  q_D \frac{(2P_d - Q)_\mu }{P_d^2 - (P_d-Q)^2} \bigg(
 \widetilde{\bar\chi}(p_\alpha ,p_\beta ,P_d-Q) - 
 \widetilde{\bar\chi}(p_\alpha ,p_\beta ,P_d) \bigg)    
 . \nn
\end{eqnarray} 
The amplitudes 
$\widetilde{\chi}(p_\alpha ,p_\beta ,P_d)$ herein  
need to be constructed from the BS amplitude of the scalar diquark by
removing the overall momentum conserving constraint $p_\alpha + p_\beta = P_d
$. With these seagull couplings, a conserved electromagnetic current operator
is obtained by including the seagull contribution
\begin{eqnarray}
  J_\mu^{sg} (p,P';k,P) \, 
&=& \\ 
&& \hskip -2.6cm 
\frac{1}{2} \biggl( M^{sg}_\mu(\eta P'+p,q',\hat\eta P - k) \, 
S^T(q') \,\widetilde{\bar\chi}^T(p_2,\hat\eta P' -p) \nn\\
&& \hskip -2.6cm - \,  \widetilde \chi(p_1,\hat\eta P-k) 
\, S^T(q) \, \bar{M}_\mu^{sg\, T}\!(-(\eta P +k),-q,\hat\eta P' -p) \,
\biggr) \; .  \nn
\end{eqnarray}
The total conserved electromagnetic current of the nucleon is therefore
given by  $ J_{\hbox{\tiny em}}^\mu :=  J^\mu_{q} \, + \, 
J^\mu_D \, + \, J^\mu_X  \, + \, J^\mu_{sg} $. 

Note that this explicit construction of the conserved current    
complies with the general gauging formalism presented in Ref.~\cite{Bla99a}.
In the reduction of Faddeev equations with separable 2-particle interactions,
the seagull couplings arise from 2-particle contributions to the bound state
current. These contributions describe the irreducible coupling of the photon
to the 2-particle scattering kernel.

Technically, the additional contributions to the electromagnetic nucleon
current arising from exchanged quark in the nucleon BSE kernel $J^\mu_X$,
as well as the seagull term   $J^\mu_{sg} $, involve two
4-dimensional loop integrations to calculate the electromagnetic form
factors from the nucleon BS amplitudes. 
As demonstrated above, this considerable extension to the Mandelstam
formalism (involving only single loop integrations) is absolutely necessary
to correctly include the non-trivial substructure of the diquark
correlations and maintain current conservation of the nucleon.   

While seagull contributions are not necessary in an approximation that employs 
point-like diquarks, as is the case in Ref.~\cite{Hel97b},
beyond-impulse contributions, such as the coupling of the exchanged-quark
to the photon are necessary! 
In the study of Ref.~\cite{Hel97b}, it was observed that neglecting this
contribution produced negligible violations to the charges of proton and
neutron and so it was dismissed as unimportant. 
However, in the case of the present study this contribution is
significant.  The reason for this is the larger value of the coupling
strength $g_s$ (obtained from the diquark normalization condition $g_s =
1/N_s^2$) used herein.   For a point-like diquark, the coupling need not be
as large.   
Furthermore, in contrast to the impulse-approximate Mandelstam currents,
which are independent of $g_s$, the contribution to the electromagnetic
current due to the exchanged quark is proportional to $g_s^2$ 
(see Eq. (\ref{J_exchange})).  
Hence, use of a smaller coupling strength $g_s^2$ in the BSE, reduces the
importance of going beyond the impulse approximation.  

As an example of the importance of the contributions beyond the impulse
terms of $J_q$ and $J_D$, we consider the results for the proton
and neutron electromagnetic charges using the amplitudes plotted in
Fig.~\ref{S1_S2} from the Mandelstam current $ J^\mu_{q} \, + \, J^\mu_D $
alone.   This leads to charges $Q_P = 0.85$ for the proton and $Q_N =
0.15$ for the neutron.
A theorem constrains the charges of the proton and neutrons to obey
$Q_P+Q_N=1$.    
The theorem relies on using $\eta = 1/3$ and is derived from the
nucleon normalization condition in Eq.~(\ref{nuc_norm}). 
However, the way it is realized here is not very satisfying. 
In Sec.~\ref{ElmFFs}, we discuss in detail the relevance of the various
contributions to the electromagnetic form factors of the proton and
neutron due to exchanged-quark-photon coupling and seagull couplings.

Having proven that the present framework conserves the electromagnetic
current, one might think that the results for the proton and neutron
charges, $Q_P$ = 1 and $Q_N = 0$, must follow trivially.
However, the verification that this framework provides the correct
charges for the proton and neutron is not entirely trivial as is demonstrated
in the next section.

For finite momentum transfer $Q = P'\! -\! P>0$, the complete
Mandelstam couplings for the diquark will still require modifications. 
If the photon is coupled to the elementary carriers of charge only, 
that is, to the quarks within the diquark (with quark charges
$q_{q\alpha}$, $q_{q\beta}$), the photon-diquark vertex will itself be of
the form,     
\begin{eqnarray}
F_\Phi^\mu(Q) &=& \\
&&\hskip -1cm  
Z^{-1} \,q_{q\alpha} \, \int \frac{d^4k}{(2\pi)^4}  \,
\tr\bigg[ S^T(- k+\hat\sigma P_d  +Q)\times \nn\\
&& 
\widetilde{\bar\chi}(k-\sigma Q,P_d+Q)   
S(k+\sigma P_d) \, \widetilde\chi(k,P_d) \times
\nn\\
&& \hskip -.5cm 
 S^T(-k+\hat\sigma P_d )
\Gamma^{\mu\, T}(-k+\hat\sigma P_d +Q, -k+\hat\sigma P_d )  \bigg] \nn \\ 
&& \hskip -1cm 
+  \,  Z^{-1} \,q_{q\beta} \, \int \frac{d^4k}{(2\pi)^4}  \,
\tr\bigg[ S^T(-k+\hat\sigma P_d ) \times \nn\\
&& 
\widetilde{\bar\chi}(k+\hat\sigma' Q,P_d+Q)
 S(k+\sigma P_d +Q) \times  \nn\\
&& \hskip -.5cm  
\Gamma^{\mu}(k+\sigma P_d  +Q,k+\sigma P_d) \,
S(k+\sigma P_d)   \,\widetilde\chi(k,P_d) \bigg]          \; . \nn
\end{eqnarray}
This gives the correct diquark charge in the limit the photon momentum $Q
\to 0$.  
In this limit, as far as the electric form factors of the nucleons are
concerned this detail is irrelevant since they are constrained to be
proportional to the charge of the nucleon. On the other hand, the anomalous
magnetic moments of the nucleons may receive additional contributions from
the quark substructure of the diquark.

%% file: Sec7.tex
\newlength{\figwidth}
\setlength{\figwidth}{0.5\linewidth}

\section{Electromagnetic Form Factors of the Nucleon}
\label{ElmFFs}

The matrix elements of the nucleon current can be para\-me\-trised as
\begin{eqnarray} 
\langle P',s'| J_{\hbox{\tiny em}}^\mu(0) | P,s \rangle &=& \\
&& \hskip -2cm \bar u(P',s') 
    \left[\gamma^\mu {\mathcal F}_1
  +\frac{i \kappa {\mathcal F}_2}{2M} \sigma^{\mu\nu} Q_\nu \right] 
  u(P,\sigma)\; . \nn
\end{eqnarray}
Here,  ${\mathcal F}_1$ and ${\mathcal F}_2$ are the Dirac charge and
the Pauli anomalous magnetic form factors, respectively \cite{Aitchison}.
$Q^\mu = P'-P$ is the spacelike momentum of the virtual photon probing
the nucleon ($-Q^2\ge 0$). Using the Gordon decomposition
\begin{eqnarray}
\bar u(P',s') \, \frac{i\sigma^{\mu\nu} Q_\nu}{2M} \,  u(P,s) &=& \\
&& \hskip -2cm 
  \bar{u}(P',s') \,  \left[\, \gamma^\mu\, - \, \frac{P_{\hbox{\tiny
BF}}^\mu} {M} \, \right] 
  u(P,s) \; , \nn
\end{eqnarray}
with the definition of the Breit momentum 
$P_{\hbox{\tiny BF}} :=(P'+P)/2$, the current can be rewritten as 
\begin{eqnarray}
\langle P',s'|J_{\hbox{\tiny em}}^\mu(0) | P,s \rangle &=& \\
&& \hskip -2cm  \bar{u}(P',s')  
   \left[ \, \gamma^\mu \, ({\mathcal F}_1+
   \kappa {\mathcal F}_2)\,  - \, \frac{P_{\hbox{\tiny BF}}^\mu}{M} \, 
 \kappa {\mathcal F}_2 \, \right] \, u(P,s) \; .\nn 
\end{eqnarray}
It is convenient in the following to introduce (matrix valued) 
matrix elements by initial and final spin-summations, 
\begin{eqnarray}
\langle P'|  \widehat J^\mu |P \rangle \, := \, 
 \langle P',s' | J^\mu | P,s \rangle \, \sum_{s,s'} 
     u(P',s') \bar{u}(P,s) \;  , \label{Jhat}
\end{eqnarray}
to remove the nucleon spinors. The frequently used Sachs electric and
magnetic form factors $G_E$ and $G_M$, are introduced via
\begin{eqnarray}
 G_E &=& {\mathcal F}_1 + \frac{Q^2}{4M^2}\kappa {\mathcal F}_2 \;
 ,\label{ge_0}  \\
 G_M &=& {\mathcal F}_1 + \kappa {\mathcal F}_2 \, . \label{gm_0}
\end{eqnarray}
These can be extracted from Eq.~(\ref{Jhat}),
\begin{eqnarray}
\langle P' | \widehat J_{\hbox{\tiny em}}^\mu(0) |P \rangle &=& \\
 && \hskip -1.5cm    \Lambda^+(P') \left[ \,  \gamma^\mu \, G_M \,   +\,
  M \, \frac{P_{\hbox{\tiny BF}}^\mu}{P_{\hbox{\tiny BF}}^2} \; (G_E-G_M) \,
  \right] \Lambda^+(P)\, , \nn 
\end{eqnarray} 
by taking traces of $\langle \widehat J_{\hbox{\tiny em}}^\mu \rangle \equiv
\langle P'| \widehat J_{\hbox{\tiny em}}^\mu(0) |P \rangle $ as follows:
\begin{eqnarray}
G_E &=& \frac{M}{2P_{\hbox{\tiny BF}}^2} \, \mbox{tr} \,
\langle \widehat J^{\hbox{\tiny em}}_\mu \rangle P_{\hbox{\tiny BF}}^\mu \; ,
\label{ge}  \\ 
G_M &=& \frac{M^2}{Q^2} \, \left( \, \mbox{tr} \, \langle \widehat
    J^{\hbox{\tiny em}}_\mu
    \rangle \gamma^\mu \, - \, \frac{M}{P_{\hbox{\tiny BF}}^2} \, \mbox{tr} \,
      \langle \widehat J^{\hbox{\tiny em}}_\mu \rangle P_{\hbox{\tiny
    BF}}^\mu \right) \; . 
\label{gm}
\end{eqnarray}
We calculate the current matrix elements using Mandelstam's approach with
the current operators defined in the previous sections, such that
\begin{eqnarray}
\langle P' |\widehat J_{\hbox{\tiny em}}^\mu(0) |P \rangle &=& 
 \label{Eq:6.10}\\
&& \hskip -1cm  \int \frac{d^4p}{(2\pi)^4}
\frac{d^4k}{(2\pi)^4} \; \bar{\psi}(-p,P') \, J_{\hbox{\tiny
em}}^\mu(p,P';k,P)  \, \psi(k,P) \nn 
\end{eqnarray}
The current operator $J_{\hbox{\tiny em}}^\mu$ consists of the four parts
which describe the coupling of the photon to quark or diquark, to the
exchanged-quark and the seagull contributions which arise from the 
coupling of the photon to the diquark BS amplitudes.
These are determined by the following kernels,
\begin{eqnarray}
  J^\mu_{q} \, &=&   q_{q} \,  \Gamma^\mu_{q} (p_q, k_q)  \;  D^{-1}(k_s)\,
                         (2\pi)^4 \delta^4(p-k-\hat\eta Q) \; , \label{jq}\\ 
  J^\mu_{D} \, &=&   q_{D}  \, \Gamma^\mu_{D} (p_s, k_s)  \;  
                  S^{-1}(k_q)\, (2\pi)^4 \delta^4(p-k+\eta Q) \; , \label{jd}\\
  J^\mu_X \,   &=&   - q_X \, \frac{1}{2} \\
 &&   \widetilde \chi(p_1,k_s) 
           \, S^T(q) \, {\Gamma^\mu_q}^T(q',q) \, S^T(q') \,
                         \widetilde{\bar\chi}^T(p_2',p_s) \; , \nn \\
  J^\mu_{sg} \, &=& \, \frac{1}{2}\,  \big( \, M_{sg}^\mu(p_q,q',k_s) \, 
S^T(q') \,\widetilde{\bar\chi}^T(p_2',p_s) \\
&& \hskip 1.5cm - \,  \widetilde \chi(p_1,k_s) 
\, S^T(q) \, \bar{M}^{\mu\, T}_{sg}\!(-k_q,-q,p_s) \,
\big) \; .  \nn 
\end{eqnarray}
The abbreviations for the various momenta are summarized in the following
table: 
\vskip -.6cm
\begin{eqnarray}
\label{mom_table} \\[-8pt]
\renewcommand{\arraystretch}{.9}
\begin{array}{>{$ \small }l<{$\hskip .25cm} | 
>{\hskip .25cm$\small }l<{$\hskip .25cm} >{\hskip .25cm $ \small }l<{$} } 
& incoming & outgoing \\
\hline \\[-8pt]
quark:   &  $\scriptstyle  k_q = \eta P + k $     
         &  $\scriptstyle   p_q = \eta P' + p $    \\
diquark: &  $\scriptstyle  k_s = \hat\eta P - k$  
         &  $\scriptstyle  p_s = \hat\eta P' - p$  \\
exchange quark: & $\scriptstyle  q = \hat\eta P -\eta P' -p-k$ 
                & $\scriptstyle  q' = \hat\eta P' -\eta P -p-k$ \\
\hline \\[-8pt]
relative momenta &($\scriptstyle  \sigma = \sigma' = 1/2$) & \\
within diquark: & $\scriptstyle  p_1 = \frac{1}{2} \, (p_q - q)  $ 
                & $\scriptstyle  p_2' =  \frac{1}{2} \, ( -k_q + q' )$ \\  
seagull quark-pair: & $\scriptstyle  p_1' = \frac{1}{2} \, (p_q - q')  $ 
                     & $\scriptstyle  p_2 = \frac{1}{2} \, (-k_q + q) $
\end{array} \nn
\end{eqnarray}
 
\begin{figure}[t]
\begin{minipage}{0.49\linewidth}
\epsfig{file=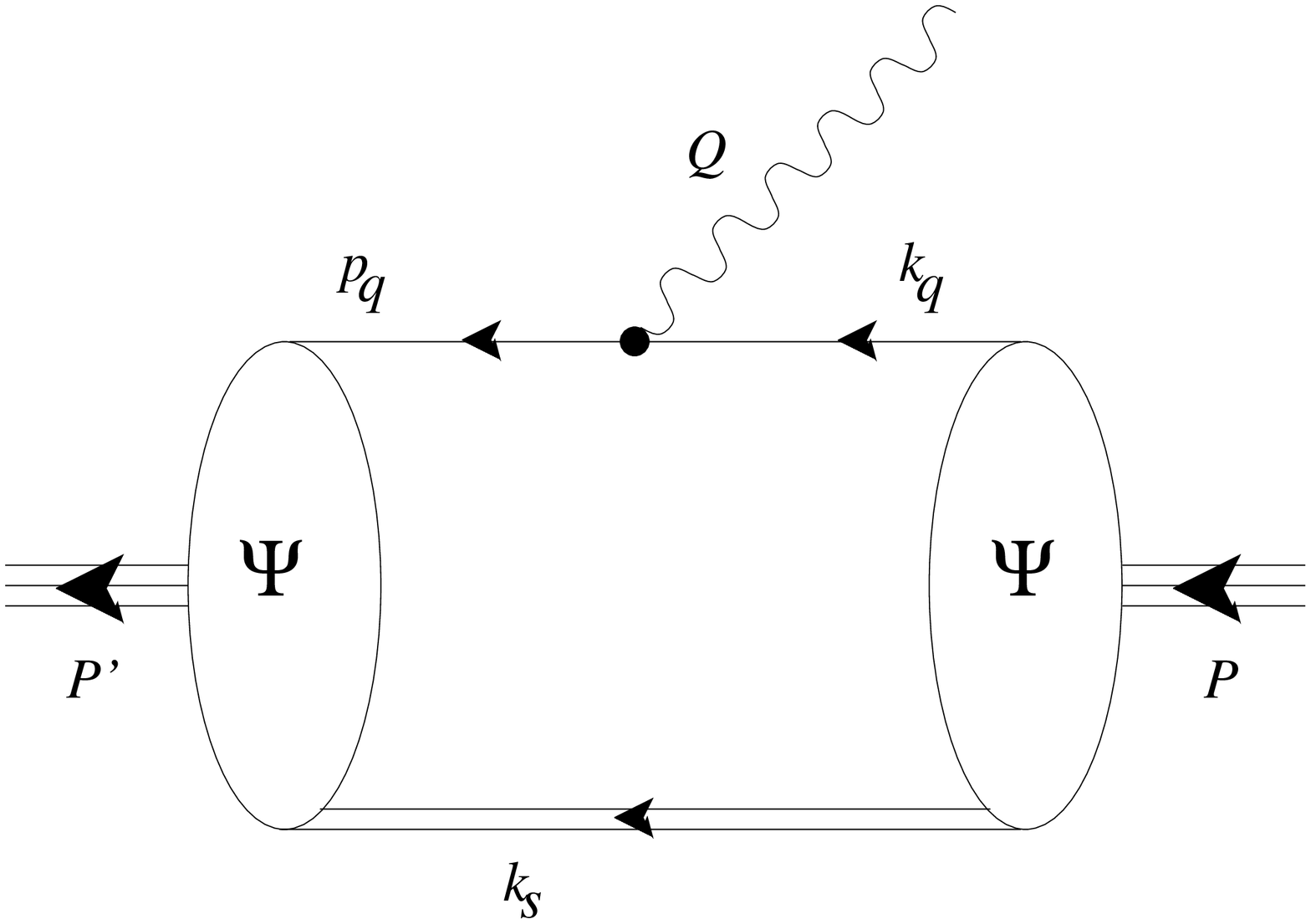, width=\linewidth}
\centerline{$\scriptstyle p_s = k_s \; , \; \; p_q = k_q + Q $.}
\end{minipage}
\hfill
\begin{minipage}{0.49\linewidth}
\epsfig{file=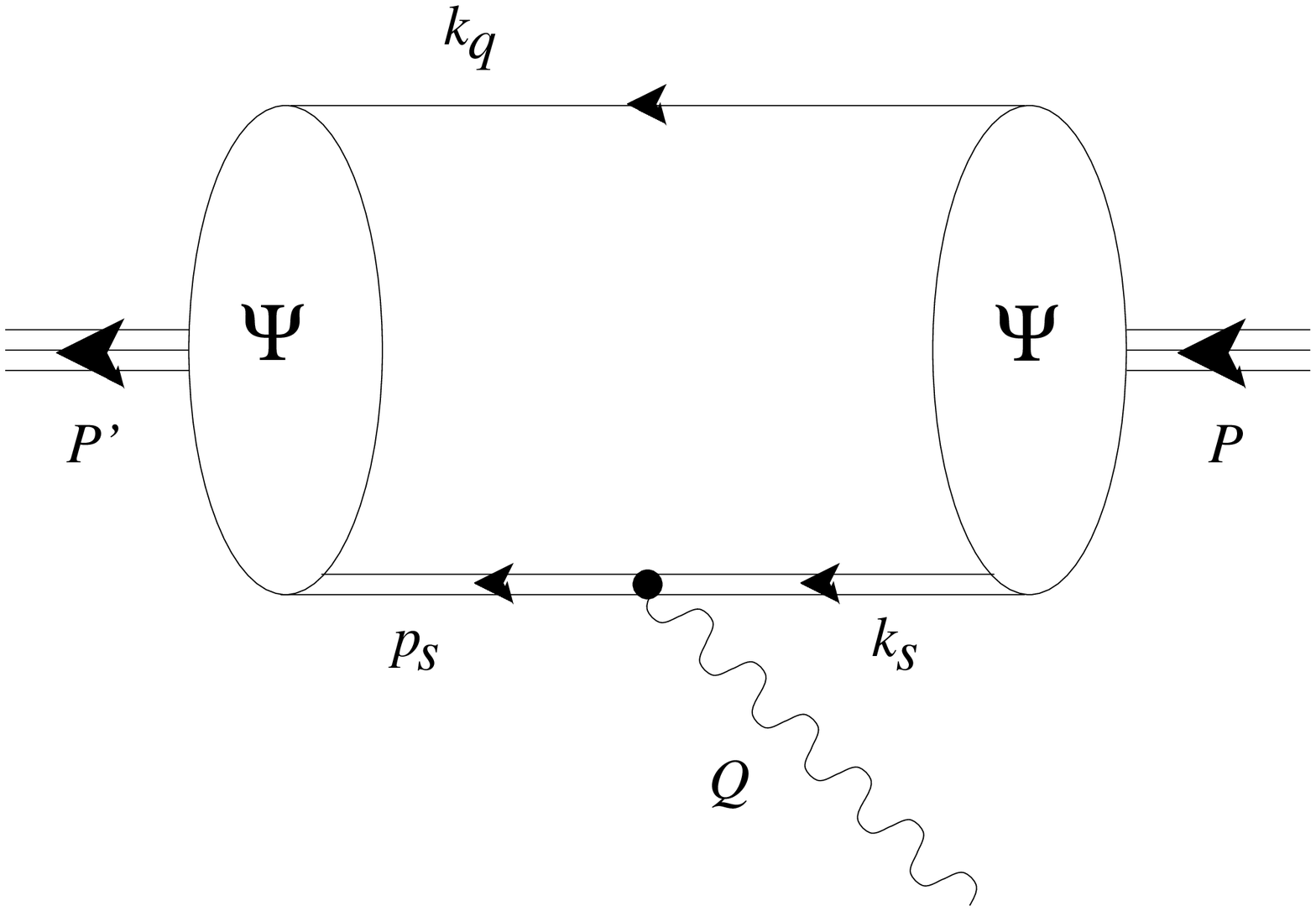, width=\linewidth}
\centerline{$\scriptstyle p_s = k_s + Q \; , \; \; p_q = k_q $.}
\end{minipage}
\caption{Impulse approximation diagrams.}
\label{IAD}
\vspace{5mm}
\end{figure}

\begin{figure}[t]
\begin{minipage}{0.49\linewidth}
\epsfig{file=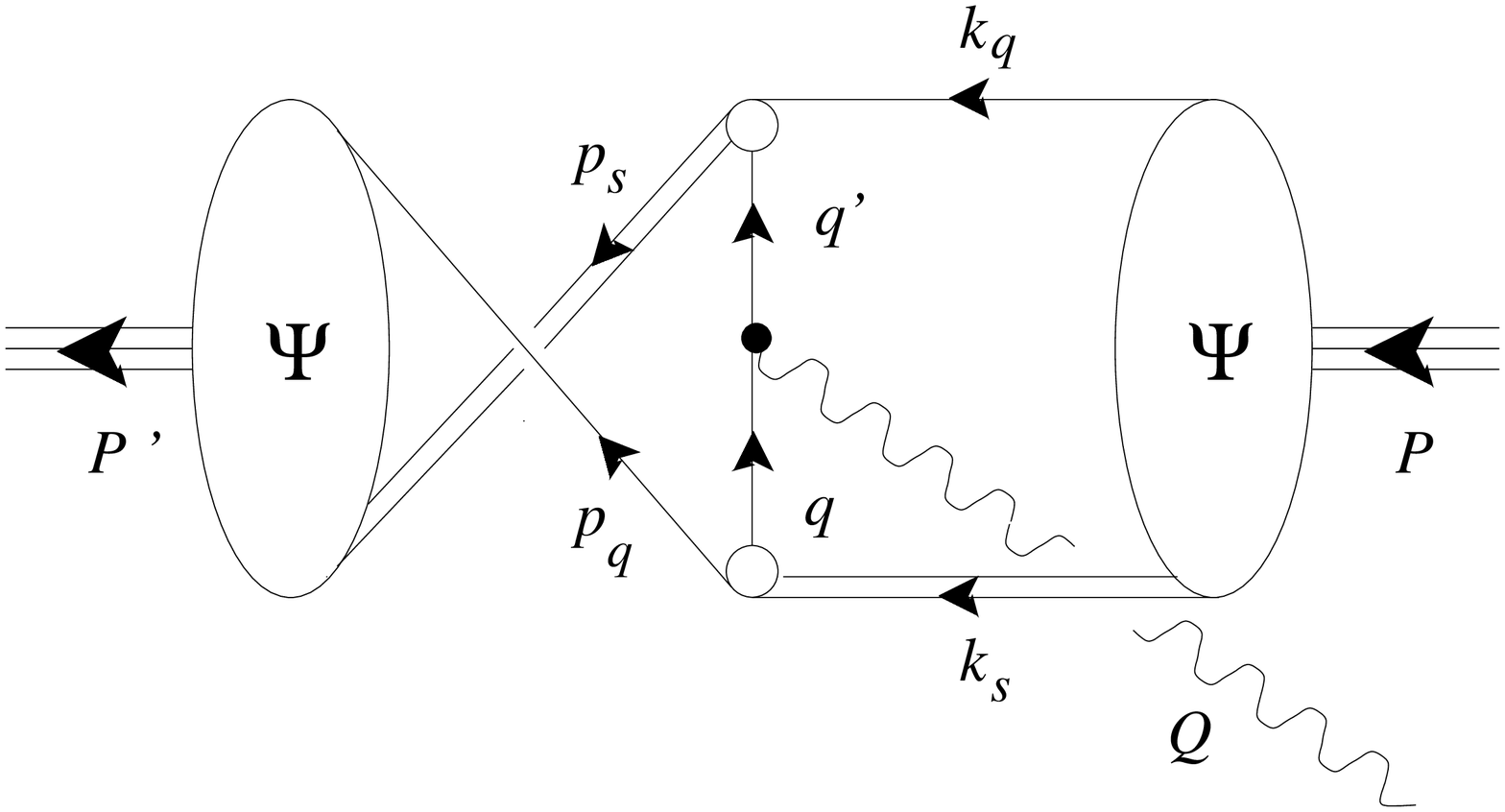,width=\linewidth}

\vspace{-.2cm}

\parbox{5.6cm}{
$\scriptstyle
    q = (\hat\eta\!-\!\eta) P_{\hbox{\tiny BF}} - p - k - Q/2$,\\
$\scriptstyle
    q'\!= (\hat\eta\!-\!\eta) P_{\hbox{\tiny BF}} - p - k + Q/2$,\\
$\scriptstyle
    p_1 = (\eta\!+\!1)Q/4-(1\!-\!3\eta)P_{\hbox{\tiny BF}}/2 + p + k/2$,\\
$\scriptstyle
    p_1' = (\eta\!-\!1)Q/4-(1\!-\!3\eta)P_{\hbox{\tiny BF}}/2 + p + k/2$,\\
$\scriptstyle
    p_2 = (\eta\!-\!1)Q/4+(1\!-\!3\eta)P_{\hbox{\tiny BF}}/2 - p/2 - k$,\\
$\scriptstyle
    p_2' = (\eta\!+\!1)Q/4+(1\!-\!3\eta)P_{\hbox{\tiny BF}}/2 - p/2 - k$.
}
\vfill

\end{minipage}
\hfill
\begin{minipage}{0.49\linewidth}
\epsfig{file=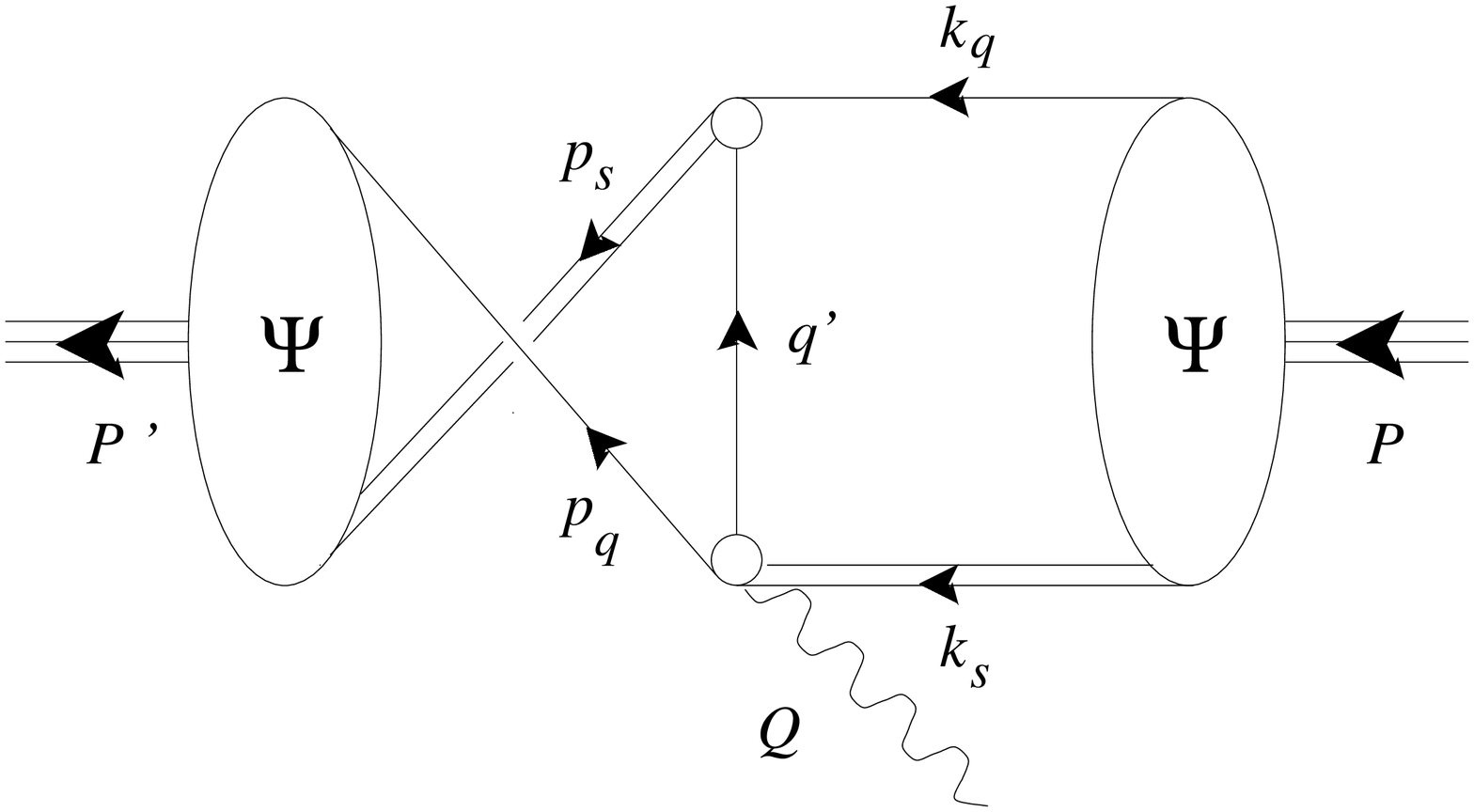,width=\linewidth} \\
\epsfig{file=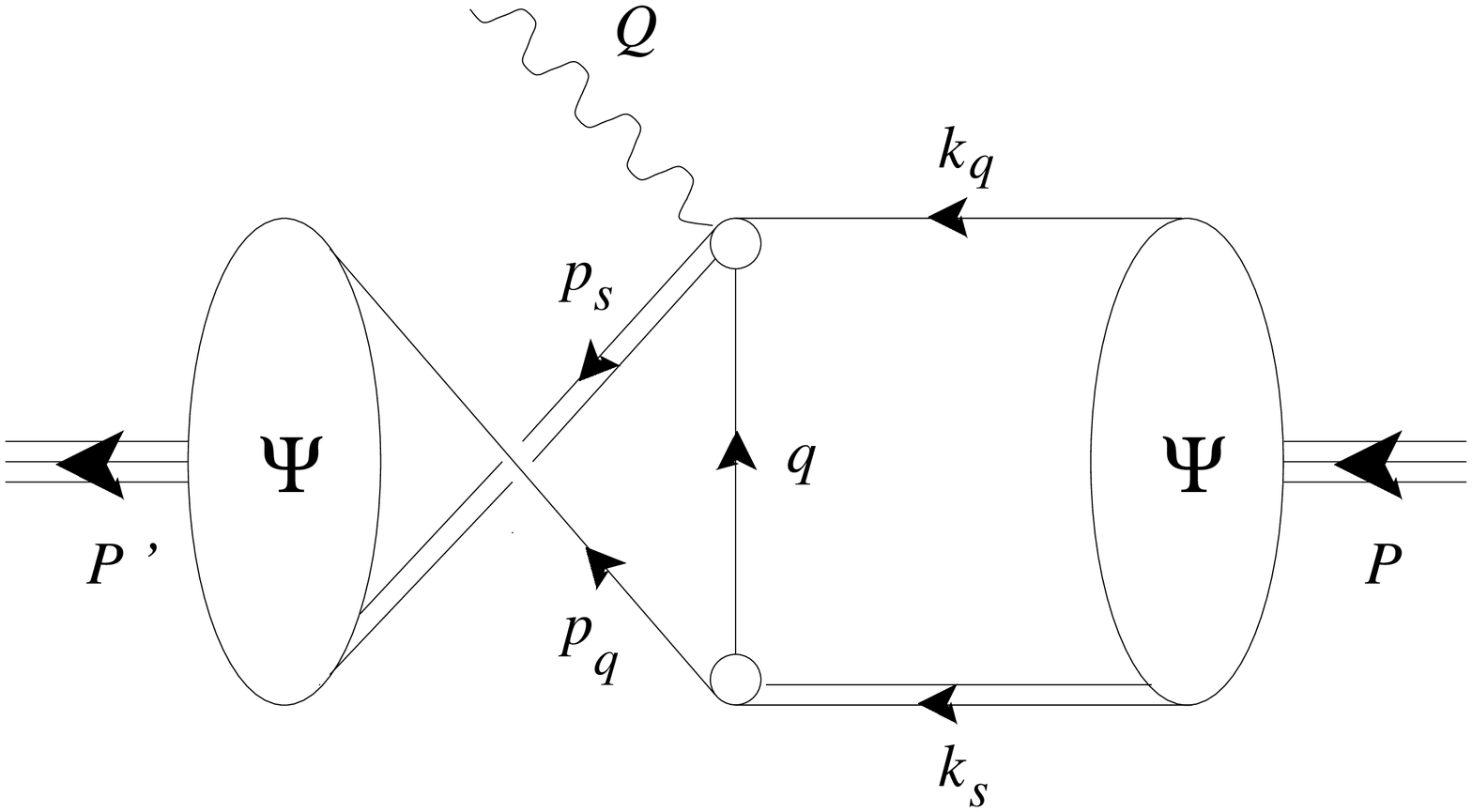,width=\linewidth}
\end{minipage}
\caption{Exchange quark and seagull diagrams.}
\label{momrout}
\end{figure}

The contributions to the form factors in the impulse approximation are
depicted in Fig.~\ref{IAD}, while the exchange-quark and seagull
contributions are shown in Fig.~\ref{momrout}.

We use $\sigma = \sigma' = 1/2$ in the diquark amplitudes. 
As discussed in Sec.~\ref{dq_corrs}, this implies that the relative momenta
$p_i$ within the diquarks are exchange symmetric and that our
parameterizations of the diquark BS amplitudes are independent of the mass of
the diquark.  We thus set, 
\begin{eqnarray}
\widetilde\chi(p_1,k_s) &\to&  \widetilde\chi(p_1^2) = \frac{\gamma_5C}{N_s}
P(-p_1^2) \; , \label{sg_momdep} \\
\widetilde{\bar\chi}(p_2',p_s) &\to&
\widetilde{\bar\chi}(p'_2 \!^2) = \frac{\gamma_5 C^{-1}\!\!}{N_s} 
P(-p'_2 \!^2) \; . \nn
\end{eqnarray} 
This also simplifies the seagull terms, as the seagull couplings to the
diquark legs do not contribute to the seagulls in this case.
The amplitudes in the brackets of the last lines of Eqs.~(\ref{sgM}) and
(\ref{sgMbar}) cancel.  

The construction of such vertex functions from the Ward-Takahashi
identities is not unique. In particular,
the forms for the irreducible seagull couplings $M^\mu_{sq}$ and $\bar
M^\mu_{sq}$ given in Eqs.~(\ref{sgM}) and (\ref{sgMbar}), respectively,
are designed for amplitudes $\widetilde\chi$ and $\widetilde{\bar\chi}$
which are functions of the scalars $p_\alpha^2$, $p_\beta^2$
and, in general, $P_d^2$.  
For $Q\to 0$, the possibility that the denominators in each of the three
terms may vanish entails that the prefactors that arise 
from expanding the amplitudes in brackets must also vanish.

\begin{eqnarray}
\widetilde\chi( (p_\alpha-Q)^2, p_\beta^2 ) - \widetilde\chi(
p_\alpha^2, p_\beta^2 ) &\to& \\
&& \hskip -1cm    - 2 (p_\alpha Q)  \,
\frac{\partial}{\partial p_\alpha^2 } \widetilde\chi( p_\alpha^2, p_\beta^2)
\; .  \nn
\end{eqnarray} 
Our present assumption on the dominant momentum dependence of these
amplitudes is slightly different though. For amplitudes 
$\widetilde\chi \equiv \widetilde\chi((p_\alpha - p_\beta)^2/4)$,
see~(\ref{sg_momdep}), the prefactors of the seagulls in the form given
in Eqs.~(\ref{sgM}) and (\ref{sgMbar}), corresponding to factors $\propto
1/(p_\alpha Q)$ and $\propto 1/(p_\beta Q)$ respectively, are not canceled in
an analogous way. To cure this, we replace the quark momenta $p_\alpha $ and
$p_\beta $ in these prefactors by (plus/minus) the relative momentum,  
$\pm (p_\alpha - p_\beta)/2 $. This yields,   
\begin{eqnarray}
iM^\mu_{sg}&=& \, {q_q} \, \frac{(4p_1'-Q)^\mu}{4p_1'Q-Q^2}
  \big(\widetilde\chi((p_1'-Q/2)^2)-\widetilde\chi(p_1' \!^2)\big) 
  \label{good_sg} \\
  & &\hskip .5cm  + {q_X}\, \frac{(4p_1'+Q)^\mu}{4p_1'Q+Q^2}
  \big(\widetilde\chi((p_1'+Q/2)^2)-\widetilde\chi(p_1' \!^2)\big) \nonumber \\
i\bar M^\mu_{sg}&=& \, {q_q}\,  \frac{(4p_2-Q)^\mu}{4p_2Q-Q^2}
  \big(\widetilde{\bar\chi}((p_2-Q/2)^2)-\widetilde{\bar\chi}(p_2^2)\big)
     \label{good_sgbar} \\
 & & \hskip .5cm + {q_X} \, \frac{(4p_2+Q)^\mu}{4p_2Q+Q^2}
  \big(\widetilde{\bar\chi}((p_2+Q/2)^2)-\widetilde{\bar\chi}(p_2^2)\big)
 \nonumber 
\end{eqnarray}
Since Ward-Takahashi identities do not completely constrain the form
of the vertices, such modification of this type is within the freedom
allowed by this ambiguity.
The forms~(\ref{good_sg},\ref{good_sgbar}) solve the
corresponding Ward identities at finite $Q$, see Eq.~(\ref{sgWI}) (with 
$ q_q + q_X = q_D$). In addition, the smoothness of the limit $ Q\to 0 $ for
the given model assumptions is ensured. In this limit,  
\begin{eqnarray} 
   iM^\mu_{sg}\,  &\to& \;  - \,  ( q_q -  q_X ) \;   p_1^\mu \;
                                         \widetilde\chi'(p_1^2) \; ,\nn \\ 
i\bar M^\mu_{sg}\,  &\to& \;  - \, (q_q -  q_X ) \;   p_2^\mu \;
                                         \widetilde{\bar\chi}'(p_2^2) \;
               . \label{diff_sg_WI} 
\end{eqnarray} 
We emphasize that this limit is unambiguous. 
It must coincide with the form required by the differential form of the
Ward identity for the seagull couplings. 
As such it restricts contributions that are both longitudinal and
transverse to the photon four-momentum $Q_{\mu}$.
The form above follows necessarily for the model diquark amplitudes
employed in the present study, and this form provides the crucial
condition on the seagull couplings that ensure charge conservation for 
the nucleon bound state. 

\noindent The nucleon charges are obtained by calculating 
\begin{eqnarray}
G_E(0)  &=&   \frac{1}{2M}  \, \int   \frac{d^4p}{(2\pi)^4}
\frac{d^4k}{(2\pi)^4} \\
&& \hskip 1.2cm   \tr \big[ \bar{\psi}(-p,P) \, P_\mu J_{\hbox{\tiny
em}}^\mu(p,P;k,P) \, \psi(k,P) \big] \; . \nn
\end{eqnarray}
The various contributions to the electromagnetic current for $Q \to 0$ ({\it
i.e.}, $P' = P$) are given by 
\begin{eqnarray}
  J^\mu_{q} \, &\to &   iq_{q}  \left(\frac{\partial}{\partial k^\mu_q}
                         S^{-1}\!(k_q)\right)  D^{-1}(k_s)\,
                         (2\pi)^4 \delta^4(p-k) , \\ 
  J^\mu_{D} \, &\to&   iq_{D}   \left(\frac{\partial}{\partial k^\mu_s}
                         D^{-1}\!(k_s)\right)     
                  S^{-1}(k_q)\, (2\pi)^4 \delta^4(p-k) , \\
  J^\mu_X \,   &\to&   - \, i q_X \, \frac{1}{2N_s^2} \,  
                      P(-p_1^2) P(-p_2^2) 
           \, \left( \frac{\partial}{\partial q^\mu} S(q) \right) , \\
  J^\mu_{sg} \, &\to& \, i (q_q - q_X) \; \frac{1}{2N_s^2}\; S(q) \\
           && \left( 
            p_{1\mu} \, P'(-p_1^2) P(-p_2^2) \, - \, p_{2\mu} \,  P(-p_1^2)
                         P'(-p_2^2) \right)  \, .  \nn 
\end{eqnarray}
Comparing this to the normalization integrals given in
Sec.~\ref{SecNucNorm}, one finds
\begin{eqnarray}
G_E(0)  &=& \\
&&  q_q \, N_q \, + \, q_D \, N_D \, + \, q_X \, N_X \, - \, (q_q
- q_X) \,  N_P \; . \nn
\end{eqnarray}
Using $q_q = 2/3$, $q_D = 1/3$, $q_X = -1/3 $ for the proton, and $q_q =
-1/3$, $q_D = 1/3$, $q_X = 2/3 $ for the neutron, together with
Eq.~(\ref{nuc_norm}), one therefore has,
\begin{eqnarray}
 1 \; &=& \, \eta N_q + (1-\eta) N_{D} + (1-2\eta) N_{X} + (1-3\eta) N_{P} \;
  ,\nn \\ 
 Q_P &=& \,  \frac{2}{3} N_q + \frac{1}{3} N_{D} -\frac{1}{3} N_{X} -  N_{P}
  \; , \label{chargeconds} \\
 Q_N &=& \, - \frac{1}{3} N_q + \frac{1}{3} N_{D} + \frac{2}{3} N_{X} +
 N_{P}\label{Qn}  \; . \nn
\end{eqnarray}
However, these three equations are not independent. 
Rewriting the normalization condition for the nucleon BS amplitudes, we
find that 
\begin{eqnarray}
 1 \; &=& \, \frac{2}{3} N_q + \frac{1}{3} N_{D} -\frac{1}{3} N_{X} -  N_{P} \\
      &&   \hskip 1cm    + \, (\eta - \frac{2}{3}) \, \big( N_q - N_D - 
                  2N_X -3 N_P \big) \; , \nn
\end{eqnarray} 
which entails that
\begin{eqnarray}
1 \, = \, Q_P \, + \, (2 - 3\eta )\,  Q_N  \; .  \label{chargesum}
\end{eqnarray} 
To verify that we do in fact obtain the correct charges of the proton and
neutron, it suffices to show that $N_q - N_D =  2 N_X + 3 N_P$; 
that is, it suffices to show that the neutron is neutral, $Q_N = 0$. 
The proof of this is straightforward and is given in
Appendix~\ref{supplCC}.    

\subsection{Numerical Computation}
\label{NC}

The numerical computation of the form factors is done in the Breit frame,
where 
\begin{eqnarray} 
 &&     Q^\mu \, = \,  (0, \vec Q ) \; , \nn\\
 &&     P^\mu \, = \,  (\omega_Q,- \vec Q/2) \; , \label{BF_def}\\
 &&     P'\ \!\!^\mu \, = \,  (\omega_Q,\vec Q/2) \; , \nn \\
 &&     P_{\hbox{\tiny BF}}^\mu  \, = \, 
             (\omega_Q, 0) \; ,   \nn 
\end{eqnarray}
with $\omega_Q = \sqrt{ M^2 + \vec Q^2/4 }$. 

The transformation of these variables to 4-dimensional Euclidean polar
coordinates follows the same prescriptions as those employed in
Sec.~\ref{QDBSE} (see Eqs.~(\ref{WickRot})), namely
\begin{eqnarray}
&& \hskip -.2cm 
\{p^2,\,  k^2, \, Q^2 \}  \, \to \, \{-p^2, \, -k^2, \, -Q^2\} \; ,   
\quad P_{\hbox{\tiny BF}}^2 \to \omega_Q^2 \; ,   \label{WickRot2} \\
&& \hskip -.2cm 
pQ \, \to \, - \, p \, |\vec Q| \, y_Q \; , \quad   kQ \, \to \, - \, k \,
|\vec Q| \, z_Q \; ,  \nn\\ 
&& \hskip -.2cm 
pP_{\hbox{\tiny BF}} \, \to \, i \omega_Q \, p\,  y_{\hbox{\tiny BF}} 
 \; , \quad
kP_{\hbox{\tiny BF}} \, \to \, i \omega_Q \,  k \, z_{\hbox{\tiny BF}} \;
,   \nn \\
&&\hskip -.2cm 
 pP' = pP_{\hbox{\tiny BF}} + pQ/2 \, 
\to \nn\\
&& \hskip 2cm 
 i \omega_Q \, p\,  y_{\hbox{\tiny BF}} \, - \, p \,  |\vec Q| \, y_Q/2
\, =: \, i \, M\, p \, y \; ,  \nn \\
&& \hskip -.2cm 
kP = kP_{\hbox{\tiny BF}} - kQ/2 \, 
\to \nn\\
&& \hskip 2cm 
 i \omega_Q \, k\,  z_{\hbox{\tiny BF}} \, + \, k \,  |\vec Q| \, z_Q/2
 \, =: \, i \, M \, k \, z \; . \nn
\end{eqnarray}   
In the presence of two independent external momenta, $Q$ and
$P_{\hbox{\tiny BF}}$, we are left with 5 independent angular variables.
Together with the absolute values of the integration momenta $p$ and $k$
the exchange-quark and seagull contributions to the form factors at finite
momentum transfer $Q$ require performing 7-dimensional integrations.
These are computed numerically using Monte Carlo integrations.

For the impulse approximation diagrams, the number of necessary
integrations collapses to three due to the momentum-conserving delta
functions in Eqs.~(\ref{jq}) and (\ref{jd}).
One of the integrations is over the absolute value of the loop momentum
$k$ and two are the angular integrations over $z_{\tiny BF}$ and $z_Q$,
the cosines of the angles between $k$ and $P_{\tiny BF}$ and $k$ and $Q$,
respectively. 

The BS amplitudes for the nucleon bound states are given in terms of the
two scalar functions $S_1(p,P)$ and $S_2(p,P) $, {\it c.f.},
Eqs.~(\ref{psiDec}) 
and~(\ref{psibarDec}) in Sec.~\ref{QDBSE}, which, we recall Eqs.~(\ref{ChebyS})
and~(\ref{ChebyM}), are expanded in terms of Chebyshev polynomials $T_n$ to
account for their dependence on the azimuthal Euclidean variable, 
\begin{eqnarray}
 S(p,y) \, \simeq \, \sum_{n=0}^{N-1} (-i)^n \, S_n(p) T_n(y) \;
 . \label{BSACE} 
\end{eqnarray} 
While the argument $y$ of the Chebyshev polynomials, the cosine between
relative and total momentum, is in $[-1,1]$ in the rest-frame of the nucleon,
this cannot be simultaneously true for the corresponding arguments in the
initial and final nucleon bound-state amplitudes at finite (spacelike)
momentum transfer $Q$. In the Breit frame, these arguments are, 
\begin{eqnarray}
     z \, &=&\, \frac{\omega_Q}{M} \,  z_{\hbox{\tiny BF}} \, - \, i \,
     \frac{1}{2} \, \frac{|\vec Q|}{M} \, z_Q \quad \hbox{and} \nn  \\  
     y &=&\, \frac{\omega_Q}{M} \,  y_{\hbox{\tiny BF}} \, +  \, i \,
     \frac{1}{2} \, \frac{|\vec Q|}{M} \, y_Q \; , \label{compl_zy}
\end{eqnarray} 
for the initial and final nucleon BS amplitudes respectively (with the angular
variables $z_Q$, $y_Q$ and $z_{\hbox{\tiny BF}}$, $y_{\hbox{\tiny BF}}$ all in
$[-1,1]$).  In order to use the nucleon amplitudes computed from the BSE in
the rest frame, analytical continuation into a complex domain is necessary.
This can be justified for the bound-state BS wave functions $\psi$ (with
legs attached). These can be expressed as vacuum expectation values of local
and almost local operators and we can resort to the domain of holomorphy of
such expectation values to continue the relative momenta of the bound-state 
BS wave function $\psi (p,P) $ into the 4-dimensional complex Euclidean
space necessary for the computation of Breit-frame matrix elements from
rest-frame nucleon wave functions. The necessary analyticity properties are
manifest in the expansion in terms of Chebyshev polynomials with complex
arguments.

There are, in general however, singularities associated with the constituent
propagators attached to the legs of the bound state amplitudes, here given by
the free particle poles on the constituent mass shells. For sufficiently small
$Q^2$ these are outside the complex integration domain. For larger $Q^2$,
these singularities enter the integration domain. As the general
analyticity arguments apply to wave functions $\psi$ rather than the truncated
BS amplitudes  $\widetilde\psi$ with 2-component structure $R(p,y)$, it is
advantageous to expand these untruncated BS wave functions directly in terms
of Chebyshev polynomials (introducing moments $R_n(p)$), 
\begin{eqnarray}
 R(p,y) \, \simeq \, \sum_{n=0}^{N-1} (-i)^n \, R_n(p) T_n(y) \; ,
 \label{BSWFCE} 
\end{eqnarray} 
and employ the analyticity of Chebyshev polynomials for the BS wave
function $R$.  This can be written in terms of the two Lorentz-invariant
functions $R_1$ and $R_2$ by 
\begin{eqnarray}
\psi(p,P) & = & D(p_s) S(p_q) \widetilde\psi(p,P)\, = \\
    && \hskip -1.5cm   D(p_s) S(p_q)  \Big( S_1(p,P)\, \Lambda^+(P) \, +\,
        S_2(p,P) \, \Xi(p,P)\, \Lambda^+(P)  \Big) \,   \nn \\
  && \hskip -.5cm =:    R_1(p,P)\, \Lambda^+(P) \, +\, R_2(p,P)  
                 \, \Xi(p,P) \, \Lambda^+(P) \; .   \nn
\end{eqnarray}
The price one must pay, however, is a considerably slower suppression of
the higher Chebyshev moments in the expansions in Eq.~(\ref{BSWFCE}) for
the BS wave functions compared to the much faster suppression observed for
the truncated amplitudes $\widetilde\psi$. 
For example, the fourth moments of the truncated nucleon amplitudes
$\widetilde\psi $ are shown in Fig.~\ref{S1_S2} of Sec.~\ref{QDBSE}.
Their magnitudes are less than 2 orders of magnitude smaller than the
leading Chebyshev moments.   One must include up to 8 Chebyshev
moments in the expansion for the untruncated BS wave functions in order to
achieve a comparable reduction.

If the truncated BS amplitudes $\widetilde\psi$ are used in the expansion,
one must account for the singularities of the quark and diquark 
legs explicitly when these enter the integration domain for some finite
values of $Q^2$. 
A naive transformation to the Euclidean metric, such as the one given by
Eqs.~(\ref{WickRot2}), is insufficient.
Rather the proper treatment of these singularities is required when they
come into the integration domain.

For the impulse-approximate contributions to the form factors 
we are able to take the corresponding residues into account explicitly in
the integration. 
Although, this is somewhat involved, it is described in
Appendix~\ref{residue}. 
For these contributions, one can compare both procedures and verify
numerically that they yield the same, unique results.
This is demonstrated in App.~\ref{residue}.

We have to resort to the BS wave function expansion for calculating the
exchange quark and seagull diagrams, however. The residue structure
entailed by the structure singularities in the constituent quark and diquark
propagators is too complicated in these cases (involving the 7-dimensional
integrations). The weaker suppression of the higher Chebyshev moments and 
the numerical demands of the multidimensional integrals thus lead to
limitations on the accuracy of these contributions at large $Q^2$ by the 
available computer resources.  

The Dirac algebra necessary to compute $G_E$ and $G_M$, according
to Eqs.~(\ref{ge}) and (\ref{gm}), can be implemented directly into our
numerical routines. 
We use the moments $S_n(p)$ or $R_n(p)$ obtained from the nucleon
BSE as described in Sec.~\ref{QDBSE}, which are real scalar functions with
positive arguments.  These functions are computed on a one-dimensional 
grid of varying momenta with typically $n_p$ = 80 points.
Then spline interpolations are used to obtain the values of these
functions at intermediate values. 
The scalar functions $S(p,P)$ and $R(p,P)$ are then easily 
reconstructed from Eqs.~(\ref{BSACE}) and (\ref{BSWFCE}), respectively. 
Then complex arguments, as given in Eqs.~(\ref{compl_zy}), appear in the
Chebyshev polynomials when the electromagnetic form factors are calculated.

In the results shown below, the 3-dimensional integrations of the impulse
approximation diagrams are performed using Gauss-Legendre or
Gauss-Chebyshev quadratures, while the 7-dimensional integrations 
necessary for the calculation of the exchange-quark and seagull
contributions are carried out by means of stochastic Monte-Carlo
integrations with 1.5$\times$10$^7$ grid points.  
We find that beyond $Q^2=3$ GeV$^2$, numerical errors for the stochastic
integrations becomes larger than 1\% of the numerical result.
This we attribute to the continuation of the Chebyshev polynomials
$T_n(z)$ to complex values of $z$ as described above.

In addition to the aforementioned complications, there is another bound on
the value of $Q^2$ above which the exchange and seagull diagrams can not be 
evaluated. It is due to the singularities in the diquark amplitudes
$\widetilde\chi (p_i)$ and in the exchange quark propagator. The rational
$n$-pole forms of the diquark amplitude,
$P_{n\mbox{\tiny-P}}(p)=(\gamma_n/(\gamma_n+p^2))^n$ for example yield the
following upper bound,
\begin{eqnarray}
 Q^2 < 4 \left( \frac{4\gamma_n}{(1-3\eta)^2} -M^2 \right) \; .
\end{eqnarray}
A free constituent propagator for the exchange quark gives the additional 
constraint,
\begin{eqnarray}
 Q^2 < 4 \left( \frac{m_q^2}{(1-2\eta)^2-M^2} \right ) \; .
\end{eqnarray}
It turns out, however, that these bounds on $Q^2$ are insignificant for the
model parameters employed in the calculations described herein.

\subsection{Results}
\label{Res}

\begin{figure*}[t]
\begin{minipage}{\linewidth}

\vspace{1cm}

\hskip -.2cm \epsfig{file=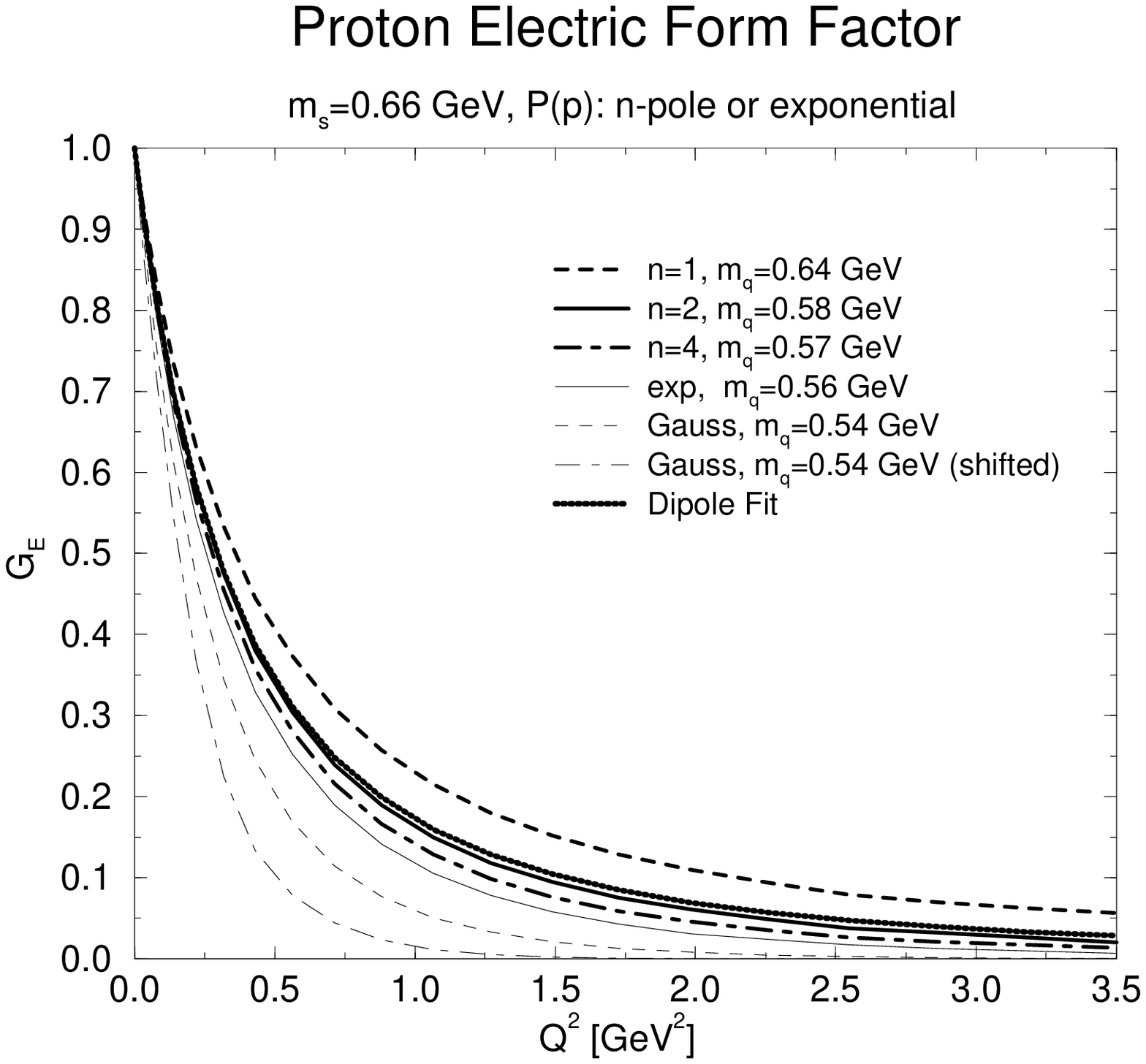,width=2\figwidth}
\hfill
\hskip -.2cm \epsfig{file=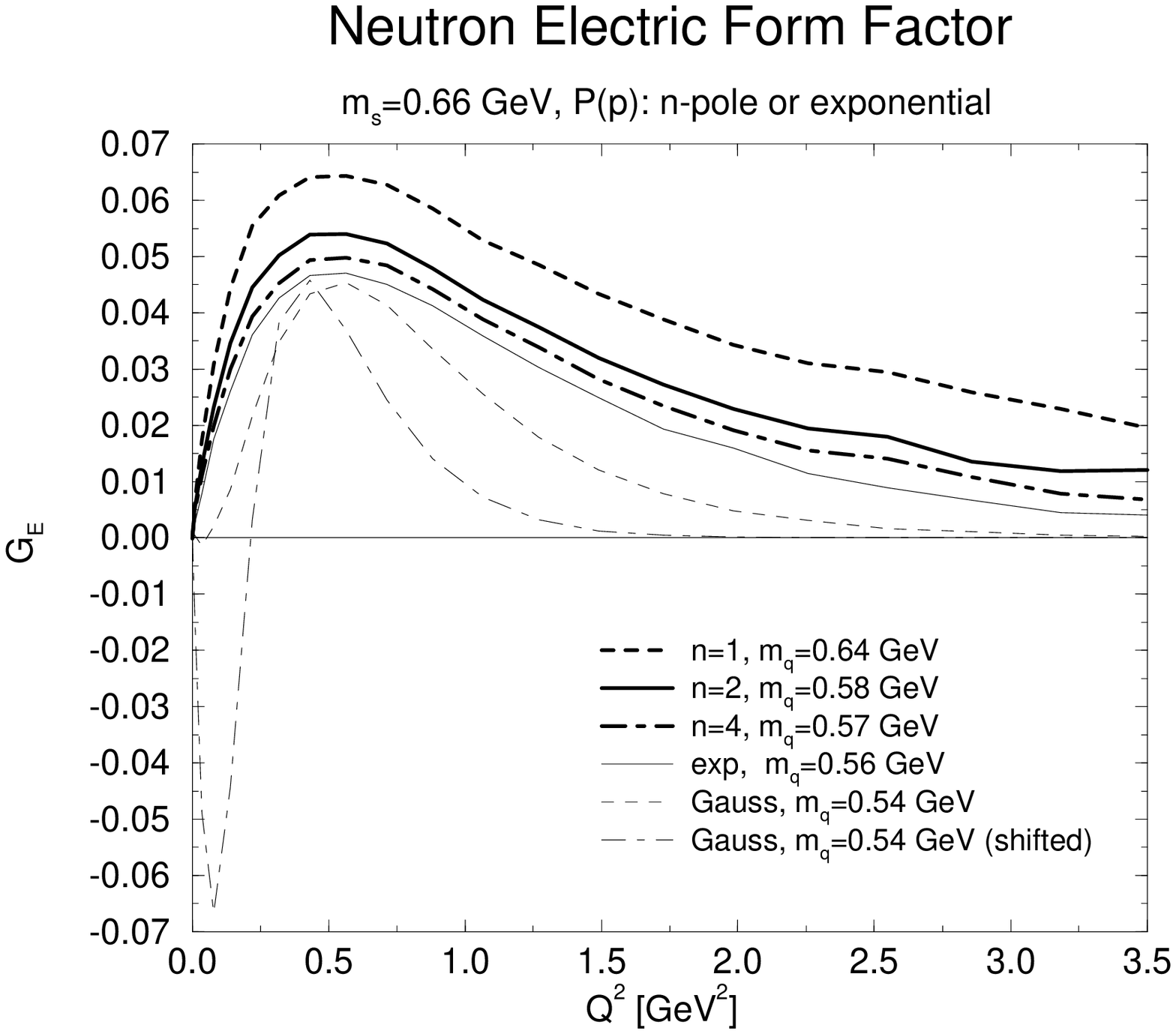,width=2\figwidth}
\end{minipage}

\bigskip

\begin{minipage}{\linewidth}
\hskip -.2cm \epsfig{file=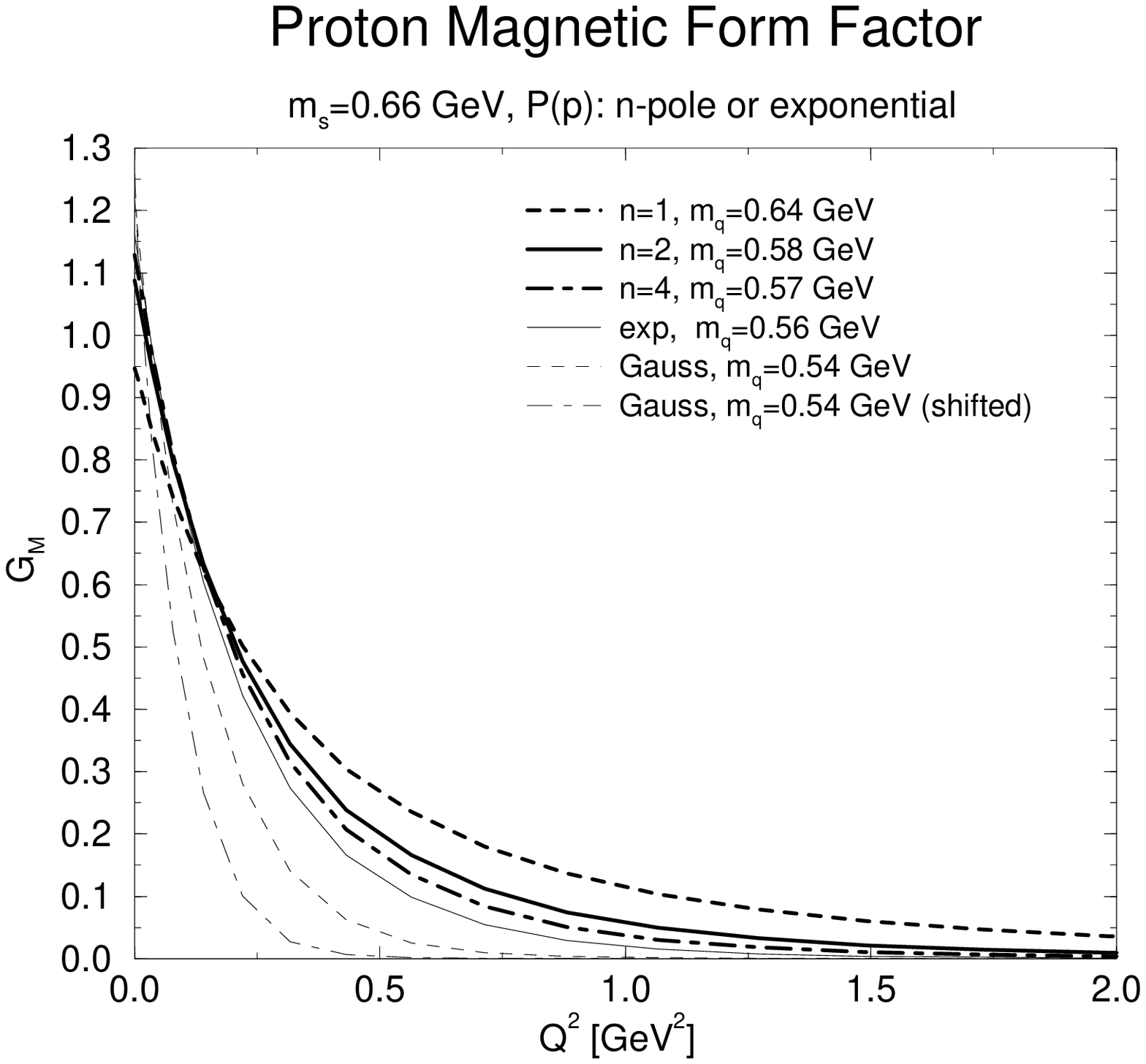,width=2\figwidth}
\hfill
\hskip -.2cm \epsfig{file=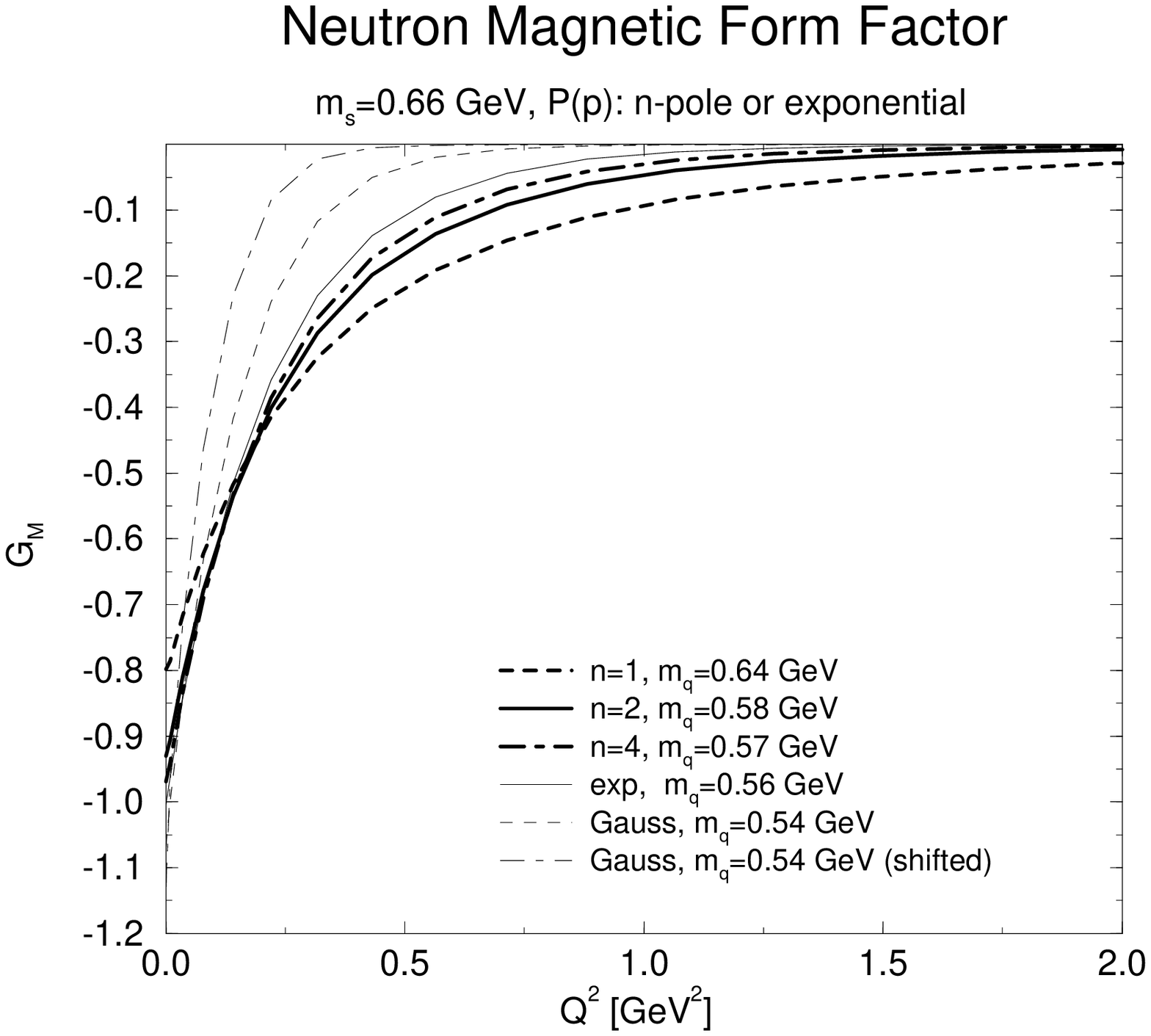,width=2\figwidth}
\end{minipage}
\refstepcounter{figure}
\centerline{\parbox{0.8\linewidth}{\small\textbf{Fig.~\thefigure.}
Nucleon electric and magnetic form-factors for fixed widths of
$S_{1,0}$. The label {\sf Gauss, shifted} refers to
a calculation with diquark vertex function $P_{\mbox{\tiny GAU}}
=\exp(-(p^2/\gamma_{\mbox{\tiny GAU}}  -1)^2)$. The parameters are 
listed in Tabs.~\protect\ref{FWHM_table} and~\protect\ref{Sgau_table} in
Sec.~\protect\ref{QDBSE}. \label{ffs}
}} 
\end{figure*}

In Fig.~\ref{ffs}, we show the electric and magnetic Sachs
form factors of the proton and neutron using the parameter sets given in 
Table~\ref{FWHM_table} in Sec.~\ref{QDBSE} which correspond to a fixed 
value for the diquark mass $m_s = $ 0.66~GeV. 
The nucleon amplitudes used in these calculations correspond to those
shown in Fig.~\ref{S1b_S2b} of Sec.~\ref{QDBSE}. 
The charge radii obtained by using the model forms of the diquark BS
amplitude given in Table~\ref{FWHM_table} are given in Table~\ref{charger}.

\begin{figure*}[t]
\begin{minipage}{\linewidth}

\vspace{1cm}

\hskip -.2cm \epsfig{file=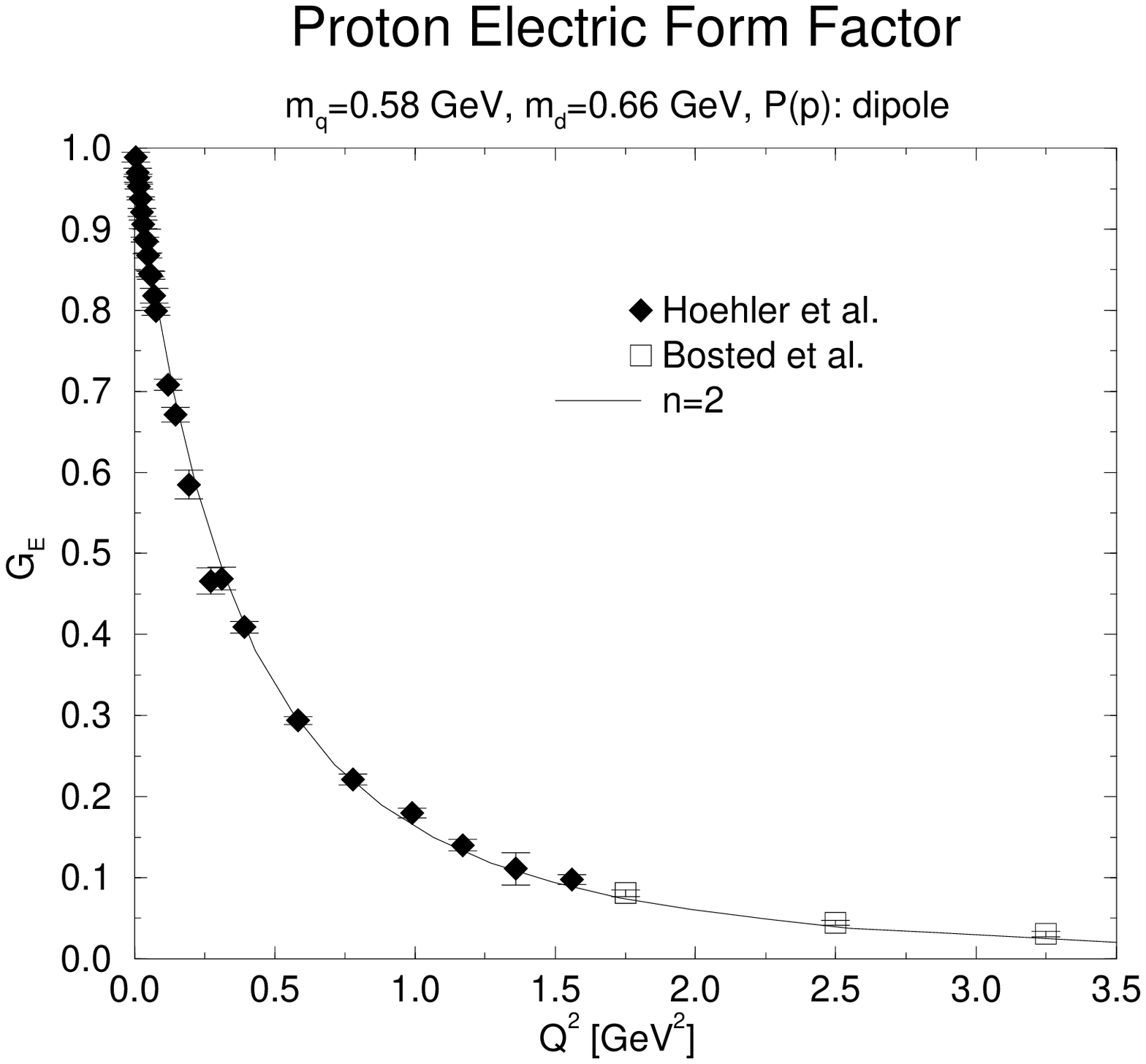,width=2\figwidth}
\hfill
\hskip -.2cm \epsfig{file=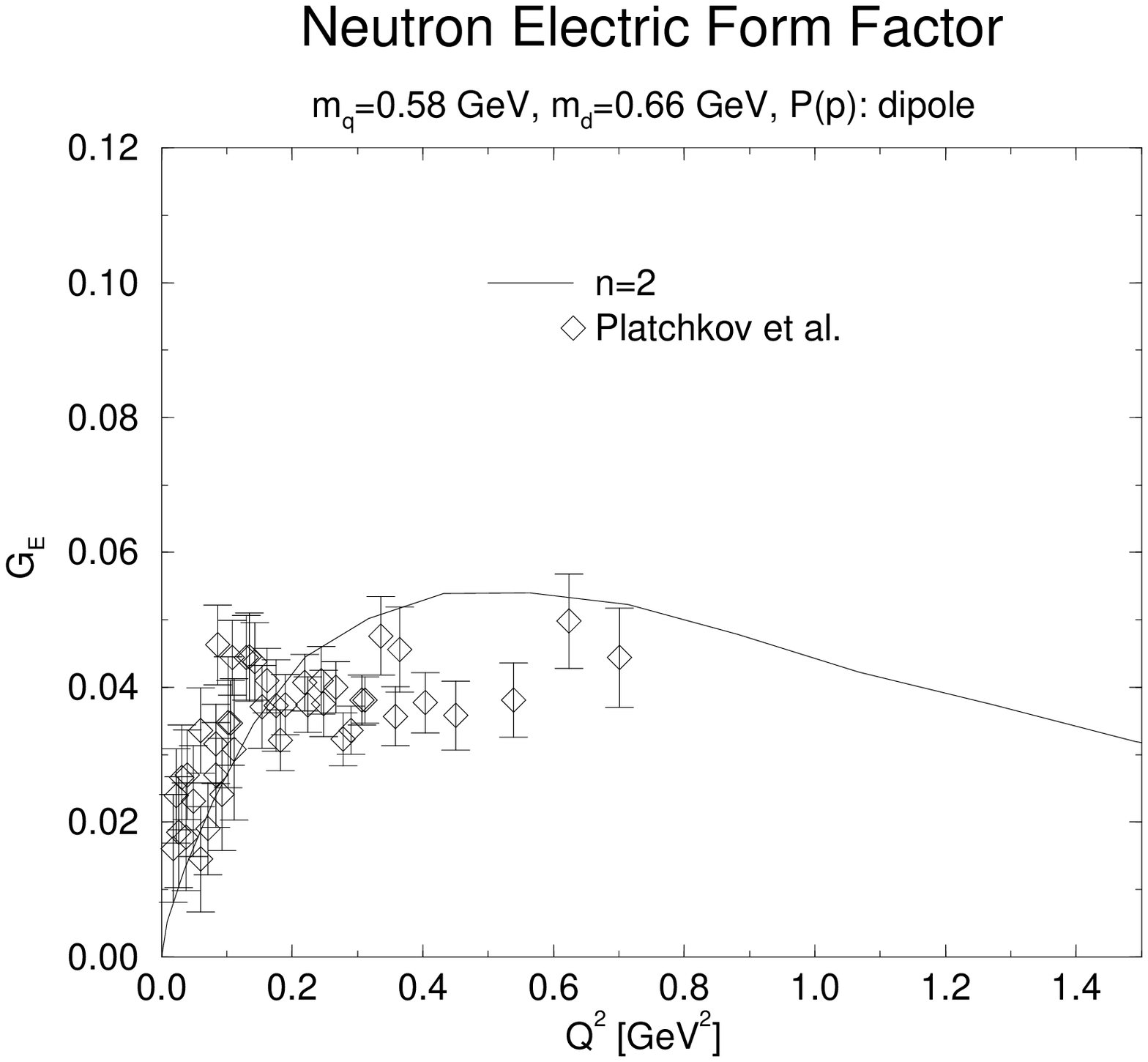,width=2\figwidth}
\end{minipage}
\refstepcounter{figure}
\centerline{\parbox{0.8\linewidth}{\small\textbf{Fig.~\thefigure.} Comparison
of the electric form factors of proton and neutron with the experimental data
of Refs.~\cite{Hoe76,Bos92} and Ref.~\cite{Pla90}, respectively. Here, the
dipole form of the diquark amplitude was used corresponding to the $n=2$
result of Fig.~\ref{ffs}.\label{ffdata}}}  
\end{figure*}

\begin{table}[b]
\begin{tabular}{ll|ll}
    & form of diquark & $r_p$ (fm)   & $r_n^2$ (fm$^2$) \\
    & amplitude $P$   & ($\pm 0.02$) & ($\pm 0.02$) \\ 
\hline 
&&& \\[-4pt]
fixed $S_{1,0}$-width : 
    & $n$=1       & 0.78 & -0.17\\ 
    & $n$=2       & 0.82 & -0.14\\
    & $n$=4       & 0.84 & -0.12\\
    & EXP         & 0.83 & -0.04\\
    & GAU         & 0.92 &  0.01\\
    & GAU shifted & 1.03 &  0.37\\[2pt]
\hline 
&&& \\[-4pt]
fixed masses:
    &$n$=1& 0.97 & -0.24\\
    &$n$=2& 0.82 & -0.14\\
    &$n$=4& 0.75 & -0.03\\
    &EXP  & 0.73 & -0.01\\   
\end{tabular}
\caption{The electric charge radii for proton and neutron for parameter sets
having either the $S_{1,0}$-width or the quark mass fixed. Error estimates
come from the uncertainty in the 7-dimensional integration. 
The corresponding experimental values are about $r_p \simeq 0.85$fm for the
proton and $r_n^2 \simeq -0.12$fm$^2$ for the neutron.
\label{charger}}
\end{table}

Examination of the charge radii given in Table~\ref{charger} reveals that
the width obtained for the nucleon BS amplitude is closely
correlated with the obtained value for the charge radius of the proton. 
The dipole, quadrupole and exponential forms for the diquark BS amplitude,
all give reasonable values for the charge radius of the proton.
The accepted value of the proton charge radius is $r_p \simeq$ 0.85~fm.
However, the width of $S_{1,0}(p)$ does not determine the behavior of the
form factors away from $Q^2=0$.  This is especially clear when the
exponential and Gaussian forms are employed for the diquark BS amplitude.
In this case, the proton electric form factor ceases to even vaguely
resemble the phenomenological dipole fit to experimental data over most of
the range of $Q^2$ shown in Fig.~\ref{ffs}. 
The neutron electric form factor is even more sensitive to the functional
form of the diquark amplitudes. The square of charge radius of the neutron
depends strongly on the chosen form of the diquark BS amplitude. 
For the exponential and Gaussian forms, the obtained value of $r^2_{n}$ is
close to zero, and it is positive when the Gaussian form of the diquark
amplitude is used with its peak away from zero ({\it i.e.}, $x_0 \not = 0$).
In Fig.~\ref{ffs}, we show the form factors that result from using the
shifted Gaussian form for the diquark BS amplitude with
$x_0/\gamma_{\hbox{\tiny GAU}} = 1$. 
In fact, the shifted-Gaussian form also produces a node in the electric form
factor of the neutron, for which there is no experimental evidence. 
We conclude that to obtain a realistic description of nucleons, 
\pagebreak[-4] one must rule out the use of forms for the diquark BS
amplitude which peak away from the origin. 

In Figure~\ref{ffdata} we compare the $n=2$ results, obtained from employing 
the diquark BS amplitude of dipole form, to the experimental data of
Refs.~\cite{Hoe76,Bos92} and Ref.~\cite{Pla90} for the electric form factors
of proton and neutron, respectively. In Ref.~\cite{Pla90}, the neutron $G_E$
is extracted from data taken on an unpolarised deuteron target and with
employing various $N$--$N$ potentials. As pointed out in
Ref.~\cite{NIKHEF}, due to possible systematic errors of this procedure,
these results should not be over-interpreted. They can serve to give us a
feeling for the qualitative behavior and rough size of the electric form
factor of the neutron, however. With resembling the phenomenological
dipole fit for the proton fairly well, as seen in Fig.~\ref{ffs}, it might
not be too surprising to discover good agreement also with the experimental
results for the proton. However, with the special emphasis of our present
study being put on charge conservation, such compelling agreement also for
finite photon momentum transfer and over the considerable range of $Q^2$ 
(from 0 up to 3.5 GeV$^2$), seems quite encouraging. Also the neutron
electric form factor compares reasonably good with the data, especially
considering that we did deliberately not put much effort in adjusting the
free parameters in our present model. 
        
The obtained magnetic moments, which range from 0.95 \dots 1.26 nuclear
magnetons for the proton and from -0.80\dots -1.13 nuclear magnetons 
for the neutron, are too small.
The accepted values for the proton and neutron magnetic moments are
2.79 and -1.91 nuclear magnetons, respectively. 
The essential reasons for this discrepancy are summarized as follows:

\begin{figure*}[t]
\begin{minipage}{\linewidth}
\vspace{1cm}
\hskip -.2cm \epsfig{file=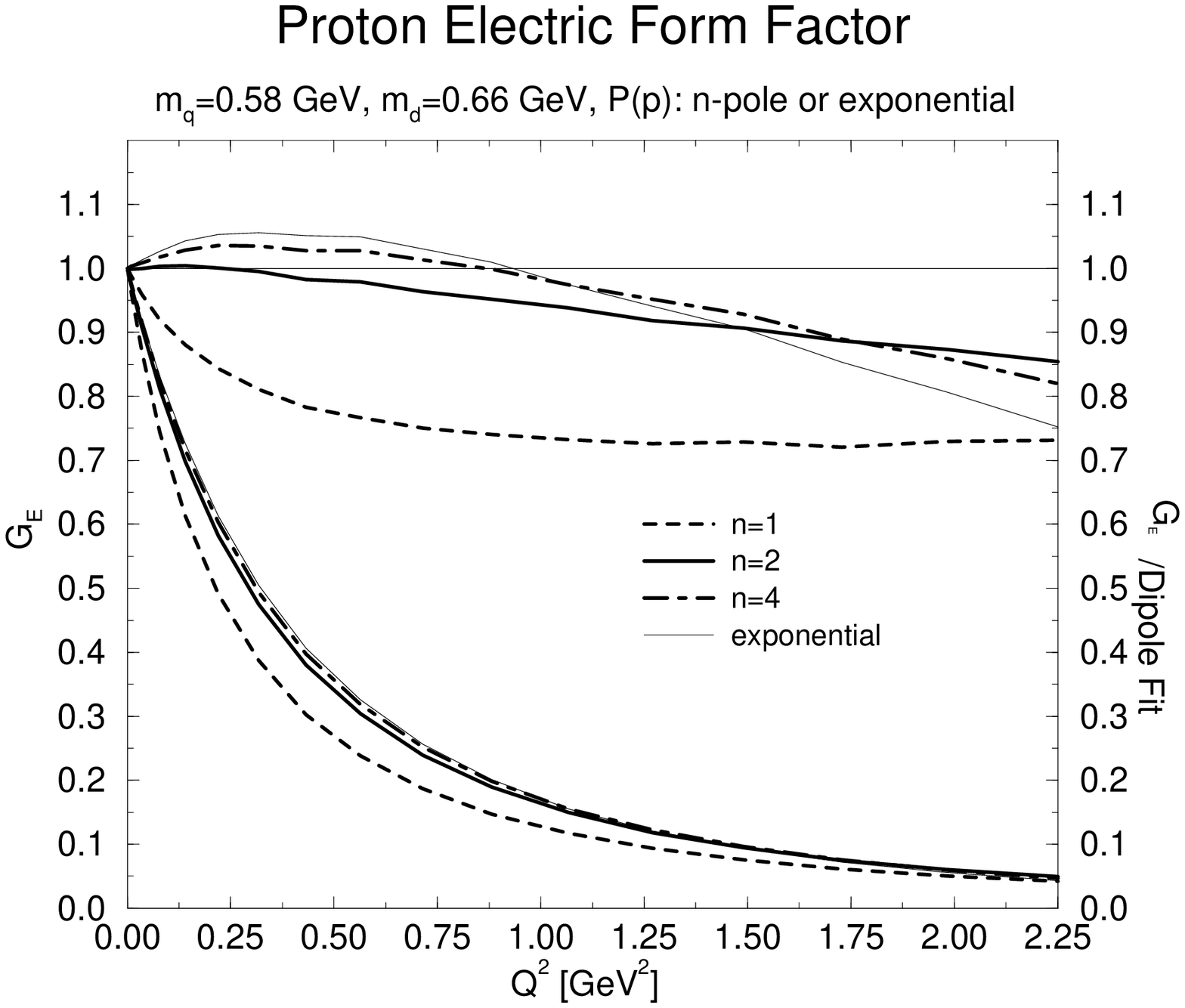,width=2\figwidth}
\hfill
\hskip -.2cm \epsfig{file=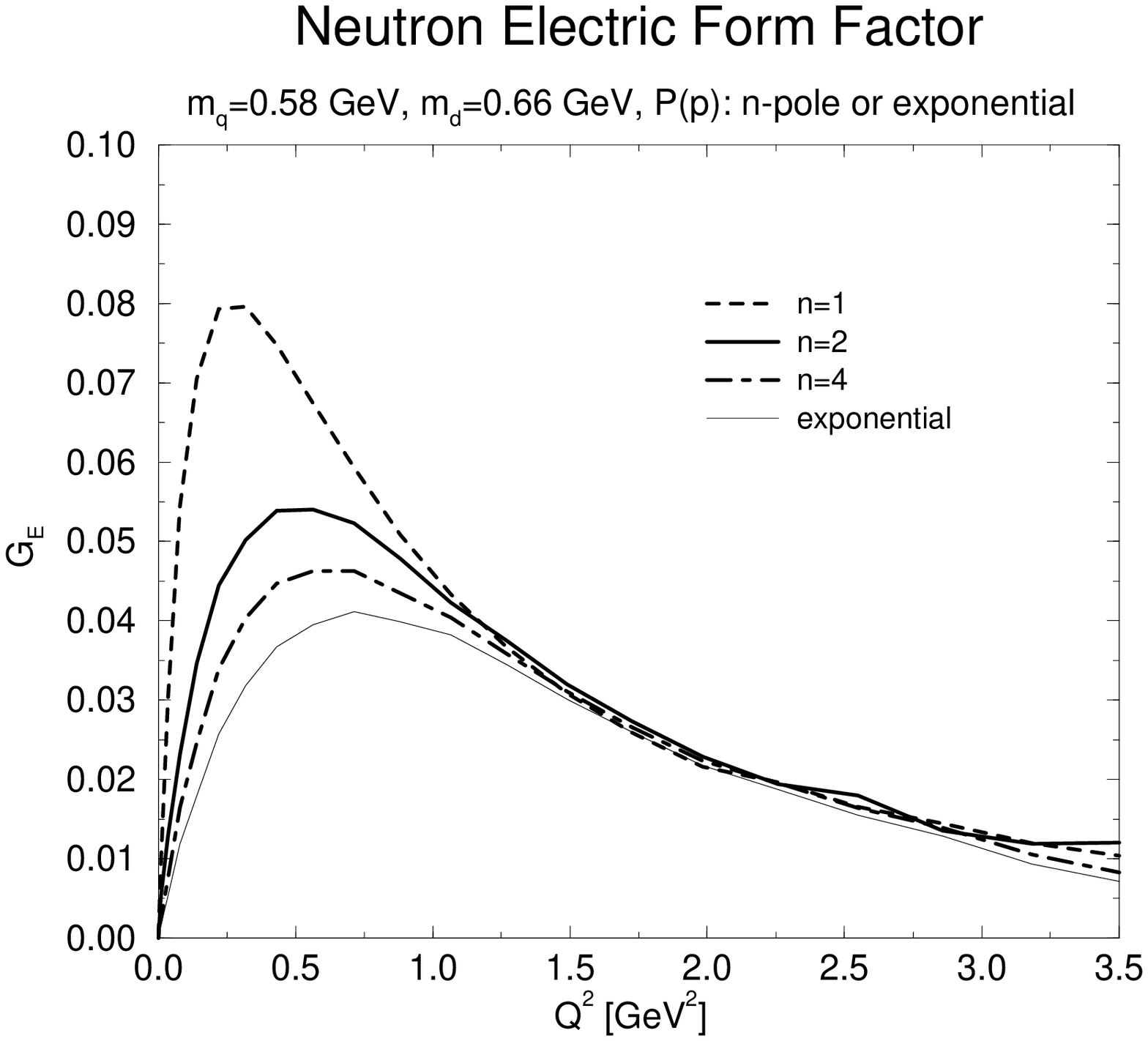,width=2\figwidth}
\end{minipage}
\refstepcounter{figure}
\centerline{\parbox{0.8\linewidth}{\small\textbf{Fig.~\thefigure.}
Nucleon electric form factors for fixed quark and
diquark masses.  
In the left plot, the bottom set of curves represent the results
obtained for the proton electromagnetic form factor $G_{E}(Q^2)$ and the
top set of curves depict the ratios of these form factors over the dipole fit
$ \big(1+Q^2/(0.84\mbox{GeV})^2\big)^{-2}$. 
\label{ffm}}}
\end{figure*}

\begin{enumerate}
\item
The next important diquark correlations which should  be included in the
present framework are those of axialvector diquarks. These are necessary for
an extension of the quark-diquark model to the decuplet baryons \cite{Oet98}.  
The contribution of their magnetic moments to the anomalous magnetic moments
of the nucleons was assessed in Ref.~\cite{Wei93}. There, the NJL model was
employed to calculate the electromagnetic form factors of on-shell diquarks,
and their influence on the nucleon magnetic moments was estimated from an
additive diquark-quark picture. The conclusion from this study was that
including the magnetic moment of the axialvector diquark alone did not
improve the nucleon magnetic moments. The additionally possible transitions
between scalar to axialvector diquarks, however, were found to raise them
substantially. Whether this finding persists in the fully relativistic
treatment, is subject to current investigations \cite{Oet00}.

Furthermore, since the axialvector diquark enhances the binding ({\it i.e.},
lowers the coupling $g_s$), it tends to lower the quark mass required to
produce the same nucleon bound state mass and as will be discussed below, a
smaller quark mass would also serve to improve the obtained values for the
magnetic moments. 
\item
In our present study, the diquark-photon vertex is that of a free scalar
particle. The contribution of the corresponding impulse approximation diagram
to the magnetic moments (in the right panel of Fig. \ref{IAD}) is small,
below 0.01 nuclear magnetons, but non-vanishing. 
Resolving the diquark substructure by coupling the photon
directly to the quarks within the diquark, and taking into account its
sub-leading Dirac structure, will dress this vertex as discussed (at the end
of Sec.~\ref{WIaS}) and it might increase that contribution. We do not expect
the gain to be substantial though. 
\item
The consistency requirement is that the strength of the 
coupling  $g_s = 1/N_s$ is given by the normalization of the diquark $N_s$. 
At present, this leads to a rather narrow nucleon BS amplitude in
combination with somewhat large values for constituent quark mass. 
Both of these effects tend to suppress the quark contribution to the form
factors arising from the impulse-approximate terms. 
If we had assumed a quark mass $m_q$ of 450~MeV, and artificially
increased the width of the diquark BS amplitude, the quark diagram alone
would easily contribute 1.3 \dots 1.5 nuclear magnetons to the magnetic
moment of the proton.   That is, a small change to the dynamics of the
diquark BS amplitude or mass of the quark can have a significant impact on
the magnetic moment of the proton.  A similar sensitivity of magnetic
moments of vector mesons to the scales in the quark propagator and bound
state vector-meson BS amplitude was also observed in Ref.~\cite{Haw99}.
\end{enumerate}

The electric form factors obtained using the fixed values for quark and
diquark masses of $m_q = 0.58$ GeV and $m_s = 0.66$ GeV,
respectively, ({\it i.e.}, for fixed values of the binding energy) are shown in
Fig.~\ref{ffm}. 
The corresponding charge radii are given in the right half of
Table~\ref{charger}.  The differences between the various diquark amplitude 
parameterizations considered herein do not lead to such dramatic differing
behaviors of the nucleon form factors in this case. 
Nevertheless, we observed that use of the exponential form of the diquark
BS amplitude still produces a proton electric form factor that falls off
too fast when compared to the phenomenological dipole fit. 
It also results in a tiny value for the square of the charge radius of the
neutron.

\begin{figure*}[t]
\begin{minipage}{\linewidth}
\vspace{1cm}
\hskip -.2cm \epsfig{file=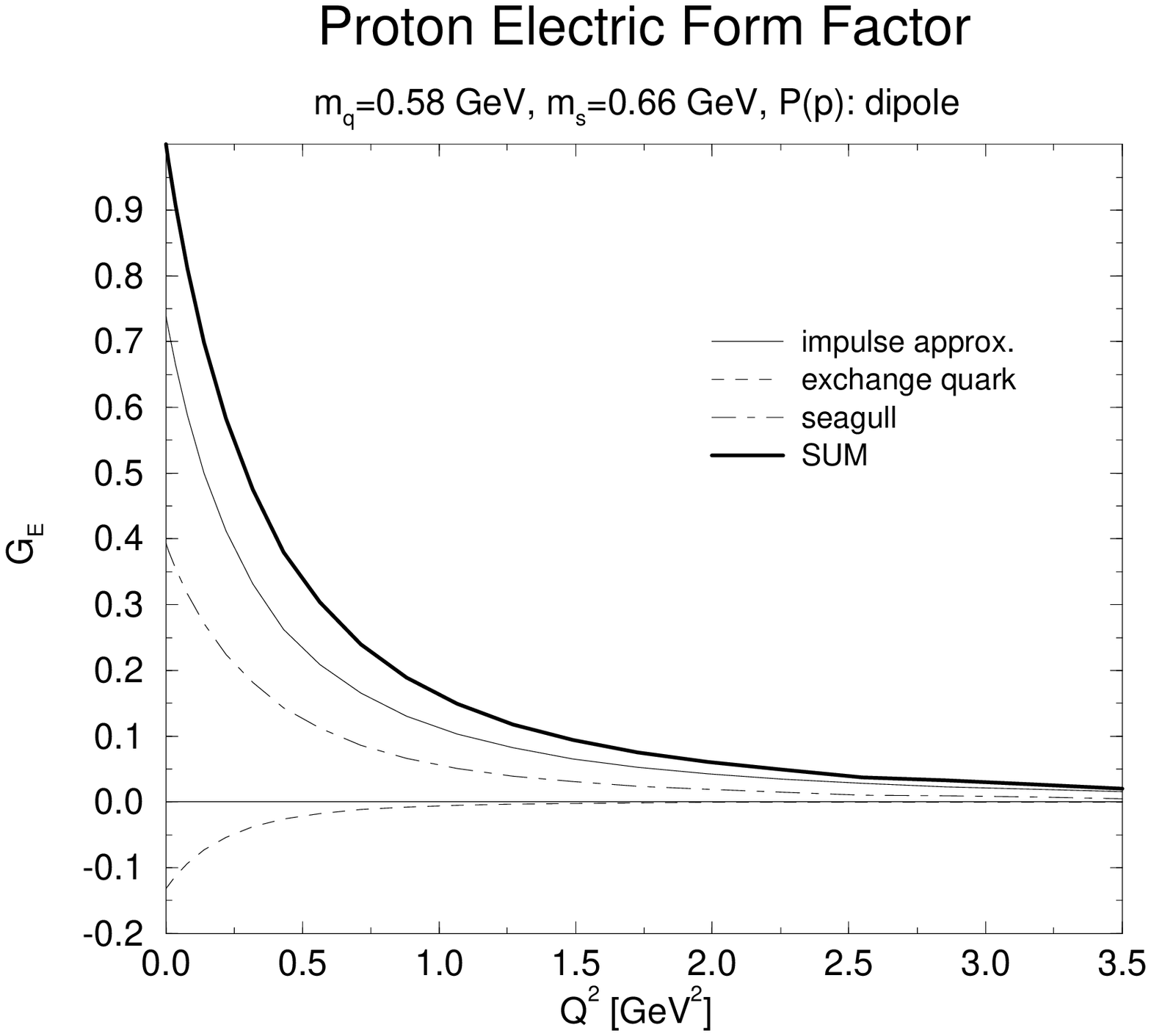,width=2\figwidth}
\hfill
\hskip -.2cm \epsfig{file=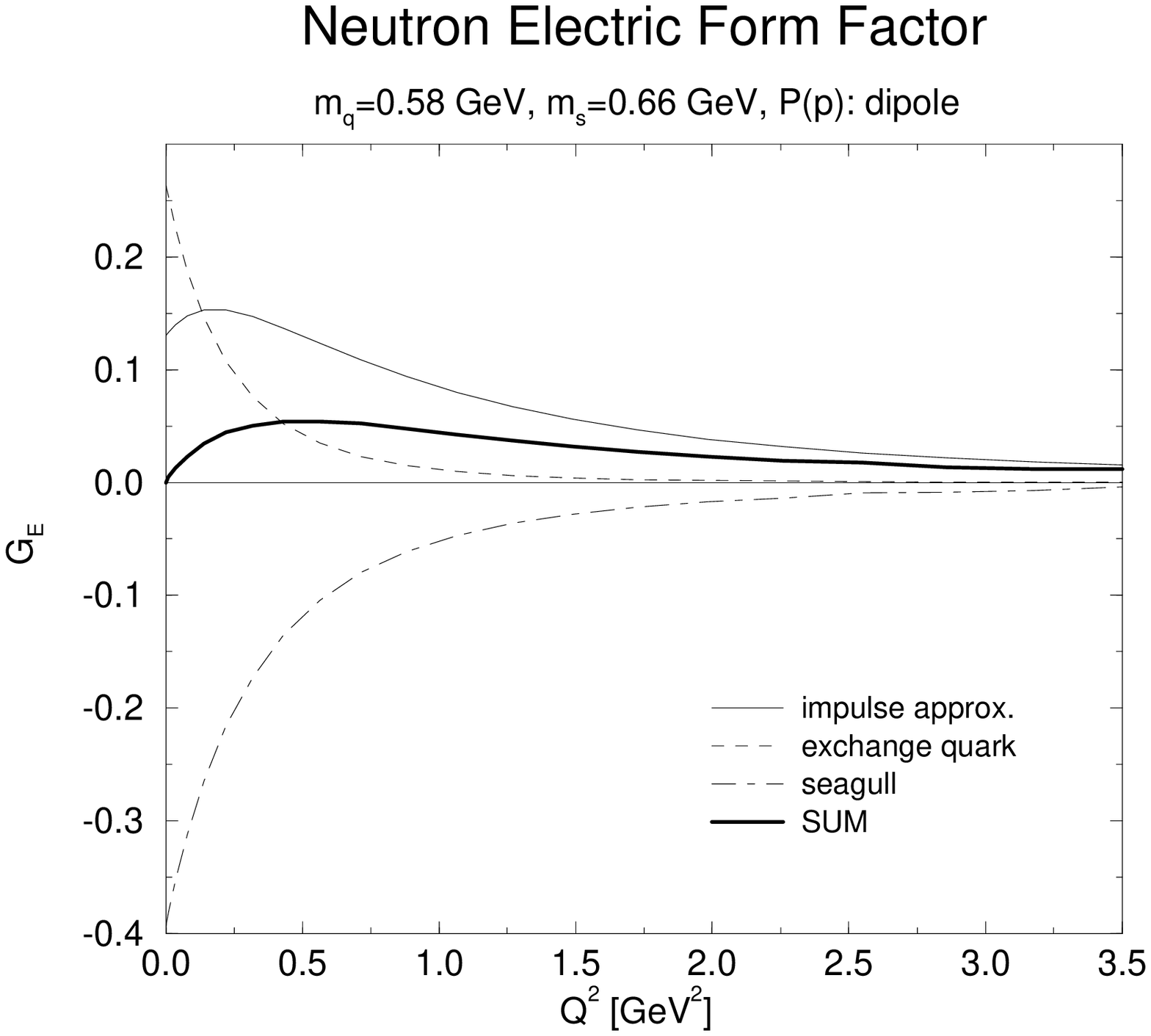,width=2\figwidth}
\end{minipage}
\refstepcounter{figure}
\centerline{\parbox{0.8\linewidth}{\small\textbf{Fig.~\thefigure.}
Contribution of the single diagrams to neutron and proton electric form
factors. The diquark amplitude is the dipole. The sum of these contributions
{\sf (SUM)} corresponds to the result of Fig.~\ref{ffdata}.\label{gedipole}}}
\end{figure*}

Finally, we compare the relative importance of the various contributions
to the form factors that arise from the impulse-approximate,
exchange-quark, and seagull diagrams. 
We separately plot each of these contributions to the total proton and
neutron electric form factors in Fig.~\ref{gedipole} for comparison. 
For the purposes of comparison, we used the parameter set for the
dipole form of the diquark BS amplitude.  (This form of the diquark BS
amplitude is used because of the excellent description it provides for the
the electric form factor of the proton.)

Here, it is interesting to note that the seagull couplings contribute
up to about 2/5 of the proton electric form factor $G_E(Q^2)$!  
It is clear that in the nucleon, these beyond-the-impulse contributions are
certainly not negligible. 
As in the previous discussion of the magnetic moments, where we observed 
a suppression of the impulse-approximate diagrams arising from the narrow
momentum distribution of the nucleon BS amplitude which was a result of
employing narrow diquark BS amplitude.
Again, we find that employing a diquark BS amplitude that is wider in
momentum space leads to a wider nucleon BS amplitude and therefore implies
that the impulse approximate contributions are more dominant than the
other contributions.   This is the case when point-like diquark BS
amplitudes are employed, such as in Ref.~\cite{Hel97b}.

For the neutron, we observe that its electric form factor arises
from a sum of large terms which strongly cancel each other to produce a
small effect.  This cancellation is the explanation for the appearance of
``wiggles'' in the results for the neutron $G_E(Q^2)$ at moderate $Q^2$
shown in Figs.~\ref{ffs}, and \ref{ffm}.   
The wiggles are artifacts of the numerical procedure employed.

%% file: Sec8.tex
\section{Conclusions}
\label{Conclusions}

We have introduced an extension of the covariant
quark-diquark model of baryon bound states.   The framework developed
herein allows for the inclusion of finite-sized diquark correlations
in the description of the nucleon bound state in a manner which preserves
electromagnetic current conservation for the first time.
For such a framework to maintain current conservation of the nucleon, it
is necessary to include contributions to the electromagnetic current which
arise from the couplings of the photon to the quark-exchange kernel of the
nucleon BSE.  
These contributions are derived from the Ward-Takahashi identities of
QED and include the coupling of the photon to the exchanged quark in the
kernel and the photon coupling directly to the BS amplitude of the
diquark (the so-called seagull contributions). 
It was shown analytically that the resulting nucleon current is conserved
and these additions are sufficient to ensure the framework provides the
correct proton and neutron charges independent of the details of the model
parameters. 

To explore the utility of this framework under the most simple model
assumptions, simple constituent-quark and constituent-diquark propagators
and one-parameter model diquark BS amplitudes were employed for the 
numerical application of the framework.
The BS amplitude for the scalar diquark was parametrized by the leading
Dirac structure and various forms for the momentum dependence of the
amplitude were investigated.
It was shown that the antisymmetry of the diquark
amplitude under quark exchange places tight constraints on the form of the
diquark BS amplitude and that the incorporation of these constraints have
the effect of removing much of the model dependence of the diquark BS
amplitude parameters from the calculation of the nucleon electromagnetic
form factors. 

Calculations of the electromagnetic form factors away from $Q^2 = 0$
require that the nucleon BS amplitudes and wave functions be boosted. 
In this Euclidean-space formulation, this amounts to a continuation
of amplitudes and wave functions into complex plane. Two procedures to
account for proper handling of poles arising from the constituent
particles in the nucleon are described and compared.  The feasibility of
each procedure is discussed and explored in detail.  
It is shown analytically and explicitly in a numerical calculation that
the two approaches produce the same results for the electromagnetic form
factors.  Thus, demonstrating that the framework properly accounts for the
non-trivial analytic structures of the constituent propagators. 

For the particular choice of simple dynamical models explored herein, the
masses of the quark and diquark propagators along with the width of the
model diquark BS amplitude are the only free parameters in the framework.
The latter width is thereby implicitly determined from fixing the nucleon mass.
It is shown, for the case of the dipole (and possibly quadrupole) form of 
the diquark BS amplitude, that these few parameters are sufficient
to provide an excellent description of the electric form factors for both
the neutron and proton.  
Other forms of the diquark BS amplitude that were explored, such as the
exponential and Gaussian forms, may be ruled out on
phenomenological grounds as they lead to nucleon electromagnetic form
factors which are inconsistent with the experimental data.

It was found that the framework is at present unable to reproduce the nucleon
magnetic moments. The calculated magnetic moments are smaller than those
obtained in experiment by about 50\%.  
Possible explanations for this are the effect of an inclusion of axialvector
diquark correlations, the addition of more complex 
structures in the diquark amplitudes, and a resolution of the
quark substructure in the diquark-photon coupling.
These improvements are the subject of work currently in progress.

In conclusion, we find that the covariant quark-diquark model of the
nucleon provides a framework that is sufficiently rich to describe the
electromagnetic properties of the nucleon.  However, to ensure that 
the framework satisfies electromagnetic current conservation one must 
go beyond the usual impulse approximation diagrams and include
contributions that arise from the photon couplings to the nucleon BSE kernel.
In a numerical application of this framework, it was found that these
contributions provide a significant part of the electromagnetic form
factors of the nucleon and can not be neglected.

%% file: Acknowledgement.tex
\section*{Acknowledgments}

The authors gratefully acknowledge valuable discussions with R.~Alkofer
and N.~Ishii. We are indebted to H.~Witala for bringing Ref.~\cite{Car98} to
our attention and to B.~Blankleider for helping to clarify the
relation of our study to the general considerations of Ref.~\cite{Bla99a}.  

The work of M.O. was supported by the Deutsche For\-schungsge\-mein\-schaft 
under contract DFG We 1254/4-1. He thanks H.~Weigel and H.~Reinhardt
for their continuing support.

The work of M.A.P. was supported by the U.S. Department of Energy
under contracts DE-FG02-87ER40365, 
DE-FG02-86ER40273 and DE-\-FG\-05-\-92ER40750,
the National Science Foundation under contract PHY9722076  and 
the Florida State University Supercomputer Computations Research Institute
which is partially funded by the Department of Energy under 
contract DE-FC05-85ER25000.

The work of L.v.S. was supported by the U.S. Department of Energy, Nuclear
Physics Division, under contract number W-31-109-ENG-38 and by the BMBF under
contract number 06-ER-809. Parts of the calculations were performed on the
IBM SP3 Quad Machine of the Center for Computational Science and Technology
at Argonne National Laboratory.

%% file: Appendix0.tex
\section{Conventions for Diquark Amplitudes}  
\label{AppDqDetails}

The BS wave functions $\chi(p,P)$ and $\bar\chi(p,P)$ of the (scalar) diquark
bound-state are defined by the matrix elements,
\begin{eqnarray}
\chi_{\alpha\beta}(x,y;\vec P) &:=& \langle q_\alpha(x) q_\beta(y) |P_+
\rangle  \\ 
\bar \chi_{\alpha\beta}(x,y;\vec P) &:=& \langle P_+ | \bar q_\alpha(x) \bar
q_\beta(y) \rangle  \; , \nn \\
&=& \left(\gamma_0 \chi^\dagger(y,x;\vec P) \gamma_0 \right)_{\alpha\beta}
\label{conj_amp}\end{eqnarray} 
Note that there is no need for time ordering here in contrast to
quark-antiquark bound states. The following normalization of the states is
used, 
\begin{eqnarray}
\langle P'_\pm | P_\pm \rangle \, = \, 2 \omega_P \, (2\pi)^3 \delta^3 (\vec P'
- \vec P) \, , \;  \omega_P^2 = \vec P^2 + m_s^2 \; ,
\end{eqnarray}
and the charge conjugate bound state being $ |P_- \rangle \, =\, C | P_+
\rangle $. The contribution of the charge conjugate bound state is included in
Eq.~(\ref{dq_pole_ms}) for $P_0 = - \omega_P $.
From invariance under space-time translations, the BS wave function has 
the general form,  
\begin{eqnarray}
\chi_{\alpha\beta}(x,y;\vec P) = e^{-iPX} \int
\frac{d^4p}{(2\pi)^4} \, e^{-ip(x-y)}     \chi_{\alpha\beta}(p,P) \, ,
\end{eqnarray} 
with  $X = (1\!-\!\sigma) x + \sigma y $,  $p := \sigma p_\alpha  -
(1\!-\!\sigma) p_\beta $, and $ \, P = p_\alpha  + p_\beta  $, where
$p_\alpha$, $p_\beta$ denote the momenta of the outgoing quarks in the
Fourier transform  $ \chi_{\alpha\beta}(p_\alpha ,p_\beta;\vec P) $ of  $
\chi_{\alpha\beta}(x,y;\vec P) $, see Fig.~\ref{dqamps}. One thus has the 
relation,
\begin{eqnarray}  
\chi_{\alpha\beta}(p,P) := \chi_{\alpha\beta}(p+ (1\!-\!\sigma) P
,-p+\sigma P ;\vec P)\big\vert_{P_0 = \omega_{\scriptscriptstyle P}} \; .
\end{eqnarray} 
In the definition of the conjugate amplitude, the convention 
\begin{eqnarray}
\bar\chi_{\alpha\beta}(x,y;\vec P) \, = \, e^{iP\bar X} \int
\frac{d^4p}{(2\pi)^4} \, e^{-ip(x-y)} \,    \bar\chi_{\alpha\beta}(p,P) \; ,
\end{eqnarray} 
with $\bar X = \sigma x + (1\!-\!\sigma) y $, ensures that hermitian
conjugation from Eq.~(\ref{conj_amp}) yields,   
\begin{eqnarray} 
\bar \chi_{\alpha\beta} (p,P)\, = \, \left( \gamma_0 \chi^\dagger(p,P)
\gamma_0 \right)_{\alpha\beta} \; . \label{hcra}
\end{eqnarray} 
In the conjugate amplitude $\bar\chi_{\alpha\beta}
(p,P) $, the definition of relative and total momenta corresponds to $p =
(1\!-\!\sigma) p'_\alpha  - \sigma p'_\beta $ and $P = - p'_\alpha  - 
p'_\beta $ for the outgoing quark momenta $p'_\alpha , p'_\beta$ in 
\begin{eqnarray}
   \bar\chi_{\alpha\beta}(p'_\alpha ,p'_\beta ;\vec P) \, =\,  
\left(\gamma_0  \chi^\dagger (-p'_\beta , -p'_\alpha ;\vec P) \gamma_0
\right)_{\alpha\beta} \; ,
\end{eqnarray} 
{\it c.f.}, Fig.~\ref{dqamps}. Note here that hermitian conjugation implies
for the momenta of the two respective quark legs, $ p_\alpha \to - p_\beta'
$, and $  p_\beta \to  - p_\alpha' $, which is equivalent
to $ \sigma \leftrightarrow (1\!-\!\sigma) $ and $P\to - P$.  Besides the
hermitian conjugation of Eq.~(\ref{hcra}), one has from the antisymmetry of
the wave function, $\chi_{\alpha\beta}(x,y;\vec P) =
-\chi_{\beta\alpha}(y,x;\vec P)$. For the corresponding functions of the
relative coordinates/momenta, this entails that $\sigma$ and $(1\!-\!\sigma)$
have to be interchanged in exchanging the quark fields,
\begin{eqnarray}
\chi(x,P) &=& \left. - \chi^T(-x,P)\right|_{\sigma
\leftrightarrow (1-\sigma)} \; , \\
\quad  \chi(p, P) &=& \left. -
\chi^T(-p,P)\right|_{\sigma \leftrightarrow (1-\sigma)} \;
. \nn 
\end{eqnarray}
This interchange of the momentum partitioning can be undone by a charge
conjugation, from which the following identity is obtained,
\begin{eqnarray} 
\chi^T(p,P) = - C \bar \chi(-p,-P) C^{-1} \; . \end{eqnarray} 
This last identity is useful for relating $\bar\chi $ to $ \chi$ in
Euclidean space. In particular, this avoids the somewhat ambiguous definition
of the conjugation following from Eq~(\ref{conj_amp}) in Euclidean space with
complex bound-state momenta.  

\input pictex/Fig2

One last definition for diquark amplitudes concerns the truncation of the
propagators $S$ attached to the quark legs, thus defining the amputated
amplitudes $\widetilde\chi$ and $\widetilde{\bar\chi}$ by 
\begin{eqnarray}
\chi_{\alpha\beta}(p,P) &=& \left( S(p_\alpha) \widetilde
\chi(p,P) S^T(p_\beta)\right)_{\alpha\beta} \; , \\
\bar\chi_{\alpha\beta}(p,P) &=& \left( S^T(-p'_\alpha) \widetilde{
\bar\chi}(p,P) S(-p'_\beta)\right)_{\alpha\beta} \; . \end{eqnarray} 
With the definitions above, the same relations hold for the amputated
amplitudes, in particular, 
\begin{eqnarray} 
\widetilde{\bar\chi}(p,P) &=& \gamma_0
\widetilde\chi^\dagger(p,P) \gamma_0 \; , \\
\widetilde\chi(p, P) &=&  \left. - \widetilde\chi^T(-p,P)\right|_{\sigma
\leftrightarrow (1-\sigma)}  \! .  \nn 
\end{eqnarray} 
In Sec.~\ref{dq_corrs} the antisymmetric Green function $G^{(0)}$ for the
disconnected propagation of identical quarks with propagator $S(p)$,
\begin{eqnarray}
G^{(0)}_{\alpha\gamma , \beta\delta}(p,q,P) \, &=& \, (2\pi)^4 \delta^4(p-q)
\\ 
&& \hskip -2cm 
S_{\alpha\beta}(\sigma P -p) \,  S_{\gamma\delta}((1\!-\!\sigma)P+p) 
\quad - \quad \mbox{crossed term}\, , \nn
\end{eqnarray}
was used to derive the normalization condition for the diquark
amplitudes. This  notation is somewhat sloppy. In particular in 
the second term proportional to $\delta^4(p+q)$, representing the crossed
propagation with exchange of the external quark lines, one may use either 
$p$ or $-q$ in the arguments of the propagators. Exchanging one for the other
is possible only with, at the same time, exchanging $\sigma \leftrightarrow
1\!-\!\sigma $ as well, however. Momentum conservation entails that incoming
and outgoing quark-pairs in successive correlation functions can only be
connected if, besides the relative momenta $p = \pm q$, also the momentum
partitioning variables of the pairs match. We can include this condition
explicitly by introducing temporarily the dimensionless $\sigma' $ and
$\sigma $ (both in $[0,1]$) for the incoming and outgoing pair respectively,
thus writing, 
\begin{eqnarray}
G^{(0)}_{\alpha\gamma , \beta\delta}(p,\sigma,q,\sigma',P) \, &=& \,
(2\pi)^4 \delta^4(p-q)  \, \delta(\sigma - \sigma')
\label{A.15}  \\ 
&& 
S_{\alpha\beta}(\sigma P -p) \,  S_{\gamma\delta}((1\!-\!\sigma)P+p) \,
 \nn \\
&& \hskip -1.5cm - \,  (2\pi)^4 \delta^4(p+q) \, \delta(\sigma
+\sigma' -1 ) \nn \\  
&& S_{\alpha\delta}(\sigma P -p) \, S_{\gamma\beta}((1\!-\!\sigma)P+p) \,\;
. \nn 
\end{eqnarray}
The inverse ${G^{(0)}}^{-1}$ can then be defined as,
\begin{eqnarray}
{G^{(0)}}^{-1}\hskip -.8cm _{\alpha\gamma , \beta\delta}(p,\sigma,q,\sigma',P)
\, &=&  \frac{1}{4} \, \Big( \,
(2\pi)^4 \delta^4(p-q)  \, \delta(\sigma - \sigma')  
\label{A.16} \\  
&& 
S_{\alpha\beta}^{-1}(\sigma P -p) \, S_{\gamma\delta}^{-1}((1\!-\!\sigma)P+p)
\nn \\ 
&& \hskip -1.5cm - \, (2\pi)^4 \delta^4(p+q) \,
\delta(\sigma +\sigma' -1 )  \nn\\
&&  S_{\alpha\delta}^{-1}(\sigma P -p) \,
S_{\gamma\beta}^{-1}((1\!-\!\sigma)P+p) \Big) \; , \nn
\end{eqnarray}
giving the exchange antisymmetric unity in the space of (identical) 2-quark
correlations upon (left or right) multiplication with $G^{(0)}$, 
\begin{eqnarray}
&& \hskip -.2cm  \int \frac{d^4k}{(2\pi)^4} \int_0^1 d\tilde\sigma    \,
G^{(0)}_{\alpha\gamma , \rho\omega}(p,\sigma,k,\tilde\sigma,P) \, 
{G^{(0)}}^{-1}\hskip -.8cm _{\rho\omega ,
\beta\delta}(k,\tilde\sigma,q,\sigma',P)  = \nn  \\
&& \hskip .2cm 
 \frac{1}{2} \bigg( \delta_{\alpha\beta}\, \delta_{\gamma\delta} \,
(2\pi)^4 \delta^4(p-q)  \, \delta(\sigma - \sigma')  \\
&& \hskip 2cm  - \,  \delta_{\alpha\delta} \,
\delta_{\gamma\beta}\, (2\pi)^4 \delta^4(p+q) \, \delta(\sigma
+\sigma' -1 )  \bigg) \; .\nn
 \end{eqnarray} 
This is the way the multiplication of 2-quark correlation functions
for identical quarks used in Sec.~\ref{dq_corrs} is understood properly.
Either Eq.~(\ref{A.15}) or Eq.~(\ref{A.16}) can be used in the derivation of
the normalization condition for the diquark amplitude $\chi$ for identical
quarks, Eq.~(\ref{dq_norm}).

\section{Supplements on the Nucleon BSE}
\label{SuppNBSE}
In this appendix we would first like to explore the possibility of having
exchange-symmetric arguments in the diquark amplitudes of the
quark-exchange kernel of the nucleon BSE. Consider the 
invariant $x_1$ for the relative momentum in the incoming diquark in the
kernel~(\ref{x_kern_par}) corresponding to the definition of the 
symmetric argument of $P(x)$ given in Eq.~(\ref{x_def}). From
Eqs.~(\ref{p_1}) and~(\ref{x_1}) one obtains (with $\sigma +\hat\sigma = 1$
and $\eta +\hat\eta = 1$ determining the momentum partitioning within the
diquark and nucleon respectively), 
\begin{eqnarray}
x_1 & = & - p^2 - \sigma\hat\sigma k^2 - pk + (1\!-\!3\eta)Pp +
(2\sigma\hat\sigma\hat\eta\!-\!\eta) Pk \nn \\
     && \hskip 2cm + (\eta (1\!-\!2\eta)-\sigma\hat\sigma\hat\eta^2) P^2 \; .
\end{eqnarray} 
One verifies readily that for $\sigma = (1\!-\!2\eta)/(1\!-\!\eta) $,
according to~(\ref{dq_mom_part}), the prefactor of the term $\propto P^2$
vanishes, and that of the term  $\propto Pk$ becomes $\hat\sigma
(1\!-\!3\eta) $ leaving only the choice $\eta = 1/3$ for a $P$-independent 
$x_1$. One might now argue that, from the antisymmetry considerations alone,
we should be free to add an arbitrary term proportional to the square of the 
total diquark momentum in the definition of $x$. We will show now that such a
redefinition cannot lead to a $P$-independent $x$ either (for values of
$\eta$ different from $1/3$).  Here, the diquark momentum
is $P_D = q+p_\alpha = \hat\eta  P - k $ and with 
\begin{eqnarray}
\hat x_1 & := & x_1 + C P_D^2 \nn \\
P_D^2 & = &   \hat\eta^2 P^2 - 2\hat\eta Pk + k^2 \; ,
\end{eqnarray} 
one finds that in order to have no terms $\propto P^2$ in  $\hat x_1 $,
\begin{eqnarray}  
C = - \hat\eta^{-2}\, (\eta (1\!-\!2\eta)-\sigma\hat\sigma\hat\eta^2) \; .
\end{eqnarray}
With this $C$, however, one has 
\begin{eqnarray}
\hat x_1 =  - p^2 - \eta \hat\eta^{-2}\!(1\!-\!2\eta) k^2 - pk +
(1\!-\!3\eta) ( Pp + \eta\hat\eta^{-1} Pk ) \, , \nn 
\end{eqnarray} 
independent of $\sigma$. This shows that the symmetric arguments of the
diquark amplitudes $\chi$ and, analogously, $\bar\chi$ can quite generally be
independent of the total nucleon momentum $P$ only for $\eta = 1/3$.

The remainder of this appendix describes the structure of the nucleon BSE for
the bound-state of scalar diquark and quark in some more detail. The form of
the bound-state pole in the scalar-fermion 4-point function,
Eq.~(\ref{nuc_pole_cont}), implies that the corresponding bound-state
amplitudes obey, 
\begin{eqnarray}
\widetilde\psi(p ,P_n) \Lambda^+(P_n) &=& \widetilde\psi(p ,P_n)\; ,\\
\Lambda^+(P_n) \widetilde{\bar\psi}(p,P_n) &=&
\widetilde{\bar\psi}(p,P_n) \; , \nn
\end{eqnarray}
with $\Lambda^+(P_n) \, =\, (\fslash P_n + M_n)/2M_n $. Therefore, the
amplitudes can be decomposed as follows:
\begin{eqnarray}
\widetilde\psi(p ,P_n)\, &=& \label{psiDec.A} \\
&& \hskip -1.5cm  S_1(p,P_n)\, \Lambda^+(P_n) \, +\,
S_2(p,P_n) \,  \Xi(p,P_n)\, \Lambda^+(P_n) \; , \nn\\
\widetilde{\bar\psi}(p,P_n) \, &=& \label{psibarDec.A} \\
&& \hskip -1.5cm  \, S_1(-p,P_n)\,  \Lambda^+(P_n) \, +\, 
S_2(-p,P_n)\,  \Lambda^+(P_n)\,  \Xi(-p,P_n) \; ,  \nn
\end{eqnarray}
with $\Xi(p,P_n) = (\fslash p - pP_n/M_n)/M_n $. This simply separates positive
from negative energy components of the amplitudes,
\begin{eqnarray}
&& \hskip -.2cm 
\fslash P_n  \, \Lambda^+(P_n) \,=\,   M_n \,  \Lambda^+(P_n) \; , \\
&& \hskip -.2cm 
\fslash P_n \, \Xi(p,P_n)\, \Lambda^+(P_n) \,=\,  - M_n \, \Xi(p,P_n)\,
\Lambda^+(P_n) \; , \nn \\  
&& \hskip -.2cm 
\hbox{and thus,} \quad  \Lambda^+(P_n)\,  \Xi(p,P_n)\,
\Lambda^+(P_n) \, = \, 0 \; . \nn
\end{eqnarray} 
One furthermore has,
\begin{eqnarray}
  \Lambda^+(P_n)\,  \Xi(p,P_n) \Xi(p,P_n)\, \Lambda^+(P_n)
 &=& \\
&& \hskip -1cm 
 \left(\frac{p^2}{M_n^2} - \frac{(pP_n)^2}{M_n^4} \right) \,
\Lambda^+(P) \; , \nn
\end{eqnarray}
which allows to rewrite the homogeneous BSE (\ref{hom_nuc_BSE}) in terms of
2-vectors $S^T(p,P_n) := (S_1(p,P_n),\, S_2(p,P_n))$, using the kernel
(\ref{x_kern_par}), 
\begin{eqnarray} 
S(p,P_n) &=& \\
&& \hskip -1.1cm 
 \frac{1}{2N_s^2}  \int \frac{d^4k}{(2\pi)^4} \; P(x_1) P(x_2)
\,  D(k_{s}) \,  T(p,k,P_n)  \, S(k,P_n) \; ,\nn
\end{eqnarray} 
with $k_s = (1\!-\!\eta)P_n-k\, $, $\, k_q = k + \eta P_n$ and 
\begin{eqnarray}
T(p,k,P_n) \, &=& \, \frac{1}{2}  
\left( \begin{array}{cc}
1 & 0 \\
0 &   ~~~~\left(\frac{p^2}{M_n^2} - \frac{(pP_n)^2}{M_n^4} \right)^{-1}
\end{array} \right) \times  \\[+12pt]
&& \hskip -2.1cm \left( \begin{array}{l}
\tr \left\{ S(q) S(k_q)  \Lambda^+(P_n)\right\} \\
\tr \left\{ S(q) S(k_q)  \Lambda^+(P_n) \Xi(p,P_n) \right\} \end{array}
\right.     \nn \\
&&  \hskip -.1cm \left. \begin{array}{r} 
  \tr \left\{ S(q) S(k_q)  \Xi(k,P_n) \Lambda^+(P_n)\right\} \\
 \tr \left\{ S(q) S(k_q)  \Xi(k,P_n) \Lambda^+(P_n) \Xi(p,P_n) \right\}
\end{array} \right) \, .  \nn
\end{eqnarray} 
After performing these traces 
the transfer to Euclidean metric introducing 4-dimensional polar variables 
is done according to the prescriptions,
\begin{eqnarray}
&&p^2,\,  k^2 \, \to \, -p^2, \, -k^2 \; , \quad P_n^2 \to M_n^2 \; , \nn \\
&& pP_n \, \to \, i M_n  p\,  y \; , \quad kP_n \, \to \, i M_n k \, z \; , 
\label{WickRot.A} \\
&& pk \, \to \, - k \, p \, u(x,y,z) \; , \nn\\
&&  u(x,y,z) \, = \, yz + x \, \sqrt{1-y^2}\sqrt{1-z^2}  \; . \nn
\end{eqnarray}   
In these variables, the nucleon BS-amplitudes are functions of the
modulus of the relative momentum and its azimuthal angle with the total
momentum. The matrix $T$ in the kernel, in addition to the moduli $p,k$ and
azimuthal angles $y,z$ of both relative momenta, also depends on the angle
$u$ between them,       
\begin{eqnarray}
&&S(p,P_n) \, \to \, S(p,y) \; ,\nn\\
&&T(p,k,P_n) \, \to\, T(p,y,k,z,u(x,y,z))\; . 
\end{eqnarray} 
The azimuthal dependence of the amplitudes is taken into account by means of
a Chebyshev expansion to order $N$, see, {\it e.g.}, Ref.~\cite{Pre94},
\begin{eqnarray}
&& S(p,y) \, \simeq \, \sum_{n=0}^{N-1} (-i)^n \, S_n(p) T_n(y) \; ,
\label{ChebyS.A} \\
&& S_n(p) \, = \, i^n \, \frac{2}{N} \sum_{k=1}^N S(p,y_k) T_n(y_k) \; ,
\label{ChebyM.A} \\ 
&& \hskip -.2cm \hbox{where the} \;\; 
y_k \, = \, \cos\left( \frac{\pi (k- 1/2)}{N} \right)   
\nn \end{eqnarray}
are the zeros of the Chebyshev polynomial of degree $N$. Here,
Chebyshev polynomials of the 1st kind are used with, for later convenience, a
somewhat non-standard normalization $T_0 := 1/\sqrt{2}$. An explicit
factor $(-i)^n$ was introduced in order to obtain real Chebyshev moments
$S_n(p)$ for all $n$. Analogous formulae are obtained for expansions in
Chebyshev polynomials of the 2nd kind, which are used in
Refs.~\cite{Hel97b,Oet98}. The nucleon BSE now reads,    
\begin{eqnarray}
S_m(p) &=& \\
&&  \ - \frac{1}{2N_s^2} \, \int \frac{k^3 dk}{(4\pi)^2} \,
 \sum_{n=0}^{N-1} \,  i^{m-n} \; T_{mn}(p,k) \, S_n(k) \; , \nn
\end{eqnarray}
with 
\begin{eqnarray}
T_{mn}(p,k) \, &=& \frac{2}{\pi} \int_{-1}^1 \, 
\sqrt{1-z^2} dz  \\
&& \hskip -1.5cm  
 \frac{1}{x_s + m_s^2}  \;  \int_{-1}^1 dx 
\,   \frac{2}{N} \sum_{k=1}^{N} \, \biggl(P(x_1) P(x_2)\biggr)_{y
= y_k} \nn\\
&& \hskip .8cm    T(p,y_k,k,z,u(x,y_k,z)) \; T_m(y_k) T_n(z) \; ,  \nn   
\end{eqnarray}
where $x_s = k^2 + 2i (1\!-\!\eta) M_n k z - (1\!-\!\eta)^2 M_n^2 $ is the
invariant momentum of the free scalar propagator $D$ of mass $m_s$.

%% file: pictex/Fig2.tex
%
\begin{figure}[t]
\vskip 1cm
\parbox{.49\linewidth}{
\begin{picture}(16000,6000)(-9000,-3100)
\drawline\fermion[\NW\REG](0,0)[2000]
\drawline\fermion[\S\REG](\pbackx,\pbacky)[\fermionlengthy]
\drawline\fermion[\S\REG](\pbackx,\pbacky)[\fermionlength]
\drawline\fermion[\NE\REG](\pbackx,\pbacky)[2000]
\drawline\fermion[\W\REG](\pfrontx,-1000)[2000]
\drawarrow[\W\ATBASE](\pmidx,\pmidy)
\drawline\fermion[\W\REG](\pfrontx,1000)[2000]
\drawarrow[\W\ATBASE](\pmidx,\pmidy)
\put(0,0){\circle*{300}}
\thinlines
\drawline\fermion[\E\REG](0,100)[2000]
\drawline\fermion[\E\REG](0,-100)[2000]
\thicklines
\drawarrow[\W\ATTIP](1000,0)
\put(-8000,4000){\framebox{$\chi_{\alpha\beta}$}}
\put(2200,-400){$P$}
\put(-3000,-2000){$p_\alpha$}
\put(-3000,1600){$p_\beta$}
\put(-4500,-1200){$x$}
\put(-4500,800){$y$}
\end{picture}}
\hfill
\parbox{.49\linewidth}{
\begin{picture}(16000,6000)(-7000,-3000)
\drawline\fermion[\NE\REG](0,0)[2000]
\drawline\fermion[\S\REG](\pbackx,\pbacky)[\fermionlengthy]
\drawline\fermion[\S\REG](\pbackx,\pbacky)[\fermionlength]
\drawline\fermion[\NW\REG](\pbackx,\pbacky)[2000]
\drawline\fermion[\E\REG](\pfrontx,-1000)[2000]
\drawarrow[\E\ATBASE](\pmidx,\pmidy)
\drawline\fermion[\E\REG](\pfrontx,1000)[2000]
\drawarrow[\E\ATBASE](\pmidx,\pmidy)
\put(0,0){\circle*{300}}
\thinlines
\drawline\fermion[\W\REG](0,100)[2000]
\drawline\fermion[\W\REG](0,-100)[2000]
\thicklines
\drawarrow[\W\ATBASE](-1000,0)
\put(-7000,4000){\framebox{$\bar\chi_{\alpha\beta}$}}
\put(-3000,-300){$P$}
\put(2400,-2000){$p'_\beta$}
\put(2400,1600){$p'_\alpha$}
\put(4000,-1200){$y$}
\put(4000,800){$x$}
\end{picture}}
\caption{Momentum definitions in diquark amplitudes.}
\label{dqamps}
\end{figure}
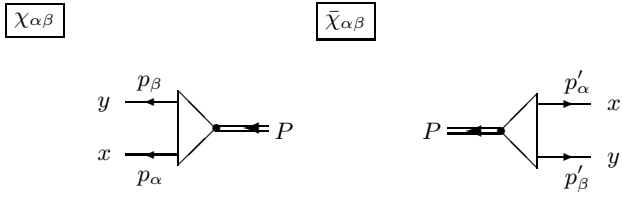

%% file: Appendix.tex
\section{Supplements on Nucleon Charge Conservation} \label{supplCC}

\newcommand{\ampu}{\widetilde{\psi}_M=S^{-1}_M D^{-1}_M \psi_M}
\newcommand{\notampu}{\widetilde{\psi}_E\not =S^{-1}_E D^{-1}_E \psi_E}

\begin{figure*}[t]
\centerline{\parbox{0.6\linewidth}{
\begin{equation*}
\begin{CD}
\int \bar\psi_M D_M^{-1}\Gamma^\mu_{q,M}\psi_M @>\text{Wick rotation}>>
\int\bar\psi_E D_E^{-1}\Gamma^\mu_{q,E}\psi_E \\
@V{\ampu}VV @VV{\notampu}V \\
\int \widetilde{\bar\psi}_M S_M\Gamma^\mu_{q,M}S_M D_M\widetilde{\psi}_M
 @>\text{Wick rotation}>>
\begin{matrix}
\int \widetilde{\bar\psi}_E S_E\Gamma^\mu_{q,E}S_E D_E\widetilde{\psi}_E
\\[2pt] 
\hskip 1.5cm +\hskip .5cm \text{residue terms} 
\end{matrix}
\end{CD}
\end{equation*}}}
\stepcounter{section}
\refstepcounter{figure}
\centerline{\parbox{0.8\linewidth}{\small\textbf{Fig.~\thefigure.} 
Interrelation of matrix elements in Minkowski and Euclidean space.
The integral sign is shorthand for the four-dimensional integration over
the relative momentum $k$, see Eq. (\ref{quark}).\label{box}}}
\end{figure*}
\addtocounter{section}{-1}

The missing step in the explicit verification of the correct charges $Q_P =
1$ for the proton and $Q_N = 0$ for the neutron is to prove that
\begin{eqnarray}
N_q - N_D \, = \, 2 N_X + 3 N_P \label{A1} 
\end{eqnarray} 
which is equivalent to $Q_N = 0$, see Eqs.~(\ref{chargeconds})
to~(\ref{chargesum}) in Sec.~\ref{ElmFFs}. To this end, note that from
Eqs.~(\ref{NqDef}) and~(\ref{NDDef}) the l.h.s above can be written as,
\begin{eqnarray}  
 N_q - N_D &=& \frac{P^\mu}{2M_n} i \int \frac{d^4k}{(2\pi)^4} \\
&& \hskip -1cm 
 \tr \left[ \bar\psi (-k,P) \, \left(\frac{\partial}{\partial
k^\mu} S^{-1}(k_q)  D^{-1}(k_s)  \right) \, \psi(k,P) \right] \nn \\
           &=&  \frac{P^\mu}{2M_n} i \int
 \frac{d^4k}{(2\pi)^4} \nn\\  
&& \hskip -1cm 
\tr \Bigg[ \bar\psi (-k,P) \,
 \left(\frac{\partial}{\partial k^\mu} S^{-1}(k_q)  D^{-1}(k_s) \, \psi(k,P)
 \, \right)    \nn\\ 
  &&  \hskip -.4cm  + \, \left( \frac{\partial}{\partial k^\mu} \bar\psi
 (-k,P) \,  S^{-1}(k_q)  D^{-1}(k_s)   \right) \, \psi(k,P) \; \Bigg] \nn\\
&& \hskip 3.5cm   - \;  \hbox{total derivative}   \; ,\nn
\end{eqnarray} 
with $k_q = \eta P +k $ and $k_s = \hat\eta P - k$ as in the previous
sections. A surface term which vanishes for normalizable BS wave functions
$\psi$ and $\bar\psi$ was not given explicitly. Using the BSEs for
$\widetilde\psi = S^{-1} D^{-1} \psi$ and $\widetilde{\bar\psi} = \bar\psi
S^{-1} D^{-1} $, {\it c.f.}, Eqs. (\ref{BSEpsi}) and (\ref{BSEpsibar}),  
in analogy to the proof of current conservation in Sec.~\ref{TECO}, one
obtains,  
\begin{eqnarray}  
 N_q - N_D &=& \frac{P^\mu}{2M_n} i \int  \frac{d^4p}{(2\pi)^4}
 \,\frac{d^4k}{(2\pi)^4}  \label{A3} \\ 
 && \hskip -1.9cm  \tr \left[ \bar\psi (-p,P)  \left( \frac{\partial}{\partial
p^\mu} K(p,k,P)   +  \frac{\partial}{\partial
k^\mu} K(p,k,P)  \right)  \psi(k,P) \right] . \nn
\end{eqnarray} 
From Eq.~(\ref{x_kern_par}) for the explicit form of the exchange kernel it
follows that   
\begin{eqnarray} 
\left( \frac{\partial}{\partial
p^\mu} \,  + \, \frac{\partial}{\partial k^\mu} \right) \, K(p,k,P) &=& -
\frac{1}{2N_s^2} \label{A4}  \\
&& \hskip -4cm  \Bigg\{ \, 2 \left( \frac{\partial}{\partial q^\mu} S(q)
\right) P(-p_1^2) P(-p_2^2)  
 +  S(q)  \nn\\
&& \hskip -3.7cm 
 \Big( \, 3 p_{1\mu} \, P_1'(-p_1^2) P(-p_2^2)
- \, 3 p_{2\mu} \,  P_1(-p_1^2) P'(-p_2^2) \, \Big) \, \Bigg\} \; ,\nn
\end{eqnarray}  
since $q = (1\!-\!2\eta )P -p-k$, $p_1 =  -(1\!-\!3\eta )P/2 + p +k/2$
and $p_2 =  (1\!-\!3\eta )P/2 - p/2 -k$. Comparing to the definitions of
$N_X$ and $N_P$, Eqs.~(\ref{NXDef}) and~(\ref{NPDef}) in Sec.~\ref{SecNucNorm} 
respectively, we see that Eq.~(\ref{A4}) inserted in Eq.~(\ref{A3}) gives 
(\ref{A1}) as required.

\section{Calculation of the Impulse Approximation Diagrams} 
\label{residue}

\stepcounter{figure}

In this appendix we discuss the difficulties in the formal
transition from the Minkowski to the Euclidean metric which are encountered
in the connection between Bethe-Salpeter amplitudes $\widetilde{\psi}$ and
Bethe-Salpeter wave functions $\psi$ in a general (boosted) frame of
reference of the nucleon bound state.

As the generic example for this discussion, we choose the first diagram in
Figure \ref{IAD} which describes the impulse-approximate contribution arising
from the coupling of the photon to the quark within the nucleon, 
$\langle \widehat J^\mu_{q}\rangle$ according to Eq.~(\ref{Eq:6.10}) with
Eq.~(\ref{jq}). Please refer to Fig.~ \ref{IAD} and Table~\ref{mom_table}   
for the momentum definitions employed herein.

In the Mandelstam formalism, such matrix elements between bound states are 
related to the corresponding BS wave functions in the in Minkowski
space, here to the nucleon BS wave functions $\psi_M$.\footnote{In the
following the subscript $_M$ stands for definitions in Minkowski space and
$_E$ for the corresponding ones in Euclidean space.}  
Upon the transition to the Euclidean metric, the corresponding
contribution to the observable, here to the nucleon form factors, 
is determined by the ``Euclidean BS wave function'' $\psi_E$. In the rest
frame of the nucleon bound state this transfer from $M \to E$ of the BS wave
functions commutes with the replacement of the wave functions by the 
truncated BS amplitudes; that is, unique results are obtained from 
the Euclidean contributions based on either employing the Minkowski space wave
functions or the BS amplitudes which are related by the truncation of the 
propagators of the constituent legs, here
$\widetilde{\psi}_M=S_M^{-1}D_M^{-1}\psi_M$,  or, {\em vice versa},
$\psi_M=S_M D_M\widetilde{\psi}_M$.  

At finite momentum transfer $Q^2$ one needs to employ BS wave functions in a
more general frame of reference, here we use the Breit frame in which neither
the incoming nor the outgoing nucleon are at rest.
As described in Sec. \ref{ElmFFs}, the ``Euclidean wave function'' $\psi_E$
in this frame is obtained from the solution to the BSE in the rest frame by
analytic continuation, in particular, by inserting complex values for the
argument of the Chebyshev polynomials, see Eqs. (\ref{compl_zy}). 
This corresponds to the transition from left to right indicated by the arrow
of the upper line in Figure~\ref{box}.

In the analogous transition on the other hand, when the truncated BS
amplitudes are employed, the possible presence of singularities in the
propagators of legs has to be taken into account explicitly. In the present
example, these are the simple particle poles of the propagators of the 
constituent quark and diquark that might be encircled by the closed path in
the $k_0$--integration. The corresponding residues have to be included in the
transition to the Euclidean metric in this case, which is indicated in the
lower line of Fig.~\ref{box}.  

The conclusion is therefore that the naive relation between BS amplitudes and
wave functions can not be maintained in the Chebyshev expansion of the 
Euclidean spherical momentum coordinates when singularities are encountered
in the truncation of the legs. Resorting to the Min\-kows\-ki space
definitions of BS amplitudes vs. wave functions, however, unique results are
obtained from either employing the domain of holomorphy of the BS wave
functions in the continuation to the Euclidean metric (with complex momenta)
or, alternatively and technically more involved, from keeping track of the
singularities that can occur in the Wick rotation when the truncated
amplitudes and explicit constituent propagators are employed. 

The rest of this section concerns the description of how to account for
these singularities which, for our present calculations employing 
constituent poles for quark and diquark, give rise to residue  
terms as indicated in the lower right corner of Fig. \ref{box}. 

To this end consider the quark contribution to the matrix elements of the
electromagnetic current which, from Eq.~(\ref{Eq:6.10}) with Eq.~(\ref{jq}),
is given by 
\begin{eqnarray} \label{quark}
 \langle \widehat J^\mu_{q} \rangle  &=&  q_q \, \fourint{k}  \\
&& \hskip -.1cm  
\widetilde{\bar\psi}(-k-\hat\eta Q,P')
 \, D(k_s)S(p_{q}) \Gamma^\mu_{quark} S(k_{q}) \widetilde\psi(k,P) \, .\nn
\end{eqnarray}
We are interested in the location of the propagator poles herein. 
For these poles, solving the corresponding quadratic equation for the zeroth
component of the relative momentum $k^0$, from Table~\ref{mom_table} and
Eqs.~(\ref{BF_def}), yields
\begin{eqnarray}
 k_q^2-m_q^2-i\epsilon=0 \;\;  \Leftrightarrow  && \\
&&\hskip -2cm  k^0_{pole,1} \, =\, 
               -\eta \, \omega_Q \, \pm \, W(m_{q},(\vec k -\eta/2 \,\vec
                                           Q)^2 ) 
                                         \nn\\
 p_q^2-m_q^2-i\epsilon=0 \; \; \Leftrightarrow  &&  \label{k2} \\
&&\hskip -2cm  k^0_{pole,2} \,=\,  -\eta\,  \omega_Q
                      \, \pm\, W(m_{q}, (\vec k -(\eta/2-1) \vec Q)^2) \nn\\
 k_s^2-m_s^2-i\epsilon=0 \; \; \Leftrightarrow  && \label{k3}  \\
&&\hskip -2cm   k^0_{pole,3}  \,  = \,  
       \hat\eta\, \omega_Q \, \pm\, W(m_s,(\vec k+ \hat\eta/2 \, \vec Q)^2) 
                   \nn 
\end{eqnarray}
with $W(m,\vec k^2) = \sqrt{\vec k^2+m^2-i\epsilon}$ (and $\eta+\hat\eta =
1$). 

\begin{figure}
\vspace{.6cm}
\hskip .2cm \epsfig{file=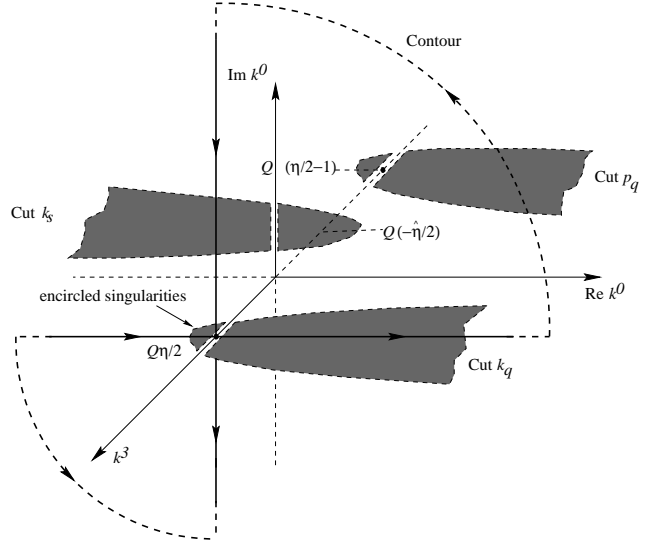,width=0.95\linewidth}
\vspace{.1cm}
\caption{Location of the relevant singularities in the impulse-approximate  
quark contribution to the form factors.
 The relative momentum $k$ is the integration variable in the loop diagram
corresponding to Eq. (\ref{quark}).}
\label{cuts}
\end{figure}

For $Q=0$, {\it i.e.}, in the rest frame of the nucleon in which the
Bethe-Salpeter equation was solved, the naive Wick rotation is justified
for $1- \frac{m_{s}}{M} < \eta < \frac{m_q}{M}$, since there always is a
finite gap between the cuts contained in the hypersurface Re$k^0$=0 of the
Re$k^0$ -- Im$k^0$ -- $\vec k$ space.
As $Q$ increases, these cuts begin are shifted along both,
the $k^3$ and the $k^0$-axis, as sketched in Figure \ref{cuts}. This 
eventually amounts to the effect that one of the two cuts arising from
each propagator crosses the Im $k^0$-axis. As indicated in the figure, 
Wick rotation $k^0 \rightarrow ik^4$ is no longer possible for arbitrary 
values of $k^3$ without encircling singularities. The corresponding residues 
thus lead to 
\begin{eqnarray} \label{residueint}
 \langle\widehat J^\mu_{q}\rangle&\rightarrow& q_q \, \int_{E}  \,
\frac{d^4k}{(2\pi)^4} \\
&& \hskip +.2cm 
 \widetilde{\bar\psi}(-k-\hat\eta Q,P') 
 \, D(k_{s})S(p_{q}) \Gamma^\mu_{q} S(k_{q}) 
 \widetilde\psi(k,P)\nonumber \\
 && \hskip -.9cm  + i \int\frac{d^3\vec k}{(2\pi)^3}  \; \theta_{\vec k} \,
 \widetilde{\bar\psi}(-k^4_{pole,1},-\vec k-\hat\eta \vec Q,P')
  D(k_s) S(p_{q}) \times \nonumber \\
 && \hskip -.1cm \Gamma^\mu_{q} 
 \text{Res}(S(k_{q})) \psi(k^4_{pole,1},\vec k,P) \nn \\[2pt]
 && + \quad\text{analogous terms for $S(p_{q})$ and $D(k_s)$} \nonumber
\end{eqnarray}
upon transforming Eq.~(\ref{quark}) to the Euclidean metric.
Here, the residue integral is evaluated
at the position of the pole in the incoming quark propagator $S(k_q)$
on the Euclidean $k^4$-axis
\begin{eqnarray}
 k_{pole,1}^4= -i\eta \sqrt{M^2+Q^2/4}+iW(m_q,(\vec k-\eta/2\, \vec Q)^2) \, ,
\nn \end{eqnarray}
where $\text{Res}(S(k_{q}))$ denotes the corresponding residue, and the
abbreviation 
\begin{eqnarray}
 \theta_{\vec k}  \equiv \theta\left(\eta \,\omega_Q -
 W(m_q,(\vec k- \eta/2 \, 
 \vec Q )^2)\right) \nn
\end{eqnarray}
was adopted to determine the integration domain for which the encircled
singularities of Figure \ref{cuts} contribute. 

Analogous integrals over the spatial components of the relative momentum
$\vec k$ arise from the residues corresponding to the poles in the outgoing 
quark propagator $S(p_q)$ and the diquark propagator $D(k_s)$ as given in
Eqs. (\ref{k2}) and (\ref{k3}).

\begin{figure}[t]
\vspace{.6cm}
\centerline{\epsfig{file=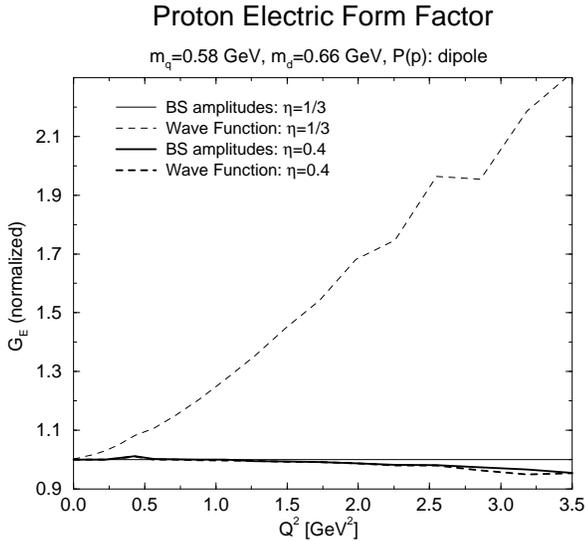,width=2\figwidth}}
\caption{The impulse-approxiamte contribution, corresponding to the diagrams
of Fig.~\protect\ref{IAD}, to the electric form factor of the proton
(employing the dipole diquark amplitude). Results of the BS wave function
calculations for the $\eta$-values $1/3$ and $0.4$ are compared to the
respective amplitude plus residue calculations. The results are normalized to
the latter with $\eta$=1/3.}\label{vw}
\end{figure}

One verifies that these cuts (as represented by the shaded areas
in Fig.~\ref{cuts}) never overlap. Pinching of the deformed contour does not
occur, since there are no anomalous thresholds for spacelike momentum
transfer $Q^2$ in these diagrams.

In Figure \ref{vw} we compare results for the electric form factor of the
proton employing the nucleon BS amplitudes, together with the procedure to
account for the singularities as outlined above,  
with the corresponding wave function calculations. The $n=2$ dipole diquark
amplitude of Sec.~\ref{QDBSE} is employed herein once more.
For $\eta = 1/3$ the Chebyshev expansion of the BS wave function to 9 orders
still turns out insufficient to provide for stable numerical results. This is
due to being too close to the range in $\eta$ that requires proper treatment of
the diquark pole contribution to the nucleon BSE which, with the present
value of\\\vfill\pagebreak\noindent 
the mass $m_s/M=0.7$ is the case for $\eta \le 1 - m_s/M = 0.3$.  
The considerably weaker suppression of higher orders in the Chebyshev
expansion of the BS wave function as compared to the expansion of the 
BS amplitude enhances the residual $\eta$-dependence of the observables
obtained from the former expansion at a given order, in particular, when it
has to reproduce close-by pole contributions in the constituent propagators.   
The impulse-approximate contributions to $G_E$ deviate substantially from
those employing the BS amplitude and residue calculations in this case. 
On the other hand, for values of the momentum partitioning variable which are
a little larger than $1/3 $ such as $\eta=0.4$ used in the other results of
Fig.~\ref{vw}, unique results are obtained from both procedures. 
Both, the BS wave function and amplitude calculation are in perfect
agreement for values of the momentum partitioning that are closer to
the middle of the range allowed to $\eta$.

For the seagull and exchange quark contributions corresponding to the
diagrams of Fig.~\ref{momrout} an analogous analysis of the singularity
structure is considerably more complicated. The explicit inclusion of their
contributions which allowed the calculation based on the BS amplitude 
expansion also of these diagrams is numerically too involved. For these
contributions to the form factors we have to resort to the BS wave function
calculations. Unlike the impulse-approximate contributions we find, however,
that the deviations in the results for the exchange quark and seagull
contributions for $\eta= 1/3$ and $\eta = 0.4$ are smaller than the numerical
accuracy of the calculations and thus negligible.  We attribute this to the
fact that the exchange quark and seagull contributions to the form
factors tend to fall off considerably faster with increasing $Q^2$ than those
of the impulse approximation. This can be seen in Figure~\ref{gedipole}.

Small residual $\eta$-dependences are observed for the momentum
transfers above 3GeV$^2$ also in the otherwise stable calculations. These give
rise to deviations in the results for the form factors of at most 4\% which 
decrease rapidly and become negligible at lower $Q^2$. 

\end{appendix}